\documentclass[11pt]{article}
	
	\newcommand{\blind}{0}
	
    \makeatletter
    \renewcommand\section{\@startsection {section}{1}{\z@}%
                                       {-3.5ex \@plus -1ex \@minus -.2ex}%
                                       {2.3ex \@plus.2ex}%
                                       {\normalfont\fontfamily{phv}\fontsize{16}{19}\bfseries}}
    \renewcommand\subsection{\@startsection{subsection}{2}{\z@}%
                                         {-3.25ex\@plus -1ex \@minus -.2ex}%
                                         {1.5ex \@plus .2ex}%
                                         {\normalfont\fontfamily{phv}\fontsize{14}{17}\bfseries}}
    \renewcommand\subsubsection{\@startsection{subsubsection}{3}{\z@}%
                                        {-3.25ex\@plus -1ex \@minus -.2ex}%
                                         {1.5ex \@plus .2ex}%
                                         {\normalfont\normalsize\fontfamily{phv}\fontsize{14}{17}\selectfont}}
    \makeatother
	
	\usepackage{amsmath}
	\usepackage{graphicx}
	\usepackage{enumerate}
	\usepackage{xcolor}
	\usepackage{natbib} 
	\usepackage{url} 
    \usepackage{appendix}
 \usepackage{hyperref}
\usepackage{bm}
\usepackage{subcaption}
\usepackage{placeins}
\usepackage{setspace,lipsum}
\usepackage{algorithm,algorithmic}
\usepackage[algo2e]{algorithm2e}

\usepackage{lipsum}
\usepackage{amsfonts}
\usepackage{graphicx}
\usepackage{epstopdf}
\usepackage{amssymb}
\ifpdf
  \DeclareGraphicsExtensions{,.pdf,.png,.jpg}
\else
  \DeclareGraphicsExtensions{}
\fi

 \newtheorem{theorem}{Theorem}
 \newtheorem{lemma}{Lemma}
\newtheorem{assumption}{Assumption}
 \newtheorem{proposition}{Proposition}
	
\begin{document}
		
		\def\spacingset#1{\renewcommand{\baselinestretch}%
			{#1}\small\normalsize} \spacingset{1}

		\if0\blind
		{
 
			\title{Sparse Sensor Allocation for Inverse Problems of Detecting Sparse Leaking Emission Sources}
			\author{Xinchao Liu$^a$, Youngdeok Hwang$^c$, Dzung Phan$^b$, Levente Klein$^b$, \\Xiao Liu$^a$ and Kyongmin Yeo$^b$ \\
			$^a$H. Milton Stewart School of Industrial and Systems Engineering,\\ Georgia Institute of Technology, Atlanta, U.S. \\
            $^b$IBM Thomas J. Watson Research Center, Yorktown Heights, U.S.\\
            $^c$Paul H. Chook Department of Information Systems and Statistics,\\ City University of New York, New York, U.S.}
            
			\date{}
			\maketitle
		} \fi
		
		\if1\blind
		{

            \title{Sparse Sensor Allocation for Inverse Problems of Detecting Sparse Leaking Emission Sources}
			\author{Author information is purposely removed for double-blind review}
			
\bigskip
			\bigskip
			\bigskip
			\begin{center}
				{\LARGE\bf \emph{IISE Transactions} \LaTeX \ Template}
			\end{center}
			\medskip
		} \fi
		\bigskip

	\begin{abstract}
This paper investigates the sparse optimal allocation of sensors for detecting sparse leaking emission sources.  Because of the non-negativity of emission rates, uncertainty associated with parameters in the forward model, and sparsity of leaking emission sources, the classical linear Gaussian Bayesian inversion setup is limited and no closed-form solutions are available. By  incorporating the non-negativity constraints on emission rates, relaxing the Gaussian distributional assumption, and considering the parameter uncertainties associated with the forward model, this paper provides comprehensive investigations, technical details, in-depth discussions and implementation of the optimal sensor allocation problem leveraging a bilevel optimization framework. The upper-level problem determines the optimal sensor locations by minimizing the Integrated Mean Squared Error (IMSE) of the estimated emission rates over uncertain wind conditions, while the lower-level problem solves an inverse problem that estimates the emission rates. Two algorithms, including the repeated Sample Average Approximation (rSAA) and the Stochastic Gradient Descent based bilevel approximation (SBA), are thoroughly investigated. It is shown that the proposed approach can further reduce the IMSE of the estimated emission rates starting from various initial sensor deployment generated by existing approaches. 
Convergence analysis is performed to obtain the performance guarantee, and numerical investigations show that the proposed approach can allocate sensors according to the parameters and output of the forward model.  
Computationally efficient code with GPU acceleration is available on GitHub so that the approach readily applicable. 
	\end{abstract}
			
	\noindent%

	\spacingset{1} 

\clearpage

\section{INTRODUCTION}

\subsection{An Overview}

Inverse modeling refers to the inference of unknown parameters of a physical system using observation data 
\citep{Houweling99,chow2008source,stockie2011mathematics,liu2023inverse}.  
Among various types of inverse problems, source estimation is an important class that can be widely found in fugitive methane gas leak source detection \citep{klein2017monitoring}, air pollution source identification \citep{hwang2019bayesian}, 
heat source localization \citep{sinsbeck2017sequential}, etc. Accurate inverse modeling hinges on where observation data are collected and
how sensors are allocated.

\vspace{-6pt}
\begin{figure}[h!]
    \centering
    \includegraphics[width=0.85\linewidth]{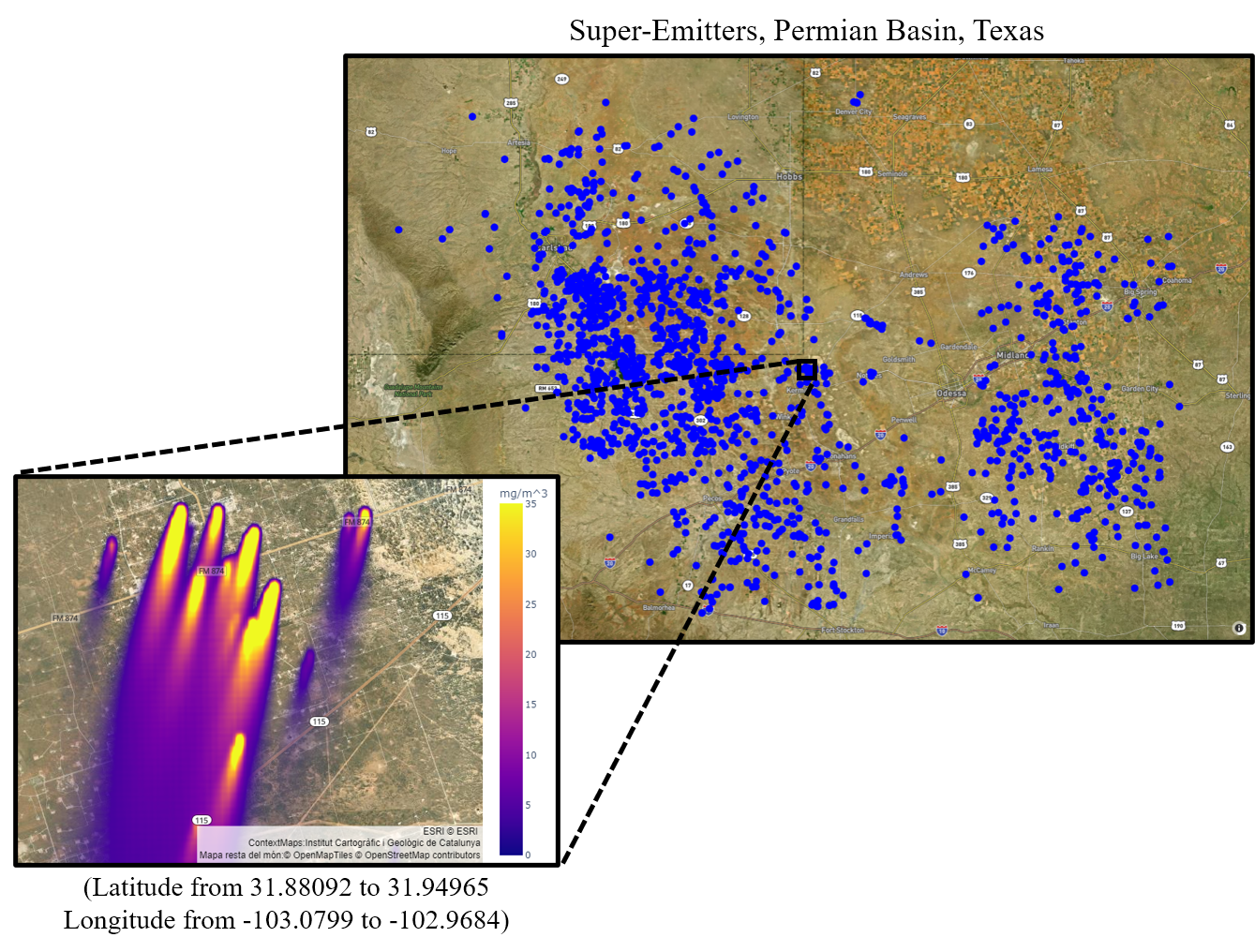}
    \vspace{-8pt}
    \caption{\textcolor{black}{$\sim$1,800 point ``super-emitters" in Permian Basin, Texas with a simulated concentration field using the Gaussian Plume model for a small region with 20 sources (the geo-referenced sources and their emission strengths are obtained from \cite{cusworth2021intermittency})}.}
    \label{fig:Permian}
\end{figure}

\vspace{-8pt}
Very often, such problems share three important characteristics. 

(i) \textit{sparsity in sensor allocation}: the number of sensors that can be placed is far less than the number of potential emission sources, meaning that it is not possible to monitor all sources individually; for example, Figure \ref{fig:Permian} shows nearly 1,800 point ``super-emitters" of methane (CH$_4$) in the Permian Basin, Texas, which cannot be monitored individually \citep{chen2022senet, cusworth2021intermittency}. The first sparsity condition naturally gives rise to an important question: when the number of sensors is far less than the number of sources, how can sensors be effectively allocated so that the emission rates can be estimated for multiple sources as accurately as possible? 

(ii) \textit{sparsity in source estimation}: only a small fraction of potential sources have leaking problems among a large number of potential emission sources. The second sparsity condition requires sparsity-promoting regularizations to be incorporated into the inverse problems where closed-formed solutions are no longer available. 

(iii) \textit{physical constraints and parameter uncertainties}: Physical constraints, such as non-negativity of emission rates, are needed when solving the inverse problem. In addition, parameter uncertainties associated with the \textit{forward} model, such as stochastic wind conditions, also need to be accounted for in sensor allocation (e.g., it makes less sense to allocate too many sensors to the upwind direction of emission sources).  

\vspace{4pt}
The considerations above motivate us to investigate a \textit{bilevel optimization} formulation of  sparse sensor allocation for inverse problems of detecting sparse leaking emission sources. 
In particular, the lower level solves an inverse problem to estimate the emission rates with non-negativity constraints on emission rates (given a candidate sensor allocation plan), whereas
the upper level chooses the sensor locations to minimize the Integrated Mean Squared Error (IMSE) of the estimated emission rates under stochastic wind conditions. A nested structure can be clearly seen, i.e. the objective function at the upper level (lower level) relies on the solutions of the lower-level inverse problem (upper-level sensor allocation). Because of the constrained inverse problem in the lower level as well as the stochastic wind condition in the upper level, the solution of this sensor allocation problem completely relies on numerical approaches, which pose non-trivial computational challenges and raise questions on the effectiveness of the numerical solutions. In this paper, we perform a comprehensive investigation on the technical details and performance of the numerical solutions, generate useful insights on sensor allocation for practice, and provide computer code that enables the users to implement the approach efficiently.   

 \vspace{-6pt}
\subsection{A Review on the Linear Gaussian Bayesian Inversion}\label{sec:LGB}

To better describe the research gaps and contributions of the current work, a review on the Linear Gaussian Bayesian inversion model is firstly presented. It is important to note that the proposed bilevel optimization framework described in this paper extends the linear Bayesian inversion model setup.  

A linear Gaussian Bayesian inverse model considers the following forward model
\begin{equation}
    \bm{\Phi}(\bm{s})=\mathcal{F}(\bm{s})\bm{\theta}+\bm{\epsilon}, \ \bm{\epsilon}\sim \mathcal{N}\big(\bm{0},\bm{\Gamma}_{\bm{\epsilon}}\big),
    \label{e:intro-observation}
\end{equation} where $\bm{\Phi}$ is the sensor observation, $\bm{s}$ represents the sensor locations, $\bm{\theta}$ denotes the emission rates, $\mathcal{F}$ is a linear parameter-to-observation mapping, $\bm{\epsilon}$ is the additive Gaussian noise with zero mean and covariance matrix $\bm{\Gamma}_{\bm{\epsilon}}$. Let $\bm{\theta}\sim \mathcal{N}(\bm{\mu}_{\text{pr}},\bm{\Gamma}_{\text{pr}})$ be the prior distribution of $\bm{\theta}$, it is well-known the posterior distribution of $\bm{\theta}$ is also Gaussian \citep{attia2023}, i.e., $\bm{\theta}\sim \mathcal{N}(\bm{\mu}_{\text{post}},\bm{\Gamma}_{\text{post}})$ where
\begin{equation}
    \begin{split}
        \bm{\mu}_{\text{post}}&=\bm{\Gamma}_{\text{post}}(\mathcal{F}^*(\bm{s})\bm{\Gamma}^{-1}_{\bm{\epsilon}}(\bm{s})\bm{\Phi}(\bm{s})+\bm{\Gamma}^{-1}_{\text{pr}}\bm{\mu}_{\text{pr}})\\
        \bm{\Gamma}_{\text{post}}&=(\mathcal{F}^*(\bm{s})\bm{\Gamma}^{-1}_{\bm{\epsilon}}(\bm{s})\mathcal{F}(\bm{s})+\bm{\Gamma}^{-1}_{pr})^{-1}
    \end{split}
    \label{e:intro-posterior}
\end{equation}
and $\mathcal{F}^*(\bm{s})$ is the adjoint of $\mathcal{F}$, and $\mathcal{F}^*(\bm{s})=\mathcal{F}^T(\bm{s})$ for linear operators. 
It is important to note that, the posterior covariance matrix $\bm{\Gamma}_{\text{post}}$ depends on the sensor location $\bm{s}$, but is independent from the prior mean and data from the forward model. Hence, the optimal sensor locations $\bm{s}$ can be chosen to 
minimize the posterior uncertainty; for example, the $A$-optimal design for the linear Gaussian case maximizes the trace of the inverse of covariance $\bm{\Gamma}_{\text{post}}$, i.e., $\bm{s}^* = \text{argmax}_{\bm{s}}\{\text{trace}(\mathcal{F}^*(\bm{s})\bm{\Gamma}^{-1}_{\bm{\epsilon}}(\bm{s})\mathcal{F}(\bm{s})+\bm{\Gamma}^{-1}_{pr})\}$. 

Let $\hat{\bm{\theta}}_{MAP}(\bm{\Phi},\bm{s})=\bm{\mu}_{\text{post}}$ be the Maximum A Posteriori (MAP) estimator of $\bm{\theta}$, the IMSE is given by 
\begin{equation}
        \Psi(\bm{s})=\mathbb{E}_{\bm{\theta}}\Big\{\mathbb{E}_{\bm{\Phi}|\bm{\theta}}\Big\{\Big\|\bm{\mu}_{\text{post}}(\bm{s})-\bm{\theta}\Big\|^2_2\Big\}\Big\}.
    \label{e:intro-risk 1}
\end{equation}

As shown in Appendix \ref{A:proof A-optimal}, the IMSE above admits a closed-form expression which can be efficiently evaluated
\begin{equation}
\Psi(\bm{s})=\Big\|\bm{\Gamma}_{\text{post}}\bm{L}^T\Big\|^2_F + \Big\|\bm{\Gamma}_{\text{post}}\mathcal{F}^*\bm{U}^T\Big\|^2_F
    \label{e:intro-risk 2}
\end{equation} 
where $\bm{L}^T\bm{L}=\bm{\Gamma}^{-1}_{pr}$, $\bm{U}^T\bm{U}=\bm{\Gamma}^{-1}_{\bm{\epsilon}}$, and $||\cdot||_F$ is the Frobenious matrix norm. 

\vspace{8pt}
However, for the inverse problem of detecting sparse leaking sources for environmental processes, the well-established $A$-optimal design has the following limitations:

$\bullet$ Firstly, the Gaussian assumption on $\bm{\theta}$ is not appropriate for modeling the uncertainty associated with emission rate which cannot take negative values. For sparse source detection, the majority of emission sources have zero or near-zero emission rates, and it is inappropriate to place a Gaussian prior for $\bm{\theta}$ centered around zero. When the Gaussian assumption is violated, the closed-form solution (\ref{e:intro-posterior}) is no longer valid. 

$\bullet$ Secondly, for the linear Bayesian inverse problem, the Maximum A Posteriori (MAP) estimator of $\bm{\theta}$, $\hat{\bm{\theta}}_{MAP}(\bm{\Phi},\bm{s})=\bm{\mu}_{\text{post}}$, is often adopted. For sparse source detection, however, this MAP estimator does not account for the sparsity of leaking sources when the majority of emission sources should have zero or near-zero emission rates. Sparsity-promoting regularizations are needed. It is also important to note that the inverse problem can be easily ill-posed given certain sensor layout, which is another reason why  regularizations are often added, e.g., the tightly coupled sets of variables \citep{herring2018lap}, the $L_1$-type prior \citep{wang2017bayesian}, the goal-oriented inversions \citep{spantini2017goal,wu2023offline}, the total variation regularization \citep{shen2002mathematical}, the fractional Laplacian \citep{antil2020bilevel}, and 
the Tikhonov regularization \citep{willoughby1979solutions,tarantola2005inverse,golub1999tikhonov}. For linear inverse problems with a squared loss function, adding the Tikhonov regularization also yields a closed-form design \citep{tarantola2005inverse,haber2009numerical,haber2012numerical}. 

$\bullet$ Finally, the classical linear Gaussian Bayesian setup above does not consider the parameter uncertainty associated with the parameter-to-observation mapping $\mathcal{F}$ in (\ref{e:intro-observation}). In practice, $\mathcal{F}$ is the forward dispersion model that depends on parameters which are rarely known precisely; for example, wind speed and direction. Parameters, like wind, are always associated with a high degree of uncertainty that could significantly alter the solution of the optimization problem. Incorporating such uncertainty into the objective function is needed to obtain more effective sensor deployment plans. 

Addressing the three issues above immediately requires extensions of the existing linear Gaussian Bayesian model setup and the standard $A$-optimal design. It also implies that numerical approaches are needed and one would need to investigate if the numerical approaches can efficiently generate meaningful solutions.

\subsection{Other Related Work}

In this section, we provide a review on other related work for sensor allocation. 
Considering discrete spatial domains, \cite{manohar2021optimal} formulated the optimal sensor allocation  as a sensor selection problem for which the best subset of sensor locations is chosen from a discrete set of potential candidates. 
This approach is closely related to the $D$-optimal design \citep{joshi2008sensor};  
for example, \cite{krause2008near} maximized the mutual information between the chosen and unselected locations, \cite{ranieri2014near} used a greedy algorithm to minimize a $D$-optimal proxy of the mean squared error, and \cite{wu2023offline} proposed a swapping greedy algorithm to minimize the expected information gain. 
Due to the combinatorial nature of the sensor selection problem, convex optimization \citep{joshi2008sensor} and heuristics \citep{yu2018scalable} have also been utilized. \cite{alexanderian2014optimal} used the $L_0$ regularization while casting the sensor placement for a Bayesian inverse problem as an $A$-optimal design problem. \cite{ruthotto2018optimal} used two separate optimal experimental design formulations to firstly determine the number of sensors with sparsity promoting regularizations, and then seek the optimal sensor locations using a relaxed interior point method. 

Considering continuous spatial domains, \cite{chepuri2014continuous} augmented the grid-based sensor allocation with continuous variables to allow off-grid sensor placement. \cite{huan2013simulation,huan2014gradient} developed gradient-based stochastic optimization methods to maximize the expected information gain while approximating forward models with polynomial chaos expansion. \cite{sharrock2022joint} presented a two-timescale continuous-time stochastic gradient descent algorithm to minimize the MSE of hidden state estimates. 

Note that the continuous-domain design problem can sometimes be converted to a discrete-domain design problem by discretizing the continuous domain and leveraging the existing  open-source tools for discrete problems, such as the `Chama' software for sensor placement optimization using impact metrics \citep{klise2017sensor}, the `Polire' software for spatial interpolation and sensor placement \citep{narayanan2020toolkit}, the `PySensors' software for selecting and placing a sparse set of sensors for classification and signal reconstruction \citep{de2021pysensors,brunton2016sparse,manohar2018data}. 


\subsection{Contributions of this Work}

To address the limitations of the linear Gaussian Bayesian inversion model and $A$-optimal design described in Section \ref{sec:LGB},  this work performs a comprehensive investigation of a bilevel optimization framework for sparse sensor allocation problem of detecting sparse emission sources. In particular, 

$\bullet$ Parameter uncertainties (e.g., stochastic wind conditions) are incorporated in the upper-level objective that involves the minimization of the overall IMSE of the inverse estimator of emission rates under stochastic wind conditions and a candidate sensor allocation plan. As a result, the closed-form expression (\ref{e:intro-risk 2}) no longer exists and the evaluation of the objective function can only be done numerically, e.g., using Sampling Average Approximation, which significantly increase the complexity of the problem. 

$\bullet$ Physical constraints (i.e., non-negative emission rates) and sparsity-promoting regularizations (i.e., elastic net) are incorporated in the lower-level inverse problem tailored for detecting sparse leaking emission sources. As a result, the closed-form solution (\ref{e:intro-posterior}) no longer exists \citep{liu2023inverse, zou2005regularization,yeo2019development}. To the best of our knowledge, there exists no prior work that explicitly tackles the constrained bilevel optimization for the optimal sensor placement problem with physically constrained and elastic-net regularized inversion estimators. 

$\bullet$ With the parameter uncertainty, physical constraints and regularizations, neither the upper-level nor the lower-level optimization problems have closed-form solutions. Hence, this paper performs theoretical convergence analysis of the stochastic optimization algorithms and shows the theoretical performance guarantee. The closed-form expression of the gradient of the upper-level objective function with respect to sensor locations is derived through chain rules.

$\bullet$ The paper presents comprehensive numerical results that generate meaningful insight for the sensor allocation problem of interest. In particular, because the bilevel optimization problem for sensor placement is usually non-convex, the solution of a first-order algorithm strongly depends on the initial design. This paper also  investigates and compares different approaches to find appropriate starting points for the stochastic optimization algorithm. The combination of appropriate initial sensor allocation (e.g. density-based space-filling design) and the proposed bilevel optimization provides a practical solution to sensor placement problems.

$\bullet$ Finally, computer code with GPU acceleration is made available to users. For the sensor allocation problem that requires computationally heavy algorithms, it is unrealistic for engineers to adopt the solution unless code is provided.

\vspace{8pt}
The remainder of this paper is organized as follows. Section \ref{s:Methodology} presents the inverse modeling and the bilevel optimization problem. Section \ref{s:bilevel} investigates two optimization algorithms for solving the proposed bilevel optimization problem and presents the convergence analysis. Numerical examples are presented in Section \ref{s:numerical} to demonstrate the performance of the proposed approach. 
Conclusions and discussions on future research are presented in Section \ref{s:conclusion}. All proofs and lengthy derivations are provided as \textbf{supplemental online materials}. 

\vspace{-10pt}
\section{Problem Setup: Optimal Sensor Allocation}\label{s:Methodology}
\vspace{-6pt}
Let $\Omega\subset \mathbb{R}^2$ be a two-dimensional rectangular spatial domain. Within $\Omega$, there exist $N_p$ \textit{potential} emission sources with known locations but unknown emission rates. Let $\theta_i\ge 0$ be the emission rate for the $i$th source, and let
$\bm{\theta}=(\theta_1, ..., \theta_{N_p})$.
Each source can only have one of two states: constant leaking (i.e. a constant and higher-than-normal emission rate) or no leaking (i.e. a constant background emission rate $\mu$ under normal operation). Without loss of generality, the background emission rate $\mu$ is set to zero throughout this paper. 
We are interested in the optimal allocations of $n$ sensors for detecting abnormal emission sources, and the number of sensors is less than (usually far less than) the number of potential sources, i.e. $n<N_p$, giving rise to the optimal sensor allocation problem. 

The forward model, which generates a steady concentration field, is given as follows,
\begin{equation}
\bm{\Phi}=G(\bm{\theta},\bm{\beta},\bm{s})+\bm{\epsilon},\quad \bm{\epsilon}\sim \mathcal{N}(\bm{0},\bm{\Gamma}_{\epsilon}),
    \label{physics}
\end{equation}

\noindent where $\bm{\Phi}\in\mathbb{R}^n$ is a vector that contains the observations from $n$ sensors, $G$ is a forward physical dispersion model, $\bm{\beta}\in\mathbb{R}^2$ is the wind vector, $\bm{s} = (\bm{s}_1,\bm{s}_2,...,\bm{s}_n)$ is the location of $n$ sensors with $\bm{s}_i=(X_{i},Y_{i})$, and $\bm{\epsilon}\sim \mathcal{N}(\bm{0},\bm{\Gamma}_{\epsilon})$ is the observation noise with $\bm{\Gamma}_{\epsilon} = \sigma^2_{\epsilon}\bm{I}$. 

In this paper, $\bm{s}$ is the decision variable and the decision space is defined by $\Omega^s=\{\bm{s}\in \Omega:\bm{s}_{L}\leq\bm{s}\leq\bm{s}_{H}\}$, where $\bm{s}_{L}$ and $\bm{s}_{H}$ respectively represent the lower and upper bounds within which sensors can be placed. 
Suppose that the emission rates can be estimated from all $N_p$ sources, $\hat{\bm{\theta}}(\bm{\Phi},\bm{\beta},\bm{s})$, the optimal $\bm{s}$ is found by minimizing the IMSE averaged over stochastic wind and emission scenarios
\begin{equation} 
    \begin{split}
        \Psi(\bm{s})&=\mathbb{E}_{\bm{\theta},\bm{\beta}}\{\mathbb{E}_{\bm{\Phi}|\bm{\theta},\bm{\beta}}\{\|\hat{\bm{\theta}}(\bm{\Phi},\bm{\beta},\bm{s})-\bm{\theta}\|^2_2\}\}\\
        &=\iiint \|\hat{\bm{\theta}}(\bm{\Phi},\bm{\beta},\bm{s})-\bm{\theta}\|^2_2 \cdot p(\bm{\Phi}|\bm{\theta},\bm{\beta},\bm{s})
         p(\bm{\theta})p(\bm{\beta})d\bm{\Phi}d\bm{\theta}d\bm{\beta},
    \end{split}
    \label{e:risk}
\end{equation}
where $p(\bm{\beta})$ and $p(\bm{\theta})$ are the prior distributions of $\bm{\beta}$ and $\bm{\theta}$. In practice, prior knowledge on $\bm{\beta}$ can be obtained from historical data or numerical weather predictions, while prior knowledge on $\bm{\theta}$ is elicited from domain experts on possible leaking scenarios. 


Direct evaluation of the objective function (\ref{e:risk}) is computationally challenging. An effective approach is to approximate (\ref{e:risk}) using Monte Carlo sample averaging as follows
\begin{equation}
\hat{\Psi}_N(\bm{s})=N^{-1}\sum_{i=1}^{N}\|\hat{\bm{\theta}}^{(i)}(\bm{s})-\bm{\theta}^{(i)}\|^2_2
    \label{e:approximate objective}
\end{equation}

\noindent where $\bm{\theta}^{(i)}$, $\bm{\beta}^{(i)}$ and $\bm{\Phi}^{(i)}$, $i=1,2,\cdots,N$, are respectively sampled from 
$p(\bm{\theta})$, $ p(\bm{\beta})$, and $ p(\bm{\Phi}|\bm{\theta}^{(i)},\bm{\beta}^{(i)},\bm{s})$, and $\hat{\bm{\theta}}^{(i)}(\bm{s})$ is the estimated $\bm{\theta}$ from $\bm{\Phi}^{(i)}$ given $\bm{s}$.

The evaluation of (\ref{e:approximate objective}) requires the estimated emission rate $\bm{\theta}^{(i)}$ from data (i.e., solving the inverse model). In this paper, we obtain $\hat{\bm{\theta}}^{(i)}$ by minimizing an elastic net loss: 
\begin{equation}
L(\bm{\theta}) = \frac{1}{2}\| G(\bm{\theta},\bm{\beta}^{(i)},\bm{s}) - \bm{\Phi}^{(i)}\|^2_{\bm{\Gamma}_{\epsilon}} 
        + \lambda_1\|\bm{\theta}\|^2_{2} + \lambda_2\|\bm{\theta}\|_{1} \text{    s.t. }\bm{\theta} \geq \bm{0},
\label{e:Inverse modeling 1}
\end{equation}
where $\| \bm{x} \|^2_{\bm{\Gamma}_{\epsilon}} = \sigma^{-2}_{\epsilon}\bm{x}^T\bm{x}$ for some vector $\bm{x}$, and $\lambda_1$ and $\lambda_2$ are the hyperparameters. 
The minimization of (\ref{e:Inverse modeling 1}) yields an MAP estimate given a prior distribution, $p(\bm{\theta};\lambda_1,\lambda_2) \propto \exp(- \lambda_1\|\bm{\theta}\|^2_{2}- \lambda_2\|\bm{\theta}\|_{1})$ for $\bm{\theta}\ge \bm{0}$ \citep{ruthotto2018optimal}. Because emission rates are non-negative, this prior distribution incorporates the truncated Gaussian ($\lambda_2=0$) and truncated Laplacian ($\lambda_1=0$) so that the prior information on $\bm{\theta}$ can be flexibly captured. The posterior distribution of $\bm{\theta}$ is given by
\begin{equation}
p(\bm{\theta}|\bm{\Phi}^{(i)},\bm{\beta}^{(i)},\bm{s})\propto p(\bm{\Phi}^{(i)}|\bm{\theta},\bm{\beta}^{(i)},\bm{s})\cdot p(\bm{\theta}|\bm{\beta}^{(i)},\bm{s})=p(\bm{\Phi}^{(i)}|\bm{\theta},\bm{\beta}^{(i)},\bm{s})\cdot p(\bm{\theta}).
\end{equation}

Because $\text{log}(p(\bm{\theta}))=c- \lambda_1\|\bm{\theta}\|^2_{2}- \lambda_2\|\bm{\theta}\|_{1}$ for $\bm{\theta}\ge \bm{0}$ with $c$ being a constant, the MAP estimate is obtained by maximizing
\begin{equation}
-\frac{1}{2}\| G(\bm{\theta},\bm{\beta}^{(i)},\bm{s}) - \bm{\Phi}^{(i)}(\bm{s})\|^2_{\bm{\Gamma}_\epsilon} 
        - \lambda_1\|\bm{\theta}\|^2_{2}- \lambda_2\|\bm{\theta}\|_{1} \textrm{ s.t. } \bm{\theta} \geq \bm{0}.
\label{MAP proof}
\end{equation}

For a linear forward process, 
$G(\bm{\theta},\bm{\beta},\bm{s}) = \mathcal{F}(\bm{\beta},\bm{s})\bm{\theta}$ 
where $\mathcal{F}(\bm{\beta},\bm{s})$ is a function of the wind vector $\bm{\beta}$ and sensor location $\bm{s}$, we obtain from (\ref{e:risk}), (\ref{e:approximate objective}) and (\ref{MAP proof}) a 
bilevel optimization problem 
\begin{subequations}
\begin{align}
        \min_{\bm{s}\in \Omega^s} \hat{\Psi}_N(\bm{s})=N^{-1}&\sum_{i=1}^{N}\|\hat{\bm{\theta}}^{(i)}(\bm{s})-\bm{\theta}^{(i)}\|^2_2
\label{optimization empirical}\\
        \text{s.t.}~     \hat{\bm{\theta}}^{(i)}(\bm{s})=\underset{\bm{\theta}}{\text{argmin}}\big\{\frac{1}{2}\|\mathcal{F}(\bm{\beta}^{(i)},\bm{s})\bm{\theta}-&\bm{\Phi}^{(i)}\|^2_{\bm{\Gamma}_{\epsilon}}+\lambda_1\|\bm{\theta}\|^2_{2} + \lambda_2\|\bm{\theta}\|_{1}: \bm{\theta}\ge\bm{0}\big\}, \quad i=0,\cdots,N-1.
        \label{optimization empirical 2}
\end{align}
\label{e:our problem}
\end{subequations}

The evaluation of the upper-level objective (\ref{optimization empirical}) requires the solution of the lower-level inverse problem (\ref{optimization empirical 2}), which is a convex Quadratic Programming (QP):
\begin{equation}
   \hat{\bm{\theta}}^{(i)}(\bm{s})=\underset{ \bm{\theta}}{\text{argmin}}\Big\{\frac{1}{2} \bm{\theta}^{T}\bm{C}^{(i)}\bm{\theta}+(\bm{d}^{(i)})^T\bm{\theta}:  \bm{\theta}\ge\bm{0}\Big\},
\label{optimization 3}
\end{equation}
where $\bm{C}^{(i)}:=\bm{C}^{(i)}(\bm{s})=\sigma^{-2}_{\epsilon}\mathcal{F}^*(\bm{\beta}^{(i)},\bm{s})\mathcal{F}(\bm{\beta}^{(i)},\bm{s})+\lambda_1\bm{I}$ is a $N_p \times N_p$ matrix, 
$\bm{d}^{(i)}:=\bm{d}^{(i)}(\bm{s})=\lambda_2\bm{1}-\sigma^{-2}_{\epsilon}\mathcal{F}^*(\bm{\beta}^{(i)},\bm{s})\bm{\Phi}^{(i)}(\bm{s})$ is a $N_p \times 1$ column vector, 
$\mathcal{F}^*$ is the complex conjugate transpose, 
and $\bm{1}$ is a $N_p$-dimensional column vector of ones.

\section{Algorithms and  Performance Analysis} \label{s:bilevel}
\vspace{-6pt}
With the parameter uncertainty, physical constraints and regularizations, neither the upper- nor the lower-level optimization problems have closed-form solutions. When $N$ is large, the computational cost of the bilevel optimization problem (\ref{e:our problem}) is non-trivial. This section investigates the repeated Sample Average Approximation (rSAA) and Stochastic Gradient Descent based bilevel approximation (SBA),  and performs theoretical convergence analysis. 
For the rSAA algorithm, we note that the global optimality is possible using existing global solvers, but it may not be ideal for large-scale problems. For the SBA algorithm, it well handles large-scale problems with parallel computing, but good initial guesses are needed due to the local solver in nature.


\vspace{-16pt}
\singlespacing
\begin{algorithm}[htb]
 \caption{Repeated SAA (rSAA) for Sensor Allocation Problem}
 \textbf{Initialization} $\{\tilde{\bm{s}}^k_{\tilde{N}}\in \Omega^s\}_{k=0}^{K-1}$ and a relatively small $\tilde{N}$
 
\For(\hfill\tcp*[f]{$K$ repeated runs \textcolor{black}{(in parallel)}}){$k=0,1,\cdots,K-1$}{
\textbf{Sample} $\{\bm{\theta}^{(i)},\bm{\beta}^{(i)},\bm{\Phi}^{(i)}\}_{i=1,\cdots,\tilde{N}}$

\textcolor{black}{\textbf{Call a \textit{global} solver for the deterministic bilevel optimization problem} $k$}

\textbf{Save} 
{$\hat{\bm{s}}^k_{\tilde{N}}$, $\hat{\Psi}^k_{\tilde{N}}:=\hat{\Psi}_{\tilde{N}}(\hat{\bm{s}}^k_{\tilde{N}})$}}

\textbf{Set} {$\hat{\bm{s}}_{\tilde{N}} =g(\hat{\bm{s}}^0_{\tilde{N}}, \hat{\bm{s}}^1_{\tilde{N}},\ldots,\hat{\bm{s}}^{K-1}_{\tilde{N}})$} (see Section \ref{s:update outer solution}) \hfill\tcp*[f]{\text{final output}}

\textbf{Return} 
$\hat{\bm{s}}_{\tilde{N}}$
  \label{algorithm 1}
 \end{algorithm}

\spacingset{1}

\vspace{-12pt}
The rSAA algorithm is summarized in Algorithm \ref{algorithm 1}. This approach involves $K$ parallel runs for $k=0,1,\cdots, K-1$. \textcolor{black}{Each run only solves a corresponding deterministic bilevel optimization problem using only a small number of $\tilde{N}$ Monte Carlo samples to speed up the computation ($\tilde{N}\ll N$).} The outputs from the $K$ repeated runs are eventually combined to obtain the final solution. Note that

$\bullet$ 
\textcolor{black}{For the $k$th run, the corresponding deterministic bilevel optimization problem is solved by a global solver \citep{liu2022optimal}. The optimal sensor location $\hat{\bm{s}}^k_{\tilde{N}}$ is found and the objective function $\hat{\Psi}^k_{\tilde{N}}:=\hat{\Psi}_{\tilde{N}}(\hat{\bm{s}}^k_{\tilde{N}})$ is evaluated for the $k$th run.}   

$\bullet$ 
After the $K$ repeated runs, the final optimal sensor location $\hat{\bm{s}}_{\tilde{N}}$ is determined from  $\hat{\bm{s}}^0_{\tilde{N}}, \hat{\bm{s}}^1_{\tilde{N}},...,\hat{\bm{s}}^{K-1}_{\tilde{N}}$. The selection of the final optimal sensor location $\hat{\bm{s}}_{\tilde{N}}$ from $\hat{\bm{s}}^0_{\tilde{N}}, \hat{\bm{s}}^1_{\tilde{N}},\ldots,\hat{\bm{s}}^{K-1}_{\tilde{N}}$ is given by a function $g$. In this paper, $g$ is chosen as the mean of $\hat{\bm{s}}^0_{\tilde{N}}, \hat{\bm{s}}^1_{\tilde{N}},\ldots,\hat{\bm{s}}^{K-1}_{\tilde{N}}$, while other choices are possible. For the selection of a sufficiently large $K$, we leave the details to Section \ref{s:convergence}.

\vspace{-16pt}
\singlespacing
\begin{algorithm}[H]
 \caption{The SGD-based Bilevel Approximation Method (SBA)}
 \textbf{Initialization} $\tilde{\bm{s}}_{\tilde{N},0}\in \Omega^s$, $\{\hat{\bm{\theta}}_{m,0}^{(i)}\in \mathbb{R}^+,\hat{\bm{\eta}}_{m,0}^{(i)}\in \mathbb{R}^+\}_{i=0,\cdots,\tilde{N}-1;m=0,\cdots,M-1}$, and the small $\tilde{N}$
 
 \For(\hfill\tcp*[f]{\textcolor{black}{upper-level problem}}) {$m = 0,1,\cdots,M-1$}{   
  {\textbf{Sample} $\{\bm{\theta}^{(i)},\bm{\beta}^{(i)},\bm{\Phi}^{(i)}\}_{i=1,\cdots,\tilde{N}}$\\
  \For(\hfill\tcp*[f]{\textcolor{black}{lower-level} problem \textcolor{black}{(in parallel)}}) {$i = 0,1,\cdots,\tilde{N}-1$}{
 \For {$j = 0,1,\cdots,J-1$}{
 \textbf{Update} $\hat{\bm{\theta}}_{m,j+1}^{(i)}, \hat{\bm{\eta}}_{m,j+1}^{(i)} \leftarrow \hat{\bm{\theta}}_{m,j}^{(i)}, \hat{\bm{\eta}}_{m,j}^{(i)}$ (see Section \ref{s:update lower-level solution})
  }}
   \textbf{Update} $\tilde{\bm{s}}_{\tilde{N},m+1} \leftarrow \tilde{\bm{s}}_{\tilde{N},m}$
   (see Section \ref{s:update outer solution})
}

\textbf{Set} 
{$\hat{\bm{s}}_{\tilde{N}} := \tilde{\bm{s}}_{\tilde{N},M}$} \hfill\tcp*[f]{final output}
}
\textbf{Return} $\hat{\bm{s}}_{\tilde{N}}$
 \label{algorithm 2}
 \end{algorithm}
 
\spacingset{1}

The SBA algorithm is summarized in Algorithm \ref{algorithm 2}. 
Unlike the rSAA, this algorithm requires only one run. Note that,

$\bullet$ 
Initial sensor locations $\tilde{\bm{s}}_{\tilde{N},0}$ are needed for initialization. Here, the first subscript $\tilde{N}$ is the number of Monte Carlo samples and the second subscript is the index of the \textcolor{black}{upper-level} iteration (``0'' corresponds to the initial setting).

$\bullet$ The upper-level optimization requires $M$ iterations ($m=0,1,\cdots,M-1$), and each iteration involves solving $\tilde{N}$ lower-level problems ($i=0,1,\cdots,\tilde{N}-1$).
Following the idea of stochastic approximation \citep{nemirovski2009robust}, the $\tilde{N}$ Monte Carlo samplings $\{\bm{\theta}^{(i)},\bm{\beta}^{(i)},\bm{\Phi}^{(i)}\}_{i=1,\cdots,\tilde{N}}$ are re-sampled for each upper-level iteration $m$. 

$\bullet$  The lower-level optimization requires $J$ iterations ($j=0,1,\cdots,J-1$) to update the estimated emission rate $\hat{\bm{\theta}}_{m,j+1}^{(i)}$ and its Lagrangian multiplier $\hat{\bm{\eta}}_{m,j+1}^{(i)}$ (see Section \ref{s:update lower-level solution}). Once the lower-level problem has been solved, each upper-level iteration updates the sensor locations $\tilde{\bm{s}}_{\tilde{N},m+1} \leftarrow \tilde{\bm{s}}_{\tilde{N},m}$ 
   (see Section \ref{s:update outer solution}). After the $M$ upper-level iterations, the optimal sensor location $\hat{\bm{s}}_{\tilde{N}} := \tilde{\bm{s}}_{\tilde{N},M}$ is found. 


In the following Sections 3.1$\sim$3.2, we provide technical details required for implementing Algorithms 2. 

\vspace{-8pt}
\subsection{Computational Details of $\hat{\bm{\theta}}_{m,j+1}^{(i)}$ and $\hat{\bm{\eta}}_{m,j+1}^{(i)}$ for the Lower-Level Problem}\label{s:update lower-level solution}

When solving the \textcolor{black}{lower-level} problem, Algorithms 2 requires the update of $\hat{\bm{\theta}}_{m,j+1}^{(i)}$ and $\hat{\bm{\eta}}_{m,j+1}^{(i)}$. For any $i=0,1,\cdots, \tilde{N}-1$,  the Lagrangian of the \textcolor{black}{lower-level} problem is 
\begin{equation}
\begin{split}
       h(\bm{s}, \bm{\theta}, \bm{\eta}) = \frac{1}{2}\bm{\theta}^T\bm{C}^{(i)}\bm{\theta}+(\textcolor{black}{\bm{d}^{(i)}})^T\bm{\theta} - \bm{\eta}^T\bm{\theta}
\end{split}
\end{equation}

\noindent
with the KKT conditions $\bm{C}^{(i)}\bm{\theta}+\textcolor{black}{\bm{d}^{(i)}}-\bm{\eta}=\bm{0}$, $\bm{\theta}, \bm{\eta} \geq \bm{0}$, and $\bm{\eta}\bm{\theta}=\bm{0}$.
The augmented primal-dual gradient algorithm can be employed to solve the \textcolor{black}{lower-level} QP problem by defining the augmented Lagrangian as \citep{meng2020aug}: 
 \begin{equation}
     h_{\gamma}(\bm{s},\bm{\theta},\bm{\eta}) = \frac{1}{2}\bm{\theta}^T\bm{C}^{(i)}\bm{\theta}+(\textcolor{black}{\bm{d}^{(i)}})^T\bm{\theta} + \sum_{b=1}^{N_p}\frac{[\gamma(-\theta_b)+\eta_b]_{+}^2 - \eta_b^2}{2\gamma},
     \label{e:Lagrangian}
 \end{equation}
 where $\gamma$ is a penalty parameter, $\theta_b$ the $b$th entry of $\bm{\theta}$, and $\eta_b$ the $b$th entry of $\bm{\eta}$.
\textcolor{black}{
\begin{proposition}
    The gradient of the augmented Lagrangian (\ref{e:Lagrangian}) with respect to $\bm{\theta}$ and $\bm{\eta}$ can be obtained as
 \begin{equation}
 \begin{split}
          \nabla_{\bm{\theta}}h_{\gamma}(\bm{s}, \bm{\theta}, \bm{\eta})&=\bm{C}^{(i)}\bm{\theta} + \textcolor{black}{\bm{d}^{(i)}} - \sum_{b=1}^{N_p}[\gamma(-\theta_b)+\eta_b]_{+}\bm{e}_b^T\\
          \nabla_{\bm{\eta}}h_{\gamma}(\bm{s}, \bm{\theta}, \bm{\eta})&=\sum_{b=1}^{N_p}\frac{1}{\gamma}([\gamma(-\theta_b)+\eta_b]_{+}-\eta_b)\bm{e}_b^T
 \end{split}
 \end{equation}
 where $\bm{e}_b$ is an $N_p$-dimensional row vector with the $b$th entry being 1 and 0 otherwise. 
 \label{p:augmented Lagrangian}
\end{proposition}}

Finally, $\hat{\bm{\theta}}_{m,j+1}^{(i)}$ and $\hat{\bm{\eta}}_{m,j+1}^{(i)}$ are updated as
\begin{equation}
    \begin{split}
&\hat{\bm{\theta}}_{m,j+1}^{(i)}=[\hat{\bm{\theta}}_{m,j}^{(i)}-\tau_{m,j}\nabla_{\bm{\theta}}h_{\gamma}(\tilde{\bm{s}}_{\tilde{N},m},\hat{\bm{\theta}}_{m,j}^{(i)},\hat{\bm{\eta}}_{m,j}^{(i)})]_+\\
&\hat{\bm{\eta}}_{m,j+1}^{(i)}=[\hat{\bm{\eta}}_{m,j}^{(i)}+\tau_{m,j}\nabla_{\bm{\eta}}h_{\gamma}(\tilde{\bm{s}}_{\tilde{N},m},\hat{\bm{\theta}}_{m,j}^{(i)},\hat{\bm{\eta}}_{m,j}^{(i)})]_+
    \end{split}
\end{equation}
where $\tau_{m,j}$ is the stepsize, and $[x]_+=x$ if $x\geq 0$ and $[x]_+=0$  if  $x< 0$. 

\subsection{Computational Details of $\tilde{\bm{s}}_{\tilde{N},m+1}$ for the Upper-Level Problem}\label{s:update outer solution}
The upper-level problem requires updating the sensor locations $\bm{s}$ given the solution of the lower-level problem. Hence, the hypergradient, i.e., the gradient of the upper-level objective function with respect to the sensor locations, is needed. Because the upper-level objective function depends on the solution of the lower-level problem, and the true optimal solution may not be found for each of the $N$ lower-level problems, approximation is needed and is given in Proposition \ref{p:hypergradient}. 

\textcolor{black}{
\begin{proposition}
The hypergradient can be approximated by
\begin{equation}
\nabla_{\bm{s}}\hat{\Psi}_{\tilde{N},m}(\bm{s})=\frac{2}{\tilde{N}}\sum_{i=1}^{\tilde{N}}(\nabla_{\bm{s}}\hat{\bm{\theta}}^{(i)})^T(\hat{\bm{\theta}}^{(i)}-\bm{\theta}^{(i)})
     \label{gradient s}
\end{equation}
\noindent where $\nabla_{\bm{s}}\hat{\bm{\theta}}^{(i)}$ is from the implicit differentiation of the lower-level optimality condition
\begin{equation}
\begin{split}
\nabla_{\bm{s}}\hat{\bm{\theta}}^{(i)} &\approx (\bm{C}^{(i)})^{-1}(-\nabla_{\bm{s}}(\bm{C}^{(i)})\hat{\bm{\theta}}^{(i)}-\nabla_{\bm{s}}\bm{d}^{(i)}+\bm{\bar{I}}^T\nabla_{\bm{s}}\bm{\bar{\eta}^{(i)}})\\
    \nabla_{\bm{s}}\bm{\bar{\eta}^{(i)}} &\approx (\bm{\bar{I}}(\bm{C^{(i)}})^{-1}\bm{\bar{I}}^T)^{-1}(\bm{\bar{I}}(\bm{C^{(i)}})^{-1}(\nabla_{\bm{s}}(\bm{C}^{(i)})\hat{\bm{\theta}}^{(i)}+\nabla_{\bm{s}}\bm{d}^{(i)})).
\end{split}   
\label{implicit gradient}
\end{equation}
where $\bm{\bar{I}}$ contains the rows of an identity matrix corresponding to the active constraints (i.e., $\theta^{(i)}=0$), and $\bm{\bar{\eta}}$ denotes the elements of $\bm{\eta}$ that correspond to the active constraints.
\label{p:hypergradient}
\end{proposition}
}

The KKT conditions associated with Proposition \ref{p:hypergradient} require the assumption of strict complementarity for (\ref{optimization 3}), i.e., for the Lagrangian multipliers $\bm{\bar{\eta}}$ that correspond to the active constraints $\bm{\bar{I}}\bm{\theta}=\bm{0}$, we have $\bm{\bar{\eta}}>\bm{0}$. 
Based on (\ref{implicit gradient}), the update equation is obtained,
\begin{equation}
\tilde{\bm{s}}_{\tilde{N},m+1}=\mathbb{P}_{\Omega^s}(\tilde{\bm{s}}_{\tilde{N},m}-\rho_m\nabla_{\bm{s}}\hat{\Psi}_{\tilde{N},m}(\tilde{\bm{s}}_{\tilde{N},m})),
\end{equation}

\noindent
where $\rho_m$ is the stepsize, $\mathbb{P}_{\Omega^s}$ denotes projection operator which projects the solution to the closest point in the feasible set $\Omega^s$ of $\bm{s}$.

\subsection{Convergence Analysis and Performance Guarantee}
\label{s:convergence}
This section presents the performance guarantee of the two algorithms by showing the upper bounds. All proofs are provided in the Appendices \ref{s:proof of theorem 2}.  

A stochastic upper bound is derived for the rSAA algorithm. To ensure $K$ is sufficiently large, the stochastic upper bound of the optimality gap can be defined as follows:
\begin{equation} \label{e:optimality gap}
\delta(K):=\Psi(\hat{\bm{s}}_{\tilde{N}})-\Psi^*,
\end{equation}
\noindent where $\Psi(\hat{\bm{s}}_{\tilde{N}})$ is the value of the objective function given $\hat{\bm{s}}_{\tilde{N}}$, and $\Psi^*$ is the true optimal value. Following \cite{shapiro2007tutorial},  $\Psi(\hat{\bm{s}}_{\tilde{N}})$ can be estimated from $N$ Monte Carlo samples, and an approximate $100(1-\alpha)\%$ confidence upper bound for $\Psi(\hat{\bm{s}}_{\tilde{N}})$ is given by
$\hat{\Psi}_{N}+z_{\alpha}\hat{\sigma}_{N}$, 
where $\hat{\Psi}_{N}(\hat{\bm{s}}_{\tilde{N}})=\frac{1}{N}\sum_{i=0}^{N-1}\hat{\Psi}^{(i)}(\hat{\bm{s}}_{\tilde{N}})$, $z_{\alpha}$ is the critical value from standard normal, and $\hat{\sigma}^2_{N}=\frac{1}{N(N-1)}\sum_{i=0}^{N-1}(\hat{\Psi}^{(i)}(\hat{\bm{s}}_{\tilde{N}})-\hat{\Psi}_{N})^2$. To derive the lower bound of $\Psi^*$, note that $\Psi^* \ge \mathbb{E}(\hat{\Psi}^k_{\tilde{N}})$, and an approximate $100(1-\alpha)\%$ lower bound for $\mathbb{E}(\hat{\Psi}^k_{\tilde{N}})$ is
$\bar{\Psi}_{\tilde{N}}-t_{\alpha}\hat{\sigma}_{\tilde{N},K}$,
where $\bar{\Psi}_{\tilde{N}}=\frac{1}{K}\sum_{k=0}^{K-1}\hat{\Psi}^k_{\tilde{N}}$, $t_{\alpha}$ is a critical value, and $\hat{\sigma}^2_{\tilde{N},K}=\frac{1}{K(K-1)}\sum_{k=0}^{K-1}(\hat{\Psi}^k_{\tilde{N}}-\bar{\Psi}_{\tilde{N}})^2$.
Hence,  a stochastic upper bound (with confidence at least $1-2\alpha$) of $\delta(K)$ is
\begin{equation}
      \Delta(K)= (\hat{\Psi}_{N}+z_{\alpha}\hat{\sigma}_{N})-(\bar{\Psi}_{\tilde{N}}-t_{\alpha}\hat{\sigma}_{\tilde{N},K}).
      \label{stochastic upper bound alpha}
\end{equation}



For the SBA algorithm, an upper bound of the hypergradient of the IMSE objective value is derived.

\begin{assumption}{(Smoothness of $\Psi$)}\label{a:smoothness}
The hypergradient $\nabla\Psi$ is Lipschitz continuous in $\bm{s}$ with a constant $\mathcal{L}_{\nabla \Psi}$, i.e., for any two sensor locations $\bm{s}_1$ and $\bm{s}_2$, 
\begin{equation}
     \|\nabla_s \hat{\Psi}(\bm{s}_2) - \nabla_{\bm{s}} \hat{\Psi}(\bm{s}_1)\|\leq \mathcal{L}_{\nabla\Psi}\|\bm{s}_2 - \bm{s}_1\|.
\end{equation}
\label{A:2}
\end{assumption}

As already discussed in (\ref{gradient s}), the solution of the \textcolor{black}{lower-level} problem affects the evaluation of the hypergradient. Let $\hat{\bm{\theta}}^{*(i)}$ and $\hat{\bm{\theta}}^{(i)}$ respectively be the true and the obtained solution of the $i$th \textcolor{black}{lower-level} problem (in many cases, $\hat{\bm{\theta}}^{*(i)} \neq \hat{\bm{\theta}}^{(i)}$), we assume that
\begin{assumption}{(\textcolor{black}{lower-level} optimality)}\label{a:lower-level gap}
The gap between $\hat{\bm{\theta}}^{*(i)}$ and $\hat{\bm{\theta}}^{(i)}$ is bounded, i.e., for some $\delta>0$, $\|\hat{\bm{\theta}}^{(i)} - \hat{\bm{\theta}}^{*(i)}\|\leq \delta$, $i=1,2,\cdots, \tilde{N}$. 
\label{A:3}
\end{assumption}
\textcolor{black}{In our numerical experiments (see Section \ref{s:numerical}), it is shown that $\delta$ is reasonably small.} Following Assumptions \ref{A:2} and \ref{A:3}, Lemma \ref{lemma1} below presents the upper bound of the approximate hypergradient (\ref{gradient s}), which is based on the obtained solution $\hat{\bm{\theta}}^{(i)}$ of the \textcolor{black}{lower-level} problem.

\begin{lemma}


For the SBA method presented in Algorithm \ref{algorithm 2}, we have
 \begin{equation}
 \begin{split}
\text{(a)}& \ \ \ \ \mathbb{E}\big(\|\nabla_{\bm{s}} \hat{\Psi}(\bm{s}; \hat{\bm{\theta}}) - \nabla_{\bm{s}} \hat{\Psi}(\bm{s};\hat{\bm{\theta}}^*)\|\big)\leq \mathcal{L}_{\Psi}\delta + \mathcal{L}_{D}\sigma\sqrt{n_{cov}\tilde{N}^{-1}},\\
\text{(b)}& \ \ \ \ \mathbb{E}\big(\|\nabla_{\bm{s}} \hat{\Psi}(\bm{s}; \hat{\bm{\theta}}) - \nabla_{\bm{s}} \hat{\Psi}(\bm{s};\hat{\bm{\theta}}^*)\|^2\big)\leq 2\mathcal{L}_{\Psi}^2\delta^2 + 2\mathcal{L}_{D}^2 \sigma^2 \tilde{N}^{-1},
 \end{split}
 \end{equation}

 \noindent
where the constant $\mathcal{L}_{\Psi}$ varying with $\Psi$ is given by Assumptions \ref{A:bounds 1}-\ref{A:data} in Appendix \ref{s:Appendix 4 lemma 1 bc}, the expectation is taken with respect to the joint distribution of wind, emission rates and observation noise,
and $\mathcal{L}_{D}$, $\sigma$ and $n_{cov}$ are constants defined in Appendix \ref{s:Appendix 4 lemma 1 bc}.
\label{lemma1}
\end{lemma}

Based on Lemma \ref{lemma1} above, we obtain the upper bound of the hypergradients given in Theorem \ref{theorem2}. The theorem requires Assumption \ref{A:gradient bound} given in Appendix \ref{s:proof of theorem 2}.

\begin{theorem}
For the SBA method presented in Algorithm \ref{algorithm 2}, 
\begin{itemize}
     \item If $\rho_m$ is a constant, i.e., $\rho_m = \rho$ and $0<\rho<\frac{2}{\mathcal{L}_{\nabla\Psi}}$, then
     \begin{equation}
                  \begin{split}
             \frac{1}{M}\sum_{k=0}^{M-1}&\mathbb{E}\Big[\|\nabla_{\bm{s}} \hat{\Psi}(s_m;\hat{\bm{\theta}}^*)\|^2\Big] \leq \frac{\mathbb{E}[\hat{\Psi}(s_0;\hat{\bm{\theta}}^*)]}{(\rho-\frac{1}{2}\rho^2 \mathcal{L}_{\nabla\Psi})M}\\ &+ \frac{\mathcal{C}_{\nabla\Psi}(\mathcal{L}_{\Psi}\delta+\mathcal{L}_{D}\frac{\sigma\sqrt{n_{cov}}}{\sqrt{\tilde{N}}})(1+\rho \mathcal{L}_{\nabla\Psi}) + \mathcal{L}_{\nabla\Psi}\rho(\mathcal{L}_{\Psi}^2\delta^2+\mathcal{L}_{D}^2\frac{\sigma^2}{\tilde{N}})}{1-\frac{1}{2}\rho \mathcal{L}_{\nabla\Psi}}. 
         \end{split}
         \label{SGD convergence 1 Theorem}
     \end{equation}  
     \vspace{-18pt}
     \item If $\rho_m$ decays with $\rho_m = \frac{\rho_0}{m+1}$, i.e., $\sum_{m=0}^{\infty}\rho_m=\infty$ and $\sum_{k=0}^{\infty}\rho_m^2 < \infty$, and we let $\bm{s}_M = \bm{s}_m$ with a probability $\frac{1}{A_M(m+1)}$, where $A_M=\sum_{m=0}^{M-1}\frac{1}{m+1}$, then
     \begin{equation}
         \lim_{M\rightarrow\infty}\mathbb{E}[\|\nabla_{\bm{s}} \hat{\Psi}(\bm{s}_M;\hat{\bm{\theta}}^*)\|^2] \leq \mathcal{C}_{\nabla\Psi}(\mathcal{L}_{\Psi}\delta + \mathcal{L}_{D}\frac{\sigma\sqrt{n_{cov}}}{\sqrt{\tilde{N}}}).
         \label{SGD convergence 2 Theorem}
     \end{equation}
\end{itemize}
\label{theorem2}
\end{theorem}

It is seen that the second term on the RHS of  (\ref{SGD convergence 1 Theorem}) goes to zero 
if $\tilde{N}\rightarrow \infty$ and $\delta \rightarrow 0$ (i.e., the approximate solution of the \textcolor{black}{lower-level} problem gets closer to the optimal solution).  
If the true optimal solution is obtained for the \textcolor{black}{lower-level} problem and a sufficiently large batch size $\tilde{N}$ is used, the first term on the RHS indicates that the solution converges to a stationary point at a rate of $M^{-1}$ if we set a constant stepsize $0<\rho<\frac{2}{\mathcal{L}_{\nabla\Psi}}$. If we adopt decaying stepsize, (\ref{SGD convergence 2 Theorem}) shows that the solution converges to a stationary point when $M$ and $\tilde{N}$ goes to infinity and the \textcolor{black}{lower-level} problem is solved to optimality (i.e., $\delta=0$).

\vspace{-6pt}
\section{Numerical Examples} \label{s:numerical}
\vspace{-6pt}
Numerical examples, as well as detailed discussions, are presented to illustrate the performance of proposed approaches. Example I is a simple illustrative example that considers the placement of one or two sensors for three emission sources only. 
In Example II, we consider a more realistic problem that involves the placement of multiple sensors for 10, 20, 50 and 100 emission sources.

A Gaussian Plume model is used as the atmospheric dispersion process, which approximates the transport of airborne contaminants due to turbulent diffusion and advection \citep{stockie2011mathematics}. \textcolor{black}{The Gaussian Plume model used for numerical and experimental examples are derived from the advection-diffusion equation which is a PDE representing the transport of a substance in 3D space. The concentration $C$ is described as by a function, $\frac{\partial C}{\partial t}+\nabla\cdot (C\bm{u})=\nabla\cdot(K\nabla C)+S$,    
where $S((x,y,z),t)$ is the emission rate of the emission source, $K((x,y,z),t)$ is the diffusion coefficient (from eddy and molecular diffusion), and $\bm{u}((x,y,z),t)$ is the wind condition.} The data is generated by the following equation, 
\begin{equation}
\Phi_i = \sum_{j=1}^{N_p}\theta_j A_j(\bm{s}_i) + \epsilon,
\end{equation}
where $\bm{s}_i$ is the location of the $i$-th sensor, $\theta_j$ is the emission rate of the $j$-th source, $\epsilon\sim \mathcal{N}(0,\sigma^2_{\epsilon})$ is the observation noise, and $A_j(\bm{s}_i)$ is the Gaussian Plume kernel,
\begin{equation}
   A_j(\bm{s}_i)=\frac{1}{2\pi K \|(\bm{s}_i-\bm{x}_j)\cdot\bm{\beta}^{\parallel}\|}\text{exp}\Big(-\frac{\big(\|(\bm{s}_i-\bm{x}_j)\cdot\bm{\beta}^{\perp}\|^2 + H_j^2\big)}{4K\|(\bm{s}_i-\bm{x}_j)\cdot\bm{\beta}^{\parallel}\|}\Big),
\end{equation}
where $K$ depends on eddy diffusivity, $H_j$ is the height of stack $j$, $\bm{x}_j$ is the location of the $j$-th emission source, $\bm{\beta}^{\perp}$ and $\bm{\beta}^{\parallel}$ are the unit vectors perpendicular and parallel to $\bm{\beta}$.

\subsection{Example I: A Simple Illustration} \label{s:toy}
\begin{figure}[h!]
    \centering
    \includegraphics[width=0.5\linewidth]{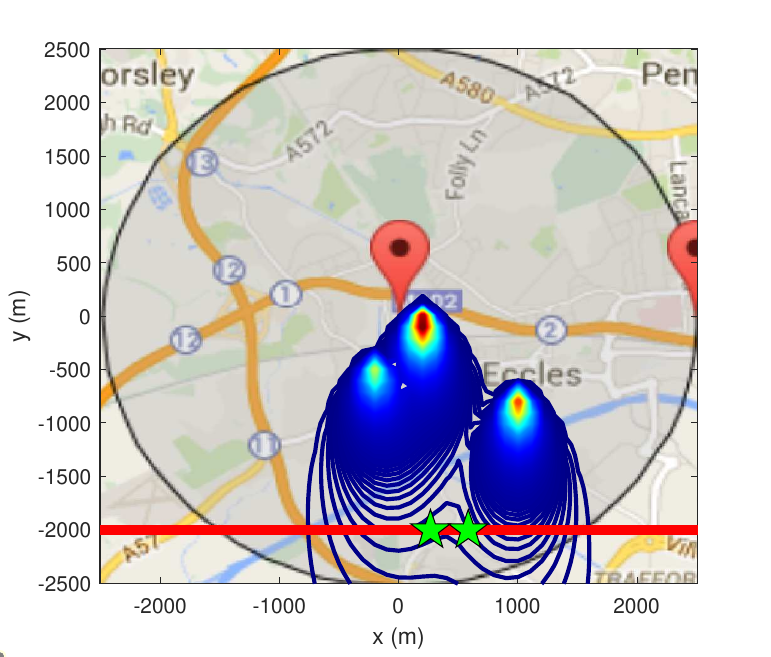}
    \vspace{-8pt}
    \caption{Placement of sensors (green stars) on the straight line (blue line).}
    \label{fig:illustration of final placement of 2 sensor}
\end{figure}

We start with a simple case for which 1 or 2 sensors are placed along a straight line for only 3 potential emission sources (see Figure \ref{fig:illustration of final placement of 2 sensor}). The wind vector is set to $\bm{\beta} = (0,-5)$, i.e., north wind, and the emission rates of the three sources are $\bm{\theta}^* = (80,60,40)$.
The standard deviation of the observation noise in (\ref{physics}) is set to $\sigma_{\epsilon} = 1$. 


\begin{figure}[h!]
     \centering
     \begin{subfigure}[b]{0.47\textwidth}
         \centering
         \includegraphics[width=0.9\textwidth]{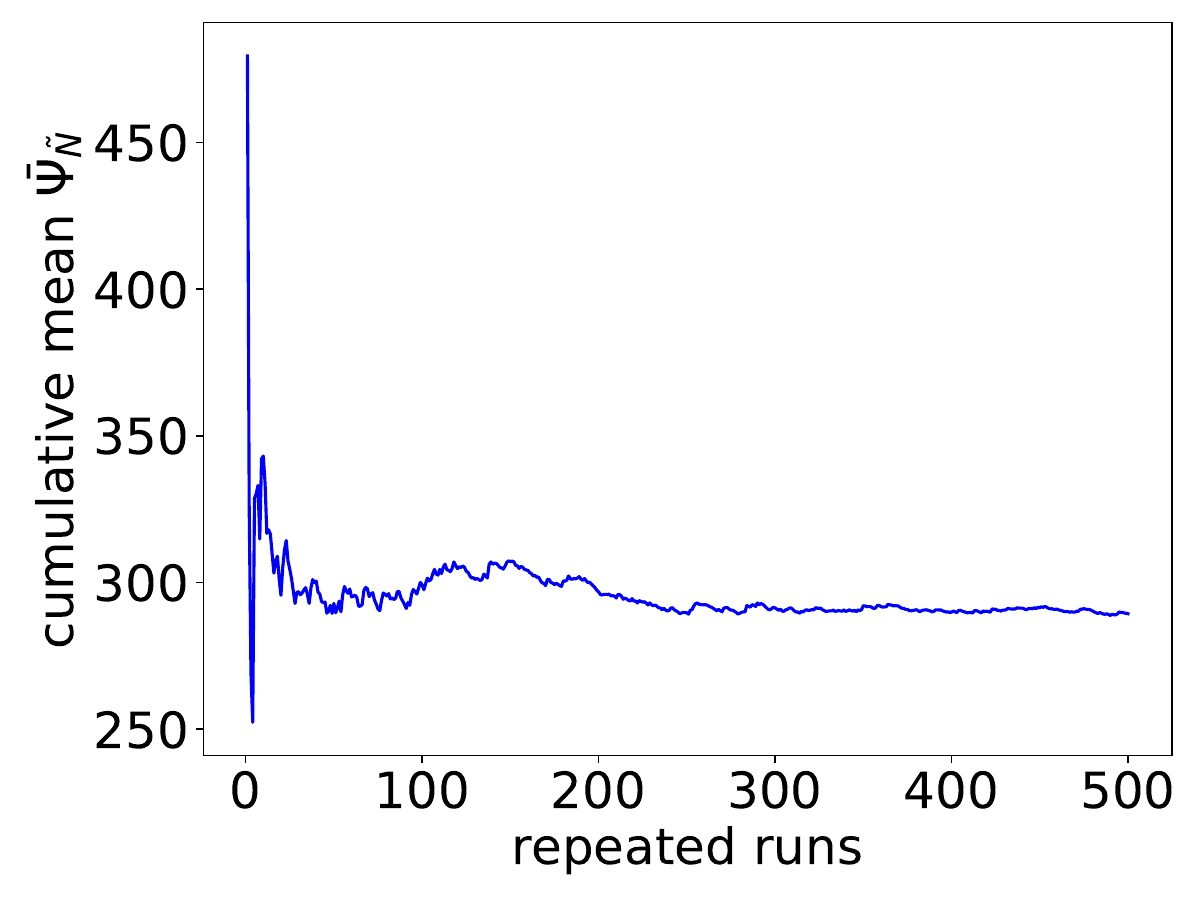}
         \vspace{-8pt}
         \caption{Cumulative mean $\bar{\Psi}_{\tilde{N}}$ in (\ref{stochastic upper bound alpha})}
         \label{fig:SAA cumulative mean}
     \end{subfigure}
     \hfill
     \begin{subfigure}[b]{0.47\textwidth}
         \centering
         \includegraphics[width=0.9\textwidth]{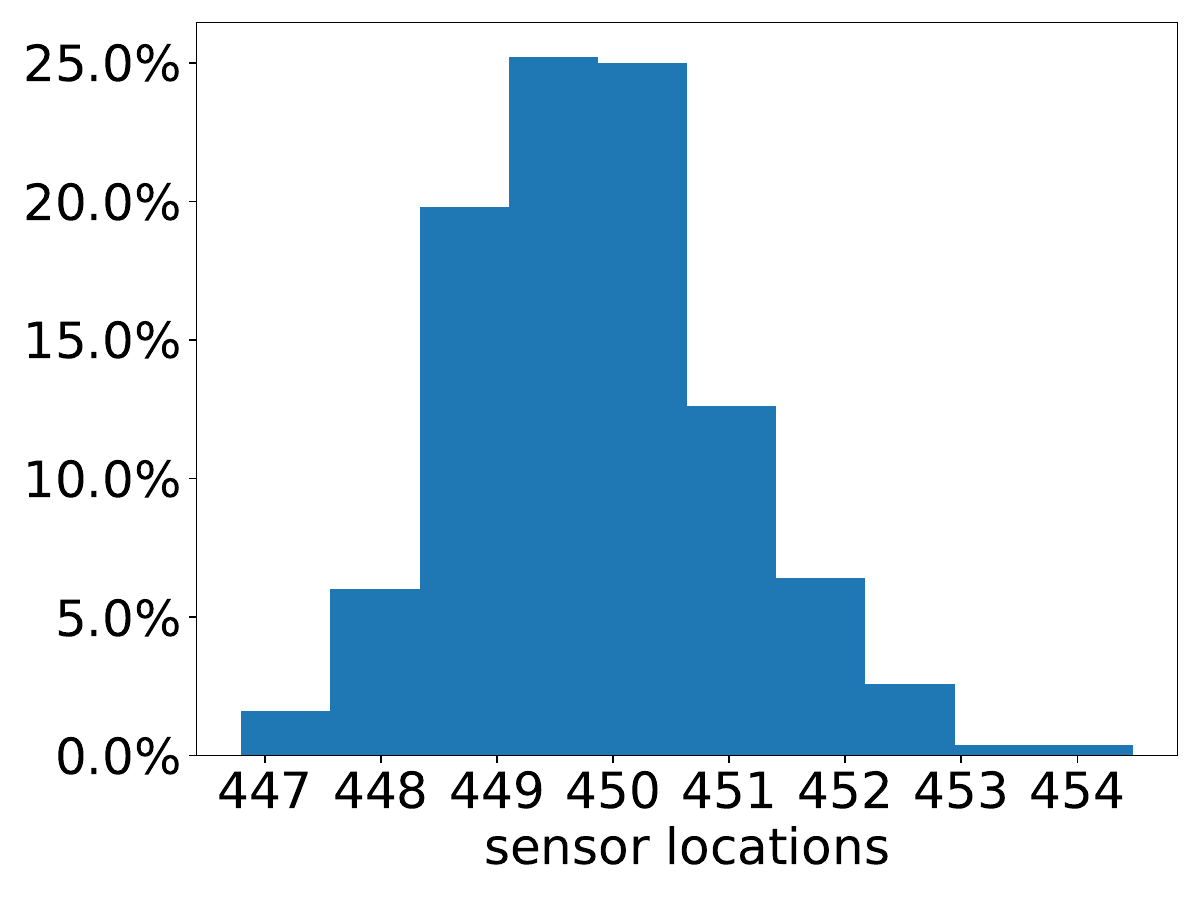}
         \vspace{-8pt}
         \caption{Histogram of sensor locations}
         \label{fig:SAA hist}
     \end{subfigure}
     \begin{subfigure}[b]{\textwidth}
         \centering
         \includegraphics[width=0.8\textwidth]{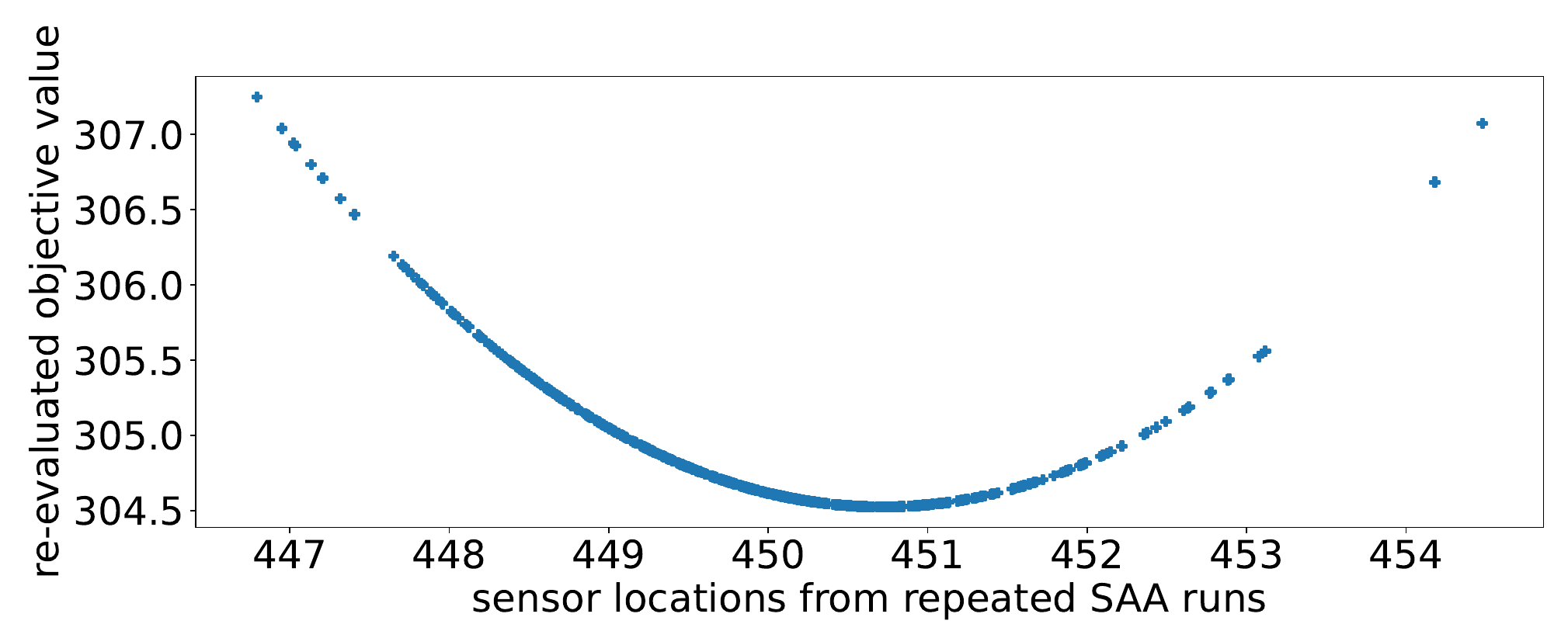}
         \vspace{-8pt}
         \caption{ Re-evaluated objective function, using a large $N=10,000$}
         \label{fig:SAA objective}
     \end{subfigure}
     \caption{Results from Example I.}
     \label{fig:sensor placement example 1 SAA}
\end{figure}

\begin{figure}[h!]
    \centering
    \includegraphics[width=0.7\linewidth]{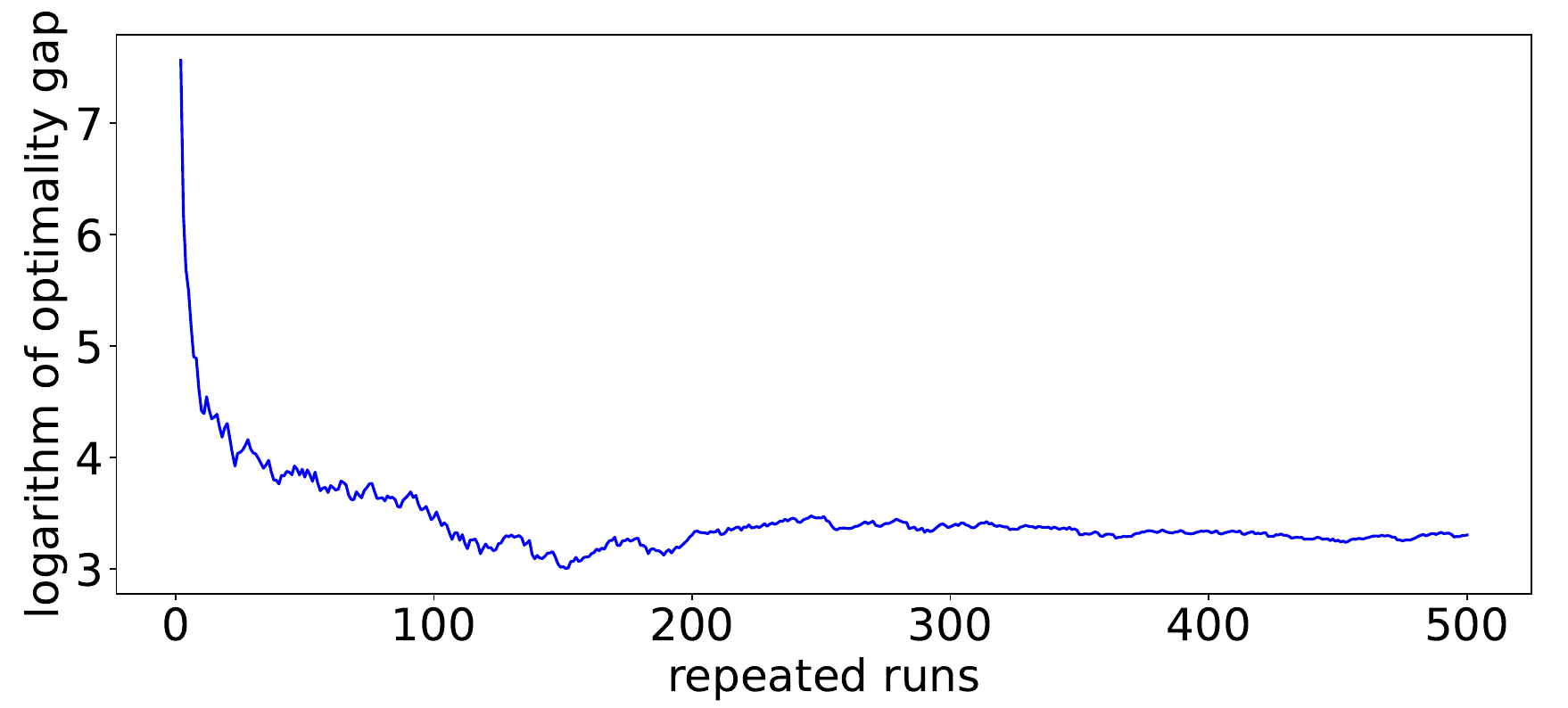}
    \vspace{-8pt}
    \caption{The stochastic upper bound defined in (\ref{stochastic upper bound alpha}), $\alpha=0.025$.}
    \label{fig:bound}
\end{figure}

We start with placing 1 sensor. Let $\lambda_1=\lambda_2=0.0001$ for the inverse model (\ref{e:our problem}),  Figure \ref{fig:sensor placement example 1 SAA} shows the results obtained from the rSAA algorithm. 
In particular, Figure \ref{fig:SAA cumulative mean} shows how the cumulative mean of the objective function changes against repeated runs (we set $\tilde{N}=5$),  which appears to converge after $K=250$ runs. 
Figure \ref{fig:SAA hist} shows the histogram of optimal sensor locations from each run, and the mean of sensor location is found to be 449.8 m.  Because we only consider the deployment of one sensor along a straight line, it is possible to re-evaluate the objective function (for validation purposes), using a large $N=10,000$, based on the optimal sensor locations from repeated runs; see Figure \ref{fig:SAA objective}. The lowest point of this curve corresponds to the true optimal solution (i.e., 450.57 m in Figure \ref{fig:SAA objective}). We see that, the solutions obtained from multiple repeated runs vary around the true optimal solution, and the average sensor location is close to the true optimal solution,  justifying the necessity of repeating SAA runs. 
Figure \ref{fig:bound} shows the (log) gap, defined in (\ref{stochastic upper bound alpha}), over repeated runs, and the convergence of the algorithm is observed. It is noted that we also run the SBA algorithm and the solution is close to $450.90$ m. Although the two algorithms achieve similar solutions, the SBA algorithm is found to be 250 times faster because only one run is needed in the SBA algorithm.





\begin{figure}[h!]
     \centering
     \begin{subfigure}[b]{0.49\textwidth}
         \centering
         \includegraphics[width=\textwidth]{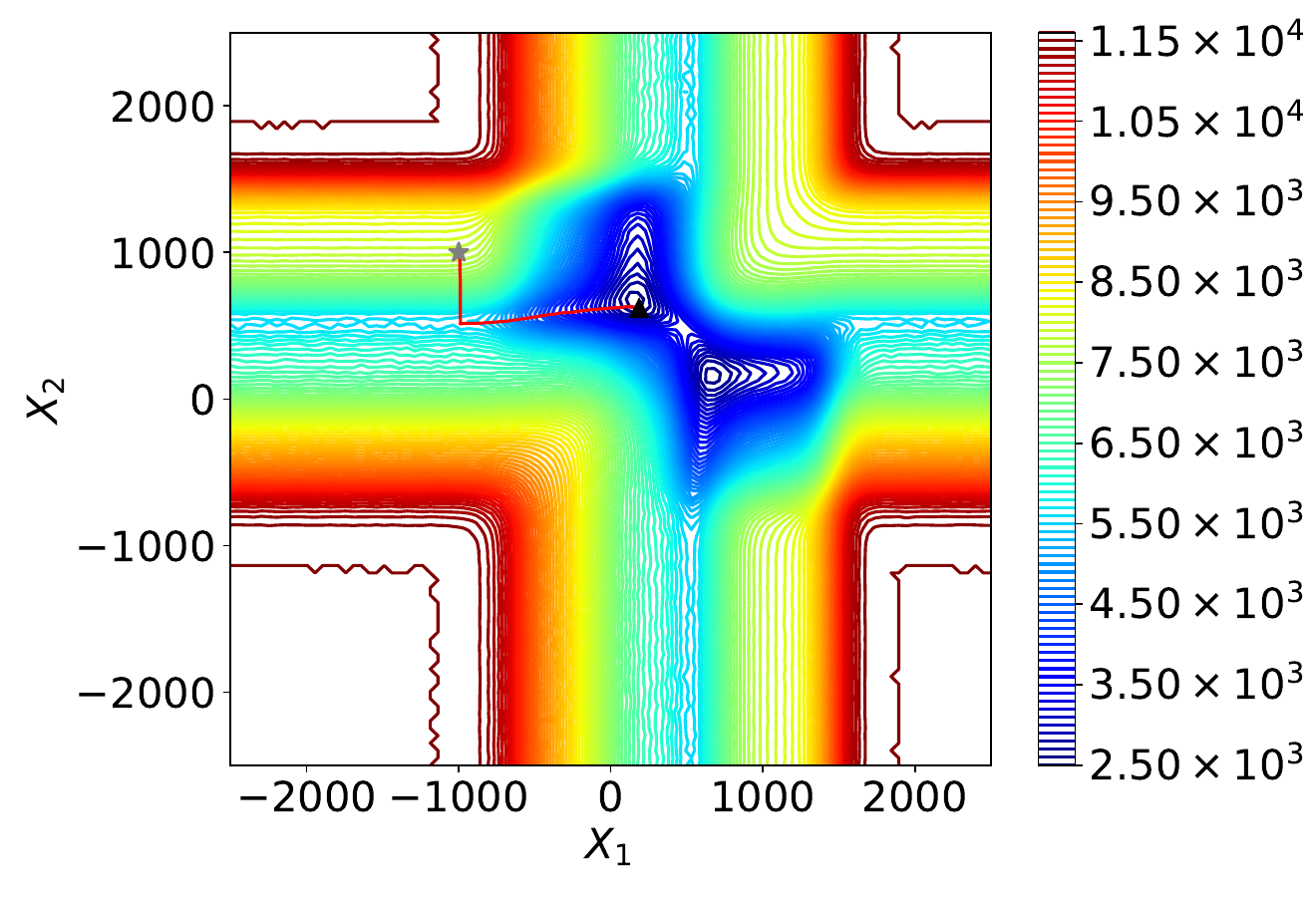}
         \vspace{-12pt}
         \caption{$\lambda_1=\lambda_2=0.01$, $J=30,000$}
         \label{fig:2 sensor 1_1}
     \end{subfigure}
     \hfill
     \begin{subfigure}[b]{0.49\textwidth}
         \centering
         \includegraphics[width=\textwidth]{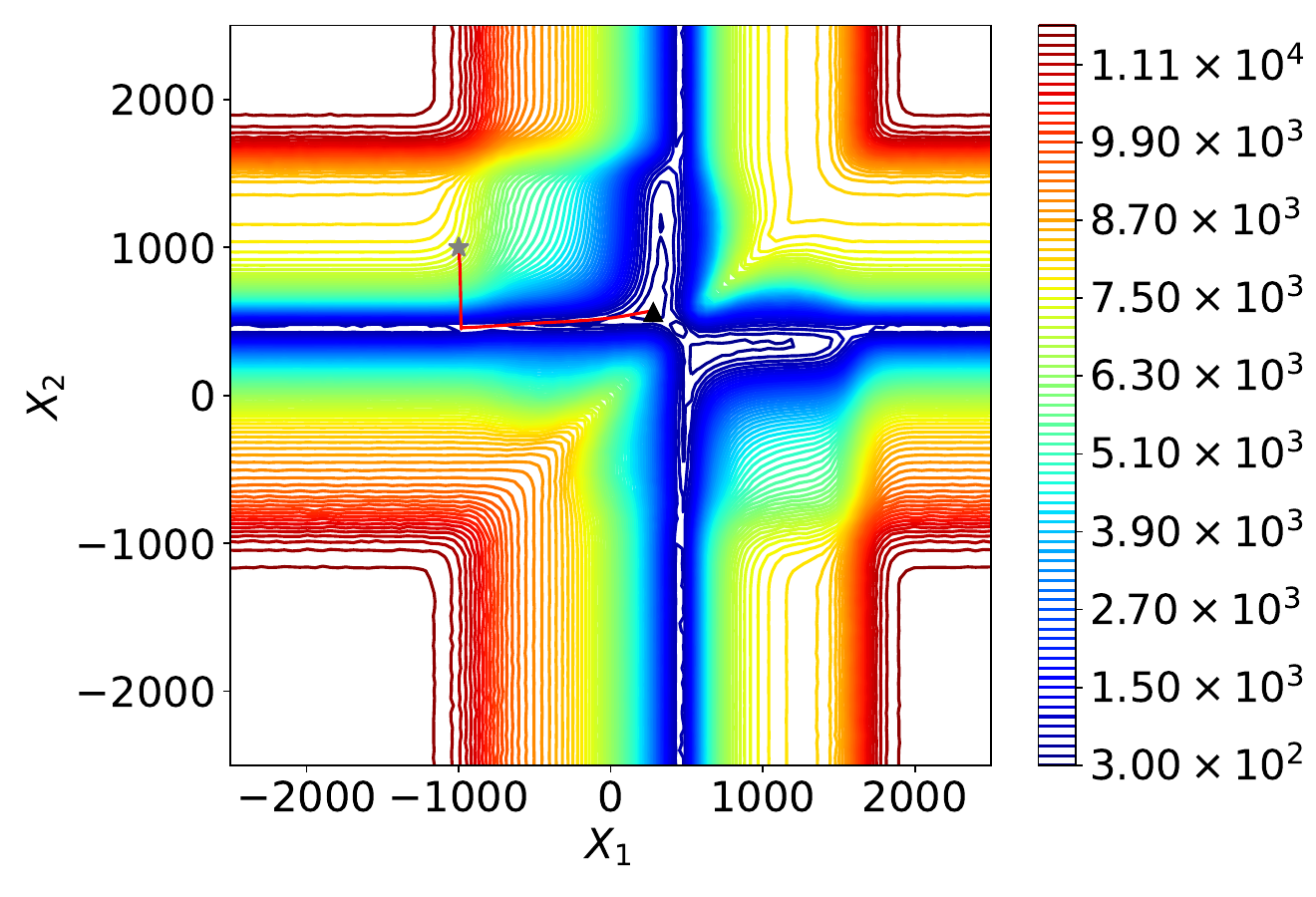}
         \vspace{-8pt}
         \caption{$\lambda_1=\lambda_2=0.001$, $J=30,000$}
         \label{fig:2 sensor 1_2}
     \end{subfigure}
     \begin{subfigure}[b]{0.49\textwidth}
         \centering
         \includegraphics[width=\textwidth]{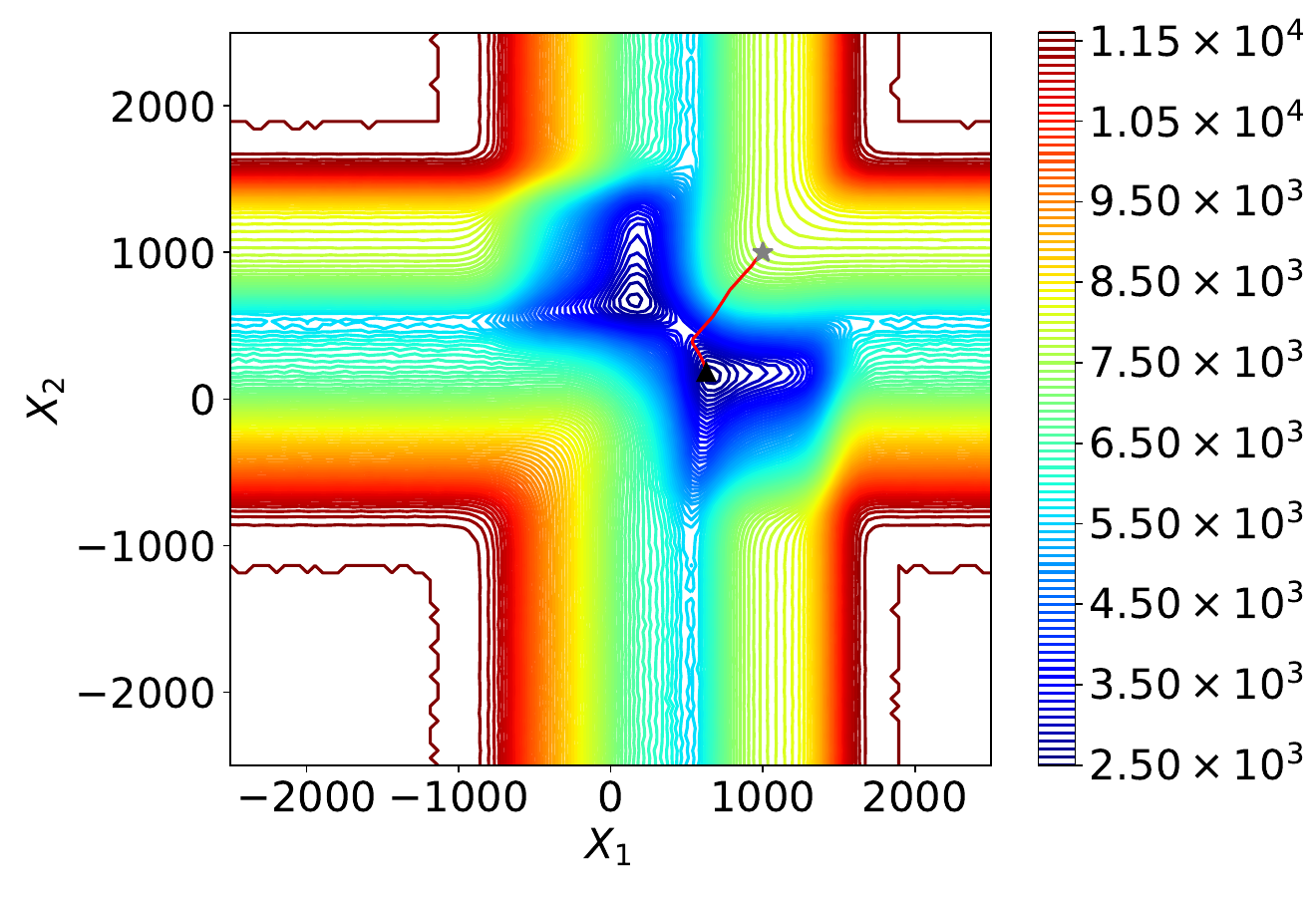}
         \vspace{-8pt}
         \caption{$\lambda_1=\lambda_2=0.01$, $J=30,000$}
         \label{fig:2 sensor 1_3}
     \end{subfigure}
     \hfill
     \begin{subfigure}[b]{0.49\textwidth}
         \centering
         \includegraphics[width=\textwidth]{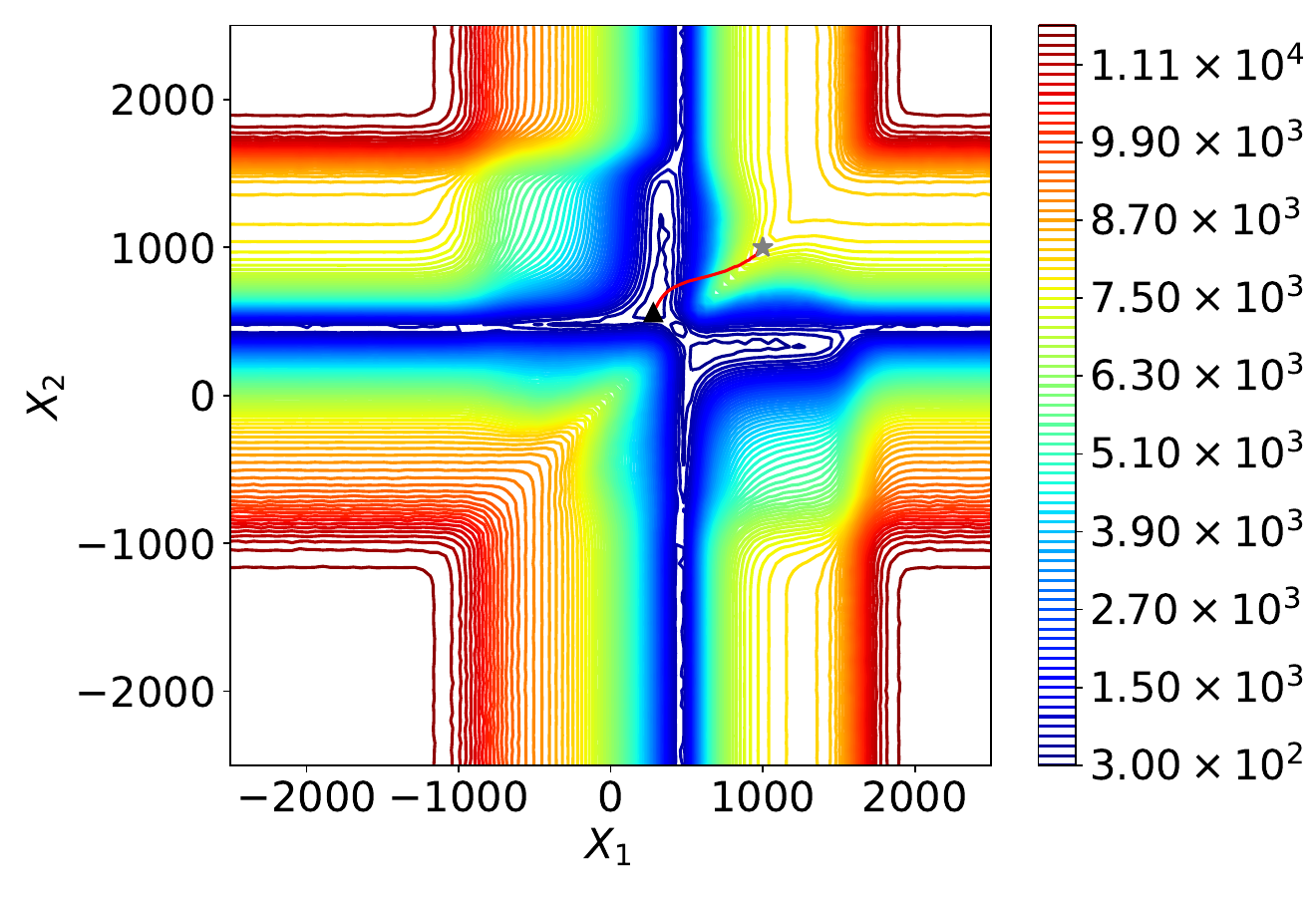}
         \vspace{-8pt}
         \caption{$\lambda_1=\lambda_2=0.001$, $J=30,000$}
         \label{fig:2 sensor 1_4}
     \end{subfigure}
     \vspace{-4pt}
     \caption{Trajectories using different initial values of  $\lambda_1$ and $\lambda_2$ ($J=30,000$).}
     \label{fig:2 sensor large q}
\end{figure}

\begin{figure}[hbt!]
     \centering
     \begin{subfigure}[b]{0.49\textwidth}
         \centering
         \includegraphics[width=\textwidth]{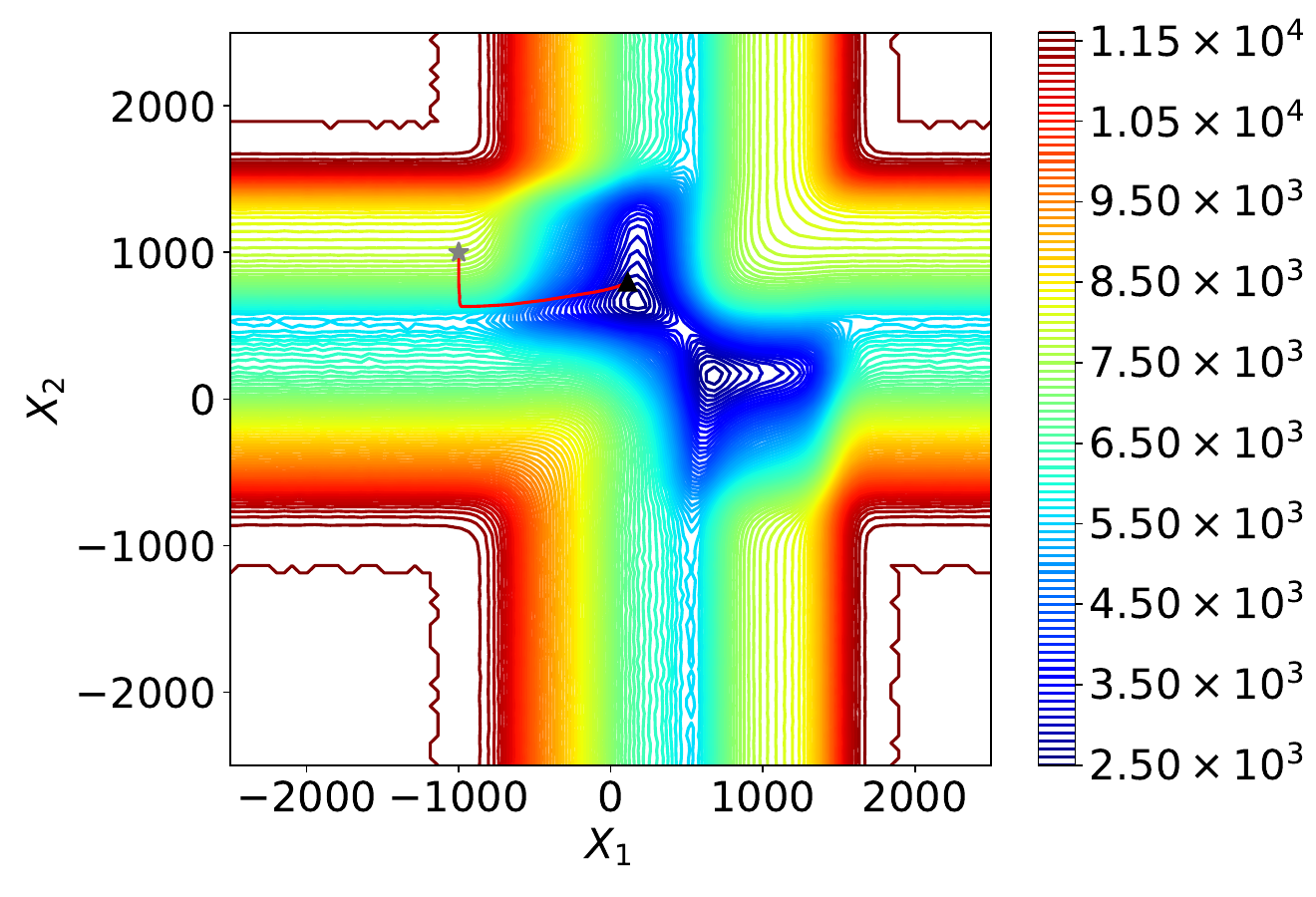}
         \vspace{-12pt}
         \caption{$\lambda_1=\lambda_2=0.01$, $J=1$}
         \label{fig:2 sensor 2_1}
     \end{subfigure}
     \hfill
     \begin{subfigure}[b]{0.49\textwidth}
         \centering
         \includegraphics[width=\textwidth]{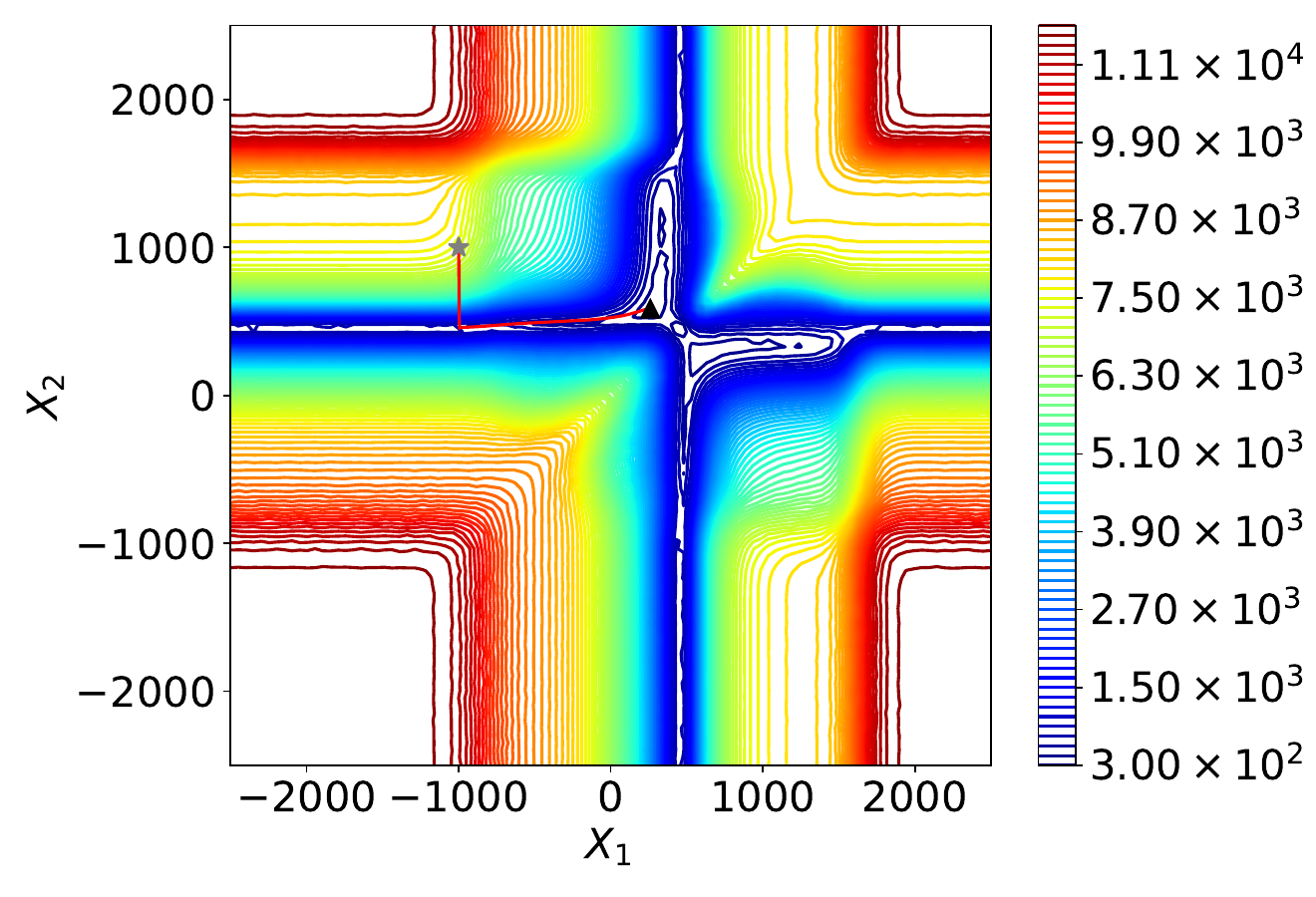}
         \caption{$\lambda_1=\lambda_2=0.001$, $J=1$}
         \label{fig:2 sensor 2_2}
     \end{subfigure}
     \vspace{-8pt}
     \begin{subfigure}[b]{0.49\textwidth}
         \centering
         \includegraphics[width=\textwidth]{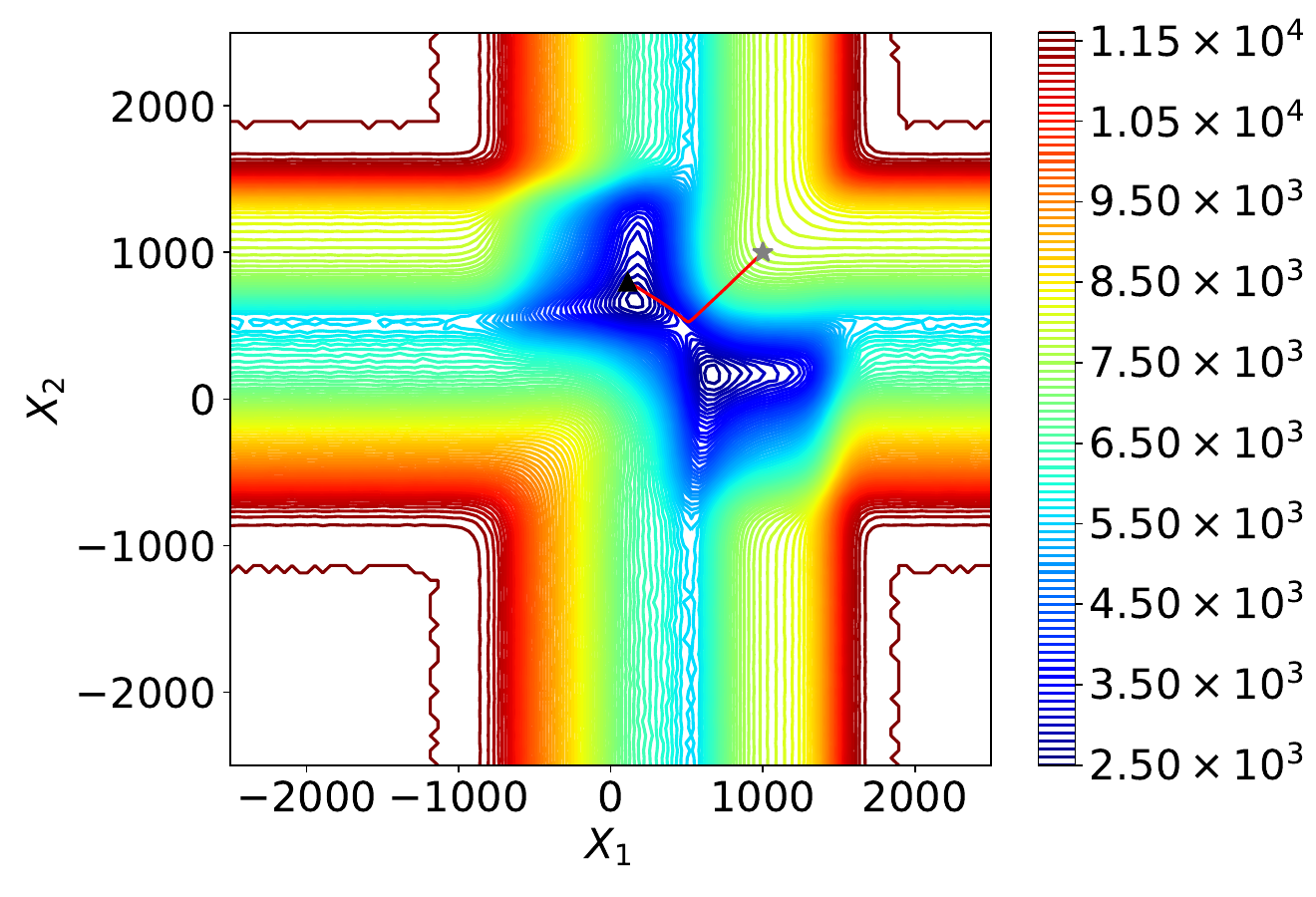}
         \vspace{-8pt}
         \caption{$\lambda_1=\lambda_2=0.01$, $J=1$}
         \label{fig:2 sensor 2_3}
     \end{subfigure}
     \hfill
     \vspace{-8pt}
     \begin{subfigure}[b]{0.49\textwidth}
         \centering
         \includegraphics[width=\textwidth]{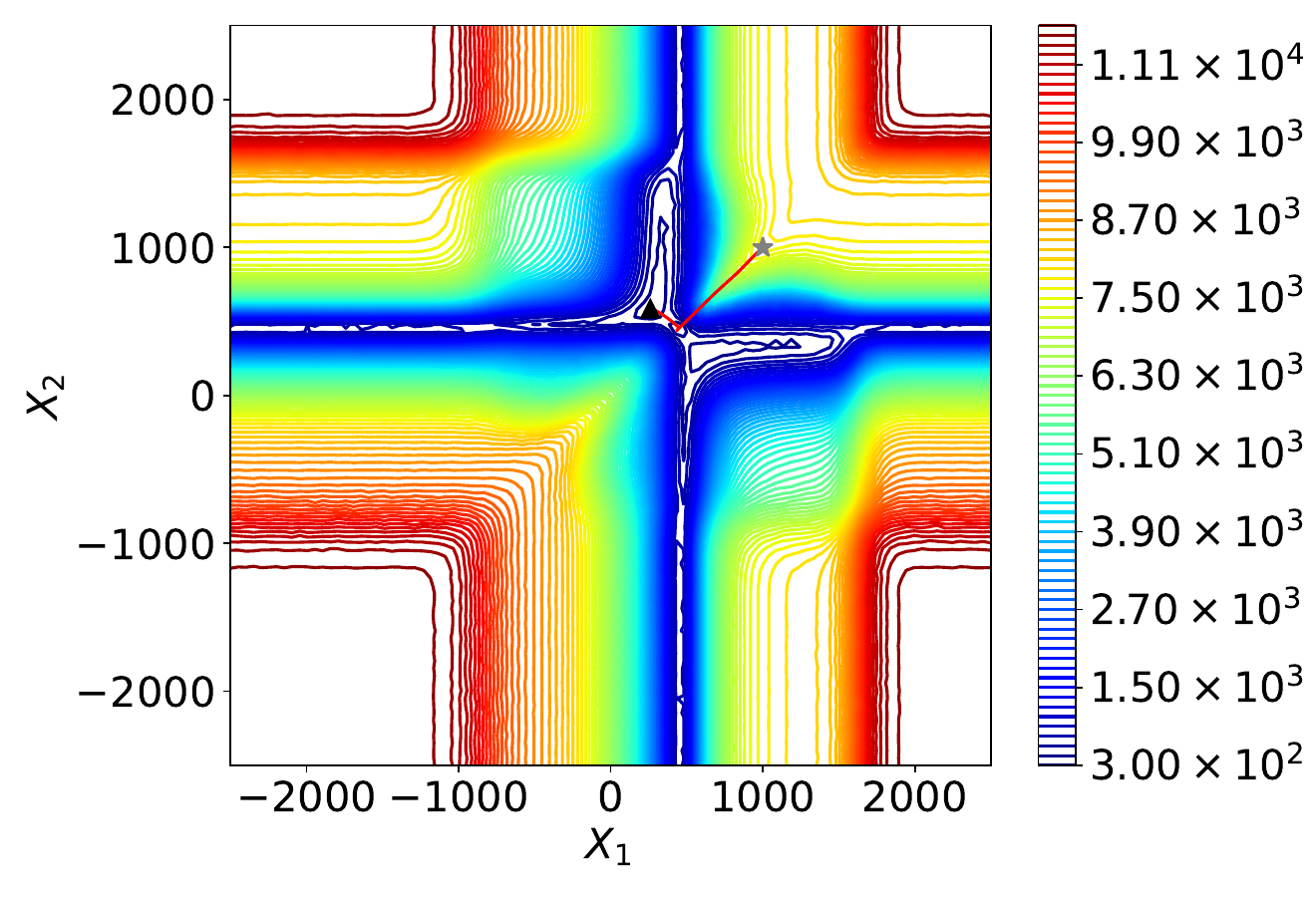}
         \vspace{-8pt}
         \caption{$\lambda_1=\lambda_2=0.001$, $J=1$}
         \label{fig:2 sensor 2_4}
     \end{subfigure}
     \vspace{-4pt}
     \caption{Trajectories using different initial values of $\lambda_1$ and $\lambda_2$ ($J=1$).}
     \label{fig:2 sensor 2}
\end{figure}

Next, we consider the placement of 2 sensors along the same line using the SBA algorithm. Figure \ref{fig:2 sensor large q} shows the trajectories of the locations of these two sensors on the straight line given different initial guesses (marked by stars). The contour in this figure is the objective function $\hat{\Psi}_N(\bm{s})$ evaluated using a large number of Monte Carlo samples for different sensor locations.  
It is seen that the sensor location goes downhill as the iteration proceeds, which demonstrates the effectiveness of the algorithm. 
We also investigate if a small $J$ can be used in Algorithm \ref{algorithm 2}, such as $J=1$, to further accelerate the \textcolor{black}{lower-level} solver. Because a smaller $J$ requires a larger $M$ for the algorithm to converge, we also double the value the $M$ when $J=1$.
The result is shown in Figure \ref{fig:2 sensor 2} with different choices of $\lambda_1$ and $\lambda_2$. 
It is seen that the SBA algorithm still works well even when $J=1$. A drawback of a small $J=1$ is that there exists an inevitable gap between the best-found solution and the true minimum (of the contour), as shown in Figure \ref{fig:2 sensor 2_1} and \ref{fig:2 sensor 2_3} when $\lambda_1=\lambda_2=0.01$. 
Finally, the optimal locations of the 2 sensors are shown in Figure \ref{fig:illustration of final placement of 2 sensor}.

\subsection{Discussions on Initial Sensor Locations}
Recall that Algorithm 2 is computationally faster than Algorithms 1, but requires initial guesses of sensor locations that affect the locally optimal solutions of sensor allocation. 
In this paper, we propose to obtain the initial sensor locations as follows:
\begin{proposition}
Assuming a Gaussian prior $\bm{\theta}\sim \mathcal{N}(\bm{\mu}_{\text{pr}},\bm{\Gamma}_{\text{pr}})$ with mean $\bm{\mu}_{\text{pr}}$ and variance $\bm{\Gamma}_{\text{pr}}$, the initial sensor locations can be chosen by minimizing
\begin{equation} \label{eq:proposition1}
   \begin{split}
        \hat{\Psi}_{\text{risk, linear, Gaussian}}(\bm{s})=\mathbb{E}_{\bm{\beta}}\{||\bm{\Gamma}_{\text{post}}\bm{L}^T||^2_F + ||\bm{\Gamma}_{\text{post}}\mathcal{F}^*\bm{U}^T||^2_F\}
   \end{split} 
\end{equation}

\noindent
where $\bm{\Gamma}_{\text{post}}$ is the posterior covariance matrix, $||\cdot||_F$ is the Frobenius norm, $\bm{L}^T\bm{L}=\bm{\Gamma}^{-1}_{pr}$, and $\bm{\Gamma}^{-1}_{\bm{\epsilon}}=\bm{U}^T\bm{U}$.
\label{p:A2}
\end{proposition}

Similar to the idea of  \cite{parise2017sensitivity,tsaknakis2022implicit}, the derivations behind Proposition \ref{p:A2} is provided in Appendix \ref{s:Appendix 1}. Note that, the objective function (\ref{p:A2}) is the objective function of the $A$-optimal design averaged over various wind conditions. For this reason, the initial sensor location obtained from Proposition 3 above is still referred to as the $A$-optimal design in this paper.

\begin{figure}[h]
    \centering
    \includegraphics[width=1\linewidth]{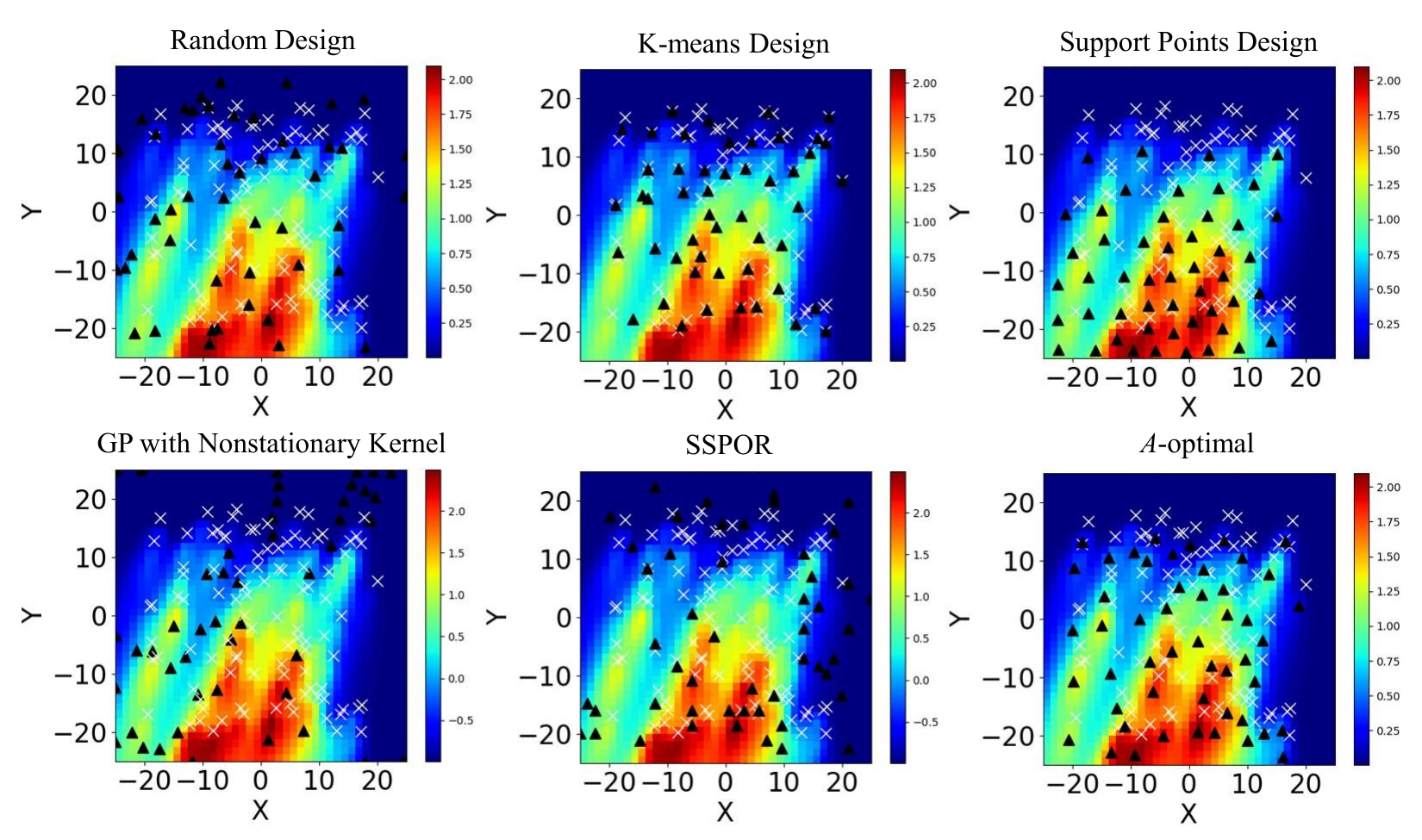}
    \vspace{-18pt}
    \caption{Initial sensor locations (indicated by ``$\blacktriangle$'') obtained from different approaches given one specific concentration field with 100 sources (indicated by ``$\times$'').}
    \label{fig:comparison of initial guess}
\end{figure}
\begin{figure}[h!]
    \centering    \includegraphics[width=0.9\linewidth]{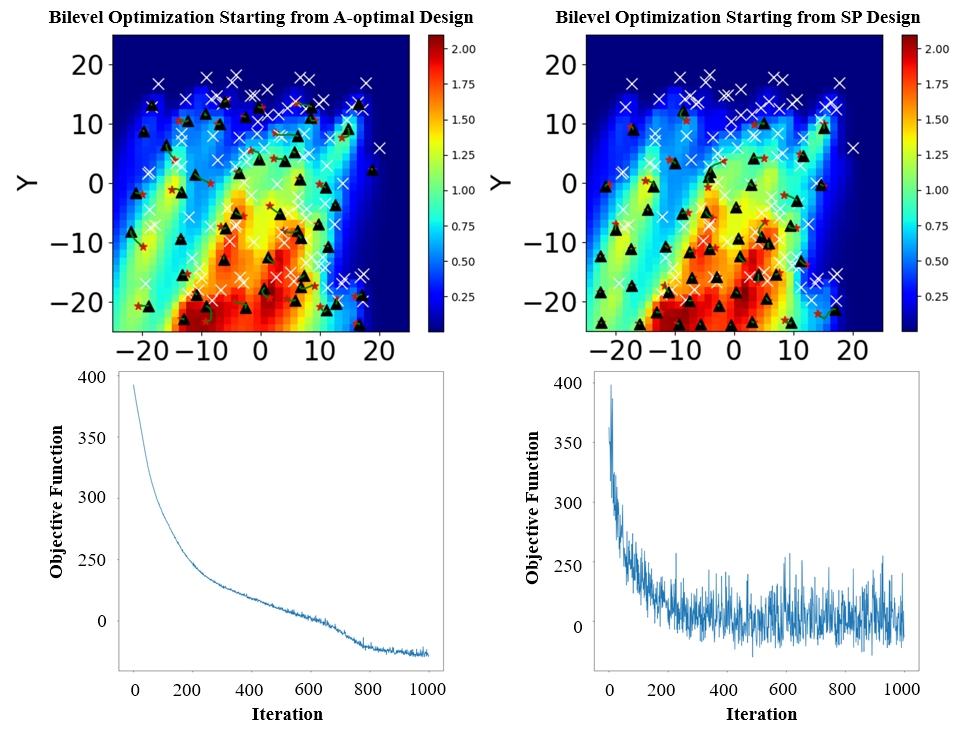}
    \vspace{-8pt}
    \caption{\textcolor{black}{The proposed approach significantly reduces the values of the objective function by moving the initial sensor locations (``$\blacktriangle$'') to the final sensor locations (``$\bigstar$'').}}
    \label{fig:comparison between A and SP}
\end{figure}

\textcolor{black}{
For comparison purposes, we also consider other possible approaches to obtain the initial conditions, including the random design, K-means design, Support Points (SP) design \citep{mak2018support}, GP with nonstationary kernel design \citep{jakkala2023efficient}, and sparse sensor placement for reconstruction (SSPOR) design \citep{manohar2018data}.}
Figure \ref{fig:comparison of initial guess} shows the initial sensor allocation (50 sensors indicated by ``$\blacktriangle$'' for 100 emission sources indicated by ``$\times$'') obtained from different approaches given a concentration field. We compare these initial designs based on two conditions: (i) sensors should not be placed at locations with zero concentration as little useful information will be collected; and (ii) sensors should not be too close to each other (i.e., the layout of sensor locations should exhibit a space-filling pattern). By comparing the six initial designs, we note that only the SP design and $A$-optimal design satisfy the two requirements. 
Hence, based on the initial designs given by the SP design and $A$-optimal design, we run the SBA algorithm to update the sensor locations and obtain the final sensor locations indicated by ``$\bigstar$'' in Figure \ref{fig:comparison between A and SP}. 
It is seen that, the proposed approach can significantly reduce the  values of the objective function starting from the initial designs, while the design initiated with the $A$-optimal design achieves a lower objective value. 
\vspace{-8pt}
\vspace{-18pt}

\subsection{Example II: Sensor Allocation Over a 2D Domain}\label{s:gaussian plume single}
\begin{figure}[h!]
    \centering
    \includegraphics[width=0.3\linewidth]{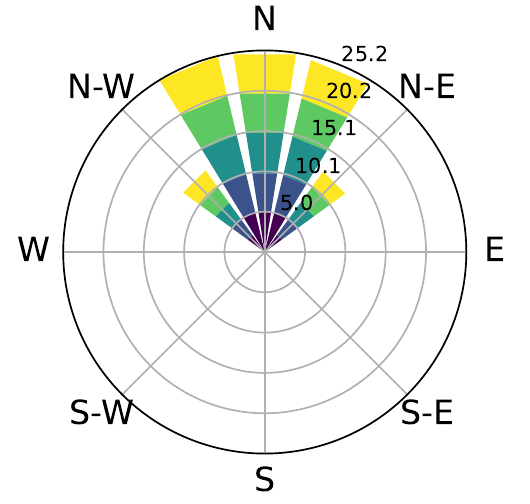}
    \vspace{-8pt}
    \caption{Wind rose plot for Example II.}
    \label{fig:wind}
\end{figure}

In Example II, a more complex problem is considered for which sensors are placed over a continuous 2D domain with 10, 20, 50 and 100 emission sources.
We start with 10 emission sources, $\left\{\bm{x}_j\right\}_{j=1...10}$, distributed over a 2D domain, $[-25,25]\times [-25,25]$. We set the source locations $\left\{\bm{x}_j\right\}_{j=1...10}$ to $\{(-15,17)$, $(-10,-5)$, $(-9,22)$, $(-5,10)$, $(5,18)$, $(5,0)$, $(8,-10)$, $(10,19)$, $(15,-10)$, $(20,5)\}$, and let the emission rate $\bm{\theta}=(\theta_1, ..., \theta_{10})$ follow a multivariate truncated (i.e., nonnegative) normal distribution obtained from a multivariate normal distribution  $ \mathcal{N}(\bm{\mu}_{\text{pr}},\bm{\Gamma}_{\text{pr}})$, where $\bm{\mu}_{\text{pr}}=(8, 10, 9, 8, 10, 9, 8, 10, 9, 10)^T$, $\bm{\Gamma}_{\text{pr}}$ is a diagonal matrix, $\sigma^2_{Pr}\bm{I}$ with $\sigma_{Pr}=20$. 
The standard deviation of the observation noise is set to $0.01$. The distribution of wind vector is shown in Figure \ref{fig:wind}, where the wind speed is uniformly sampled between $[1,2]$, and the wind direction is sampled between north-west and north-east.
The SBA algorithm is used to find the optimal sensor locations. For the \textcolor{black}{lower-level} problem, we let $\lambda_1=\lambda_2=0.01$, and $J=2000$. The learning rate $\tau_{m,j}=0.0005$ for any $m$ and $j$. For the \textcolor{black}{upper-level} loop, the learning rate $\rho_m=0.00005$ for any $m$. The re-sampling size $\tilde{N}$ is set to 100. 

\begin{figure}[h!]
     \centering
     \begin{subfigure}[b]{0.47\textwidth}
         \centering
         \includegraphics[width=0.8\textwidth]{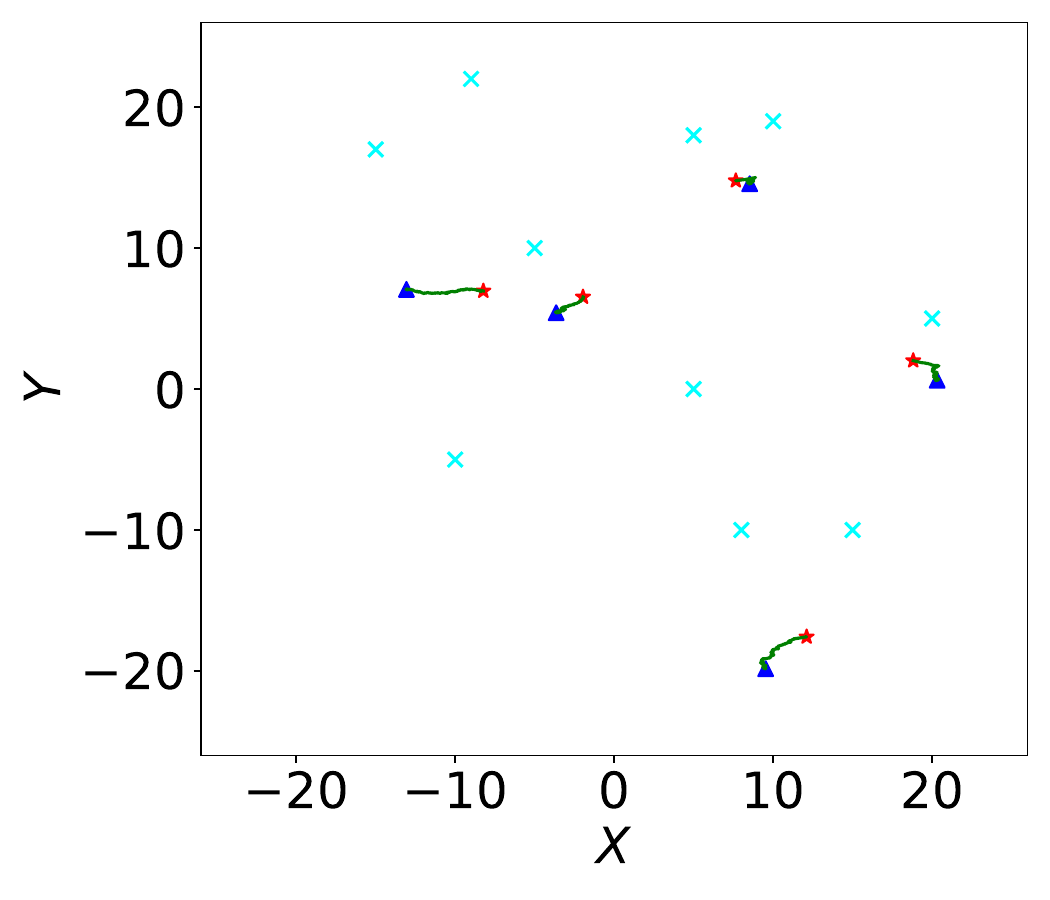}
         \vspace{-8pt}
         \caption{updates of sensor locations ($M$=300)}
         \label{fig:5 sensor updates}
     \end{subfigure}
     \hfill
     \vspace{-8pt}
     \begin{subfigure}[b]{0.47\textwidth}
         \centering
         \includegraphics[width=0.8\textwidth]{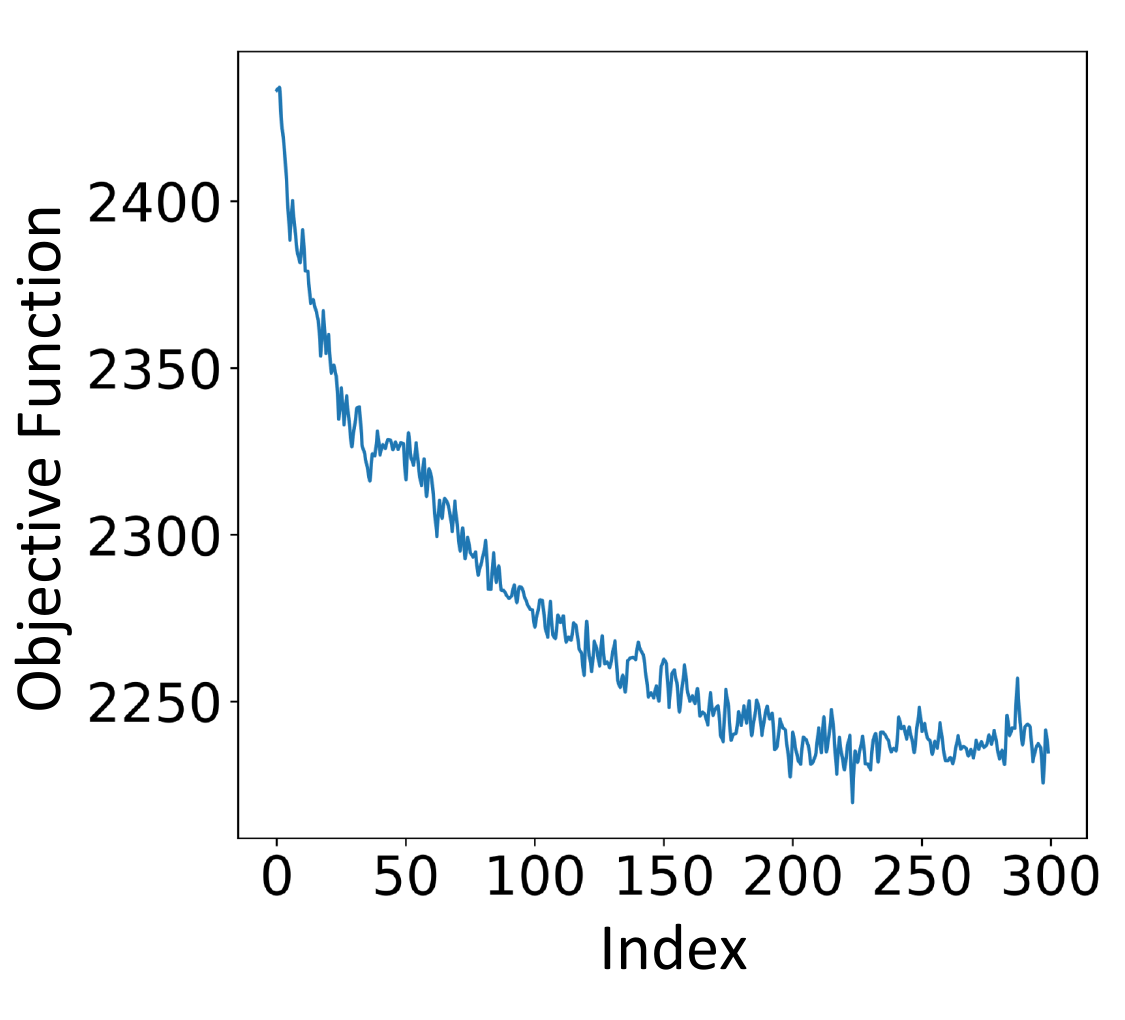}
         \vspace{-8pt}
         \caption{objective value along iterations}
         \label{fig:5 sensor convergence}
     \end{subfigure}
     \caption{Deployment of 5 sensors for 10 emission sources (initial location: $\blacktriangle$; final location: $\bigstar$; sources: $\times$).}
     \label{fig:sensor placement example 1}
\end{figure}

\begin{figure}[h!]
     \centering
     \begin{subfigure}[b]{0.47\textwidth}
         \centering
         \includegraphics[width=0.8\textwidth]{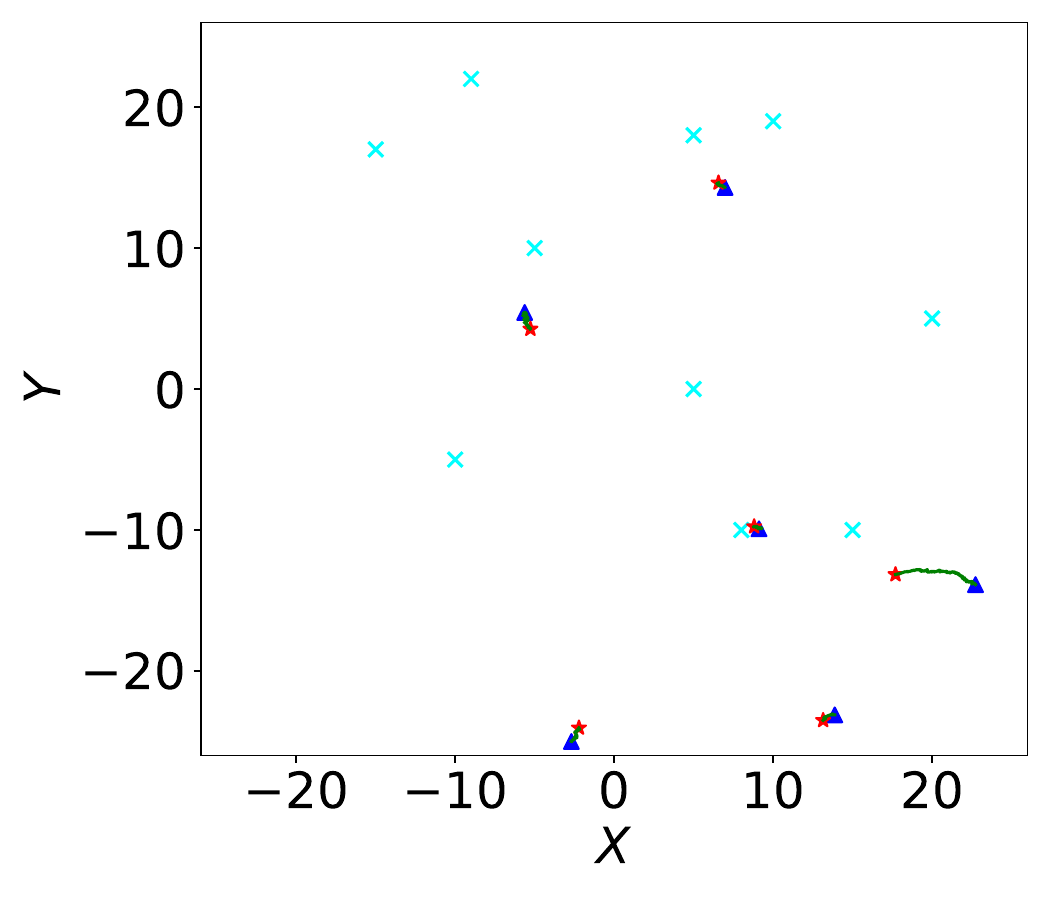}
         \vspace{-8pt}
         \caption{updates of sensor locations ($M$=300)}
         \label{fig:6 sensor updates}
     \end{subfigure}
     \hfill
     \vspace{-8pt}
     \begin{subfigure}[b]{0.47\textwidth}
         \centering
         \includegraphics[width=0.8\textwidth]{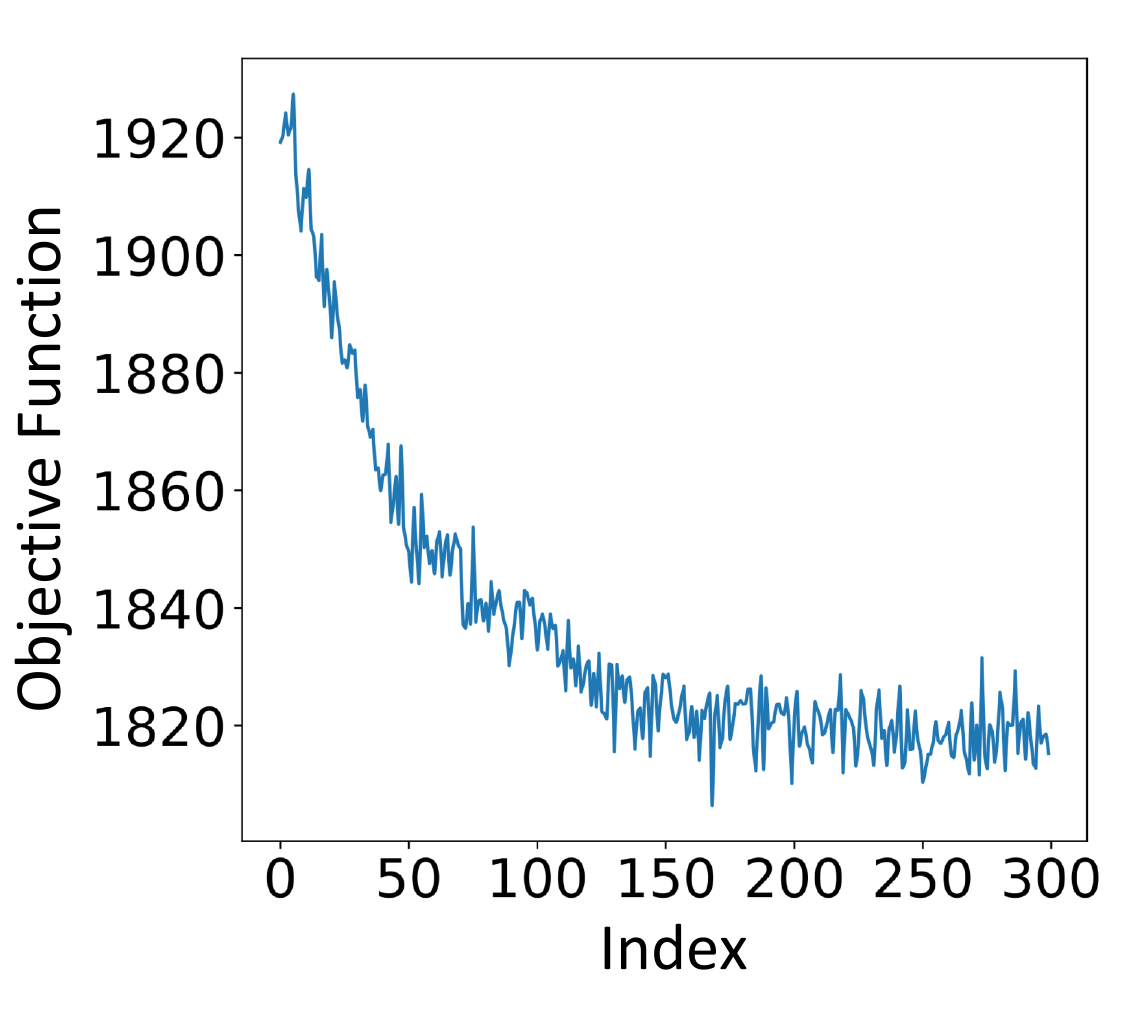}
         \vspace{-8pt}
         \caption{objective value along iterations}
         \label{fig:6 sensor convergence}
     \end{subfigure}
     \caption{Deployment of 6 sensors for 10 emission sources (initial location: $\blacktriangle$; final location: $\bigstar$; sources: $\times$).}
     \label{fig:sensor placement example 2}
\end{figure}


\begin{figure}[h!]
     \centering
     \begin{subfigure}[b]{0.47\textwidth}
         \centering
         \includegraphics[width=0.8\textwidth]{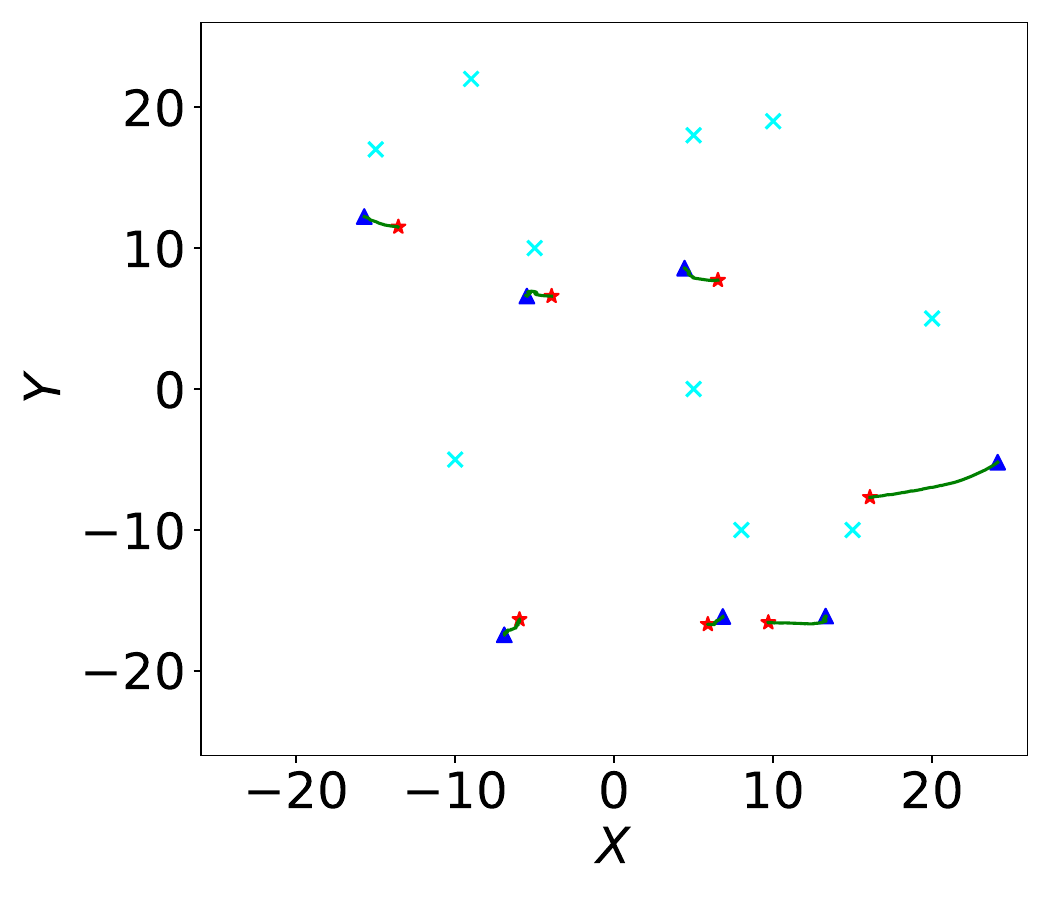}
         \vspace{-8pt}
         \caption{updates of sensor locations ($M$=300)}
         \label{fig:7 sensor updates 1}
     \end{subfigure}
     \hfill
     \begin{subfigure}[b]{0.47\textwidth}
         \centering
         \includegraphics[width=0.8\textwidth]{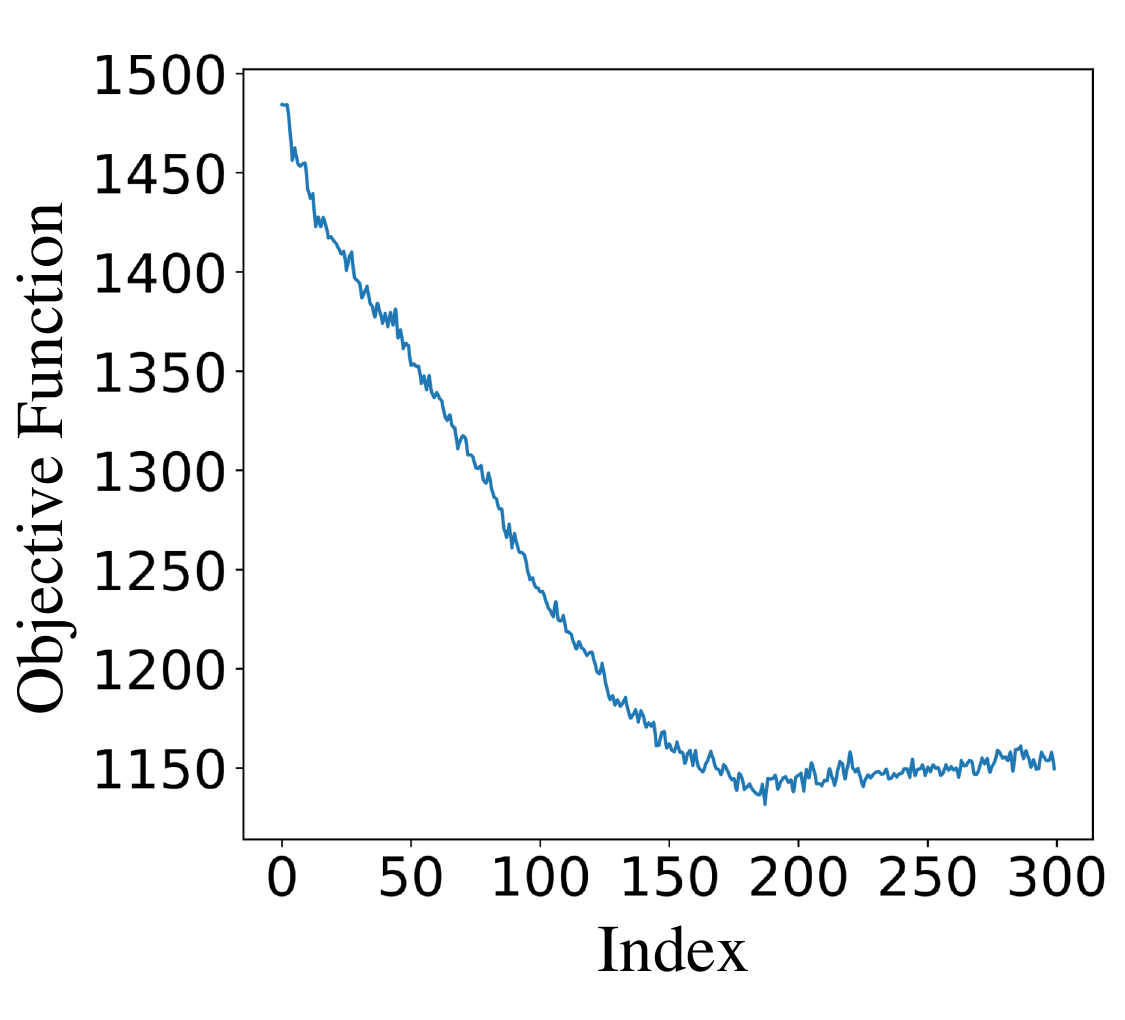}
         \vspace{-8pt}
         \caption{objective value along iterations}
         \label{fig:7 sensor convergence 1}
     \end{subfigure}
     \begin{subfigure}[b]{0.47\textwidth}
         \centering
         \includegraphics[width=0.8\textwidth]{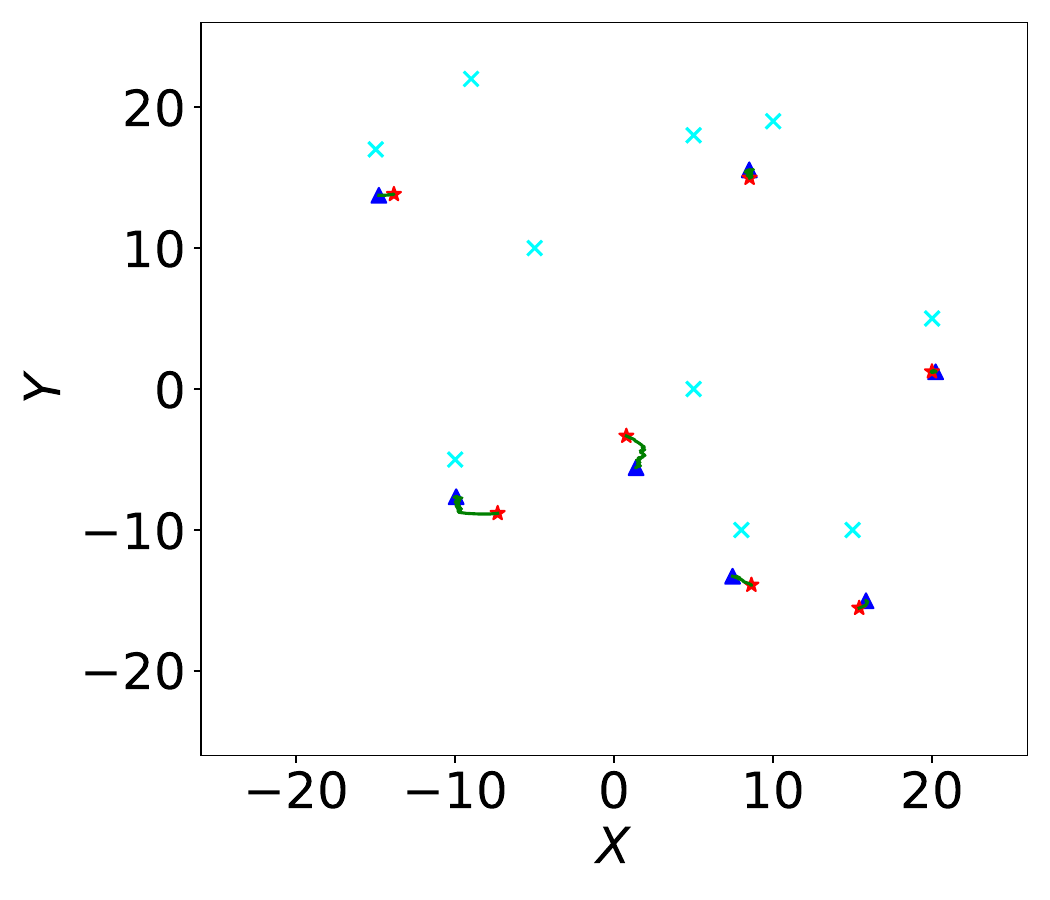}
         \vspace{-8pt}
         \caption{updates of sensor locations ($M$=300)}
         \label{fig:7 sensor updates 2}
     \end{subfigure}
     \hfill
     \begin{subfigure}[b]{0.47\textwidth}
         \centering
         \includegraphics[width=0.8\textwidth]{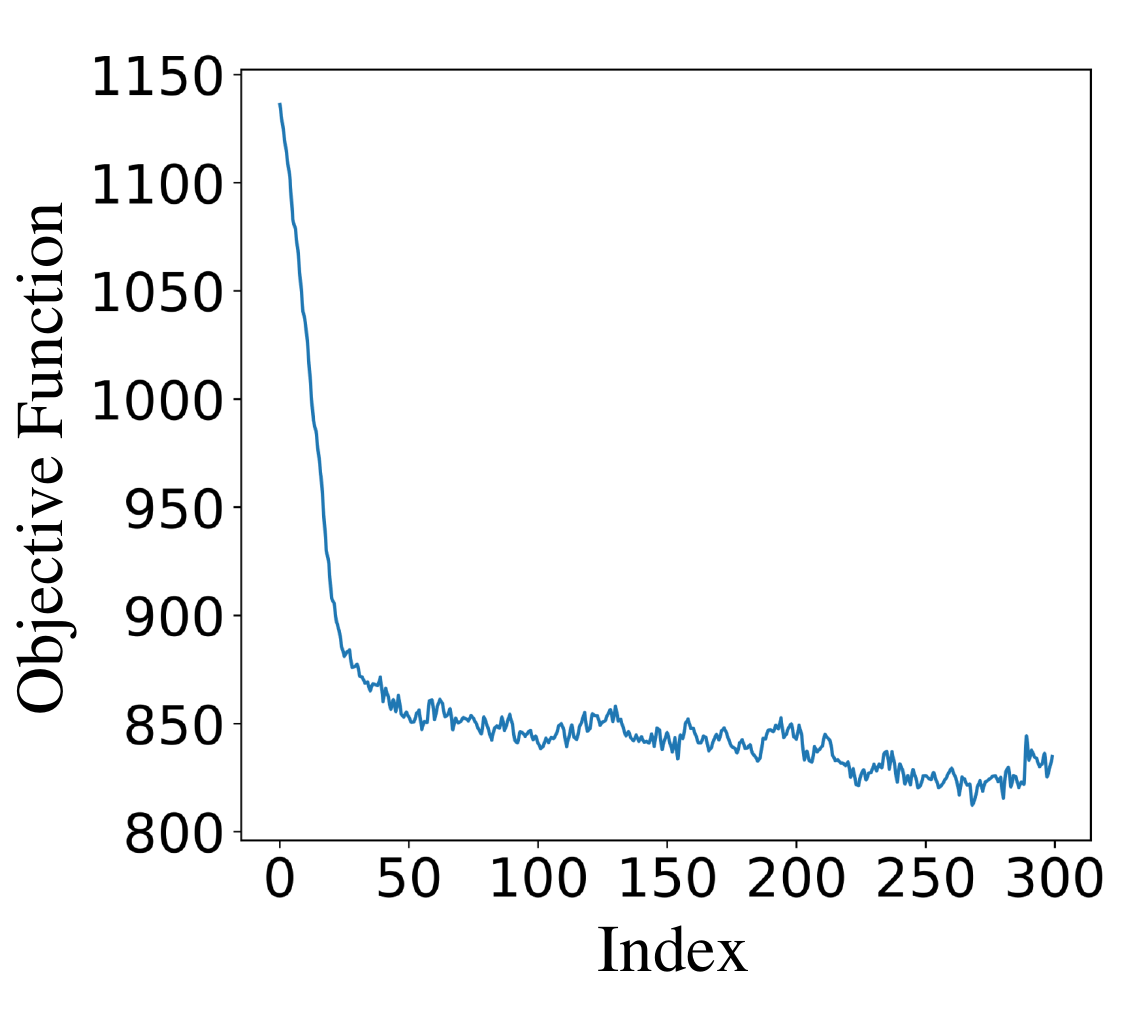}
         \vspace{-8pt}
         \caption{objective value along iterations}
         \label{fig:7 sensor convergence 2}
     \end{subfigure}
     \vspace{-6pt}
     \caption{Allocation of 7 sensors for 10 emission sources with different initial guesses.}
     \label{fig:different initial guess}
\end{figure}


The locations of sensors and the corresponding objective values along the iterations are shown Figures \ref{fig:sensor placement example 1} and \ref{fig:sensor placement example 2}. In these figures, the objective value is re-evaluated with large Monte Carlo samples (i.e., 100,000 samples) for each iteration step, and the iteration number $M$ for the \textcolor{black}{upper-level problem} is chosen according to the computing budget.
We see that the SBA algorithm is able to iteratively optimize sensor allocation with decreasing objective values. In Appendix \ref{s:Appendix 5}, we present more results on different scenarios of the number of sensors, number of emission sources, initial sensor locations, and \textcolor{black}{lower-level} problem iteration limit $J$. 

It is also worth noting that the final sensor locations highly depend on the initial guess. In Figure \ref{fig:7 sensor updates 1} and \ref{fig:7 sensor updates 2}, we generate different initial sensor locations, and obtain different final designs. The solutions reach different local optimums (or saddle points) due to different initial sensor locations, and the objective value also converges differently to the corresponding local minimum as shown in Figure \ref{fig:7 sensor convergence 1} and \ref{fig:7 sensor convergence 2}. 

The \textcolor{black}{lower-level} iteration number $J$ also affects the final designs of sensor locations. A small $J$ affects the choice of the upper-level learning rate $\rho_m$ and upper-level iteration number $M$. Based on our numerical experiments, a small $J$ reduces the total computational time but may cause oscillation along iterations if  the same upper-level learning rate is used. For example, we compare $J=2000$ and $J=200$ for the 7-sensor placement task, as shown by Figure \ref{fig:7 sensor updates 1} and \ref{fig:7 sensor update small q_200} (in the Appendix \ref{s:Appendix 5}). Both settings converge to local optimums but a `ziggy' movement of sensor locations is observed when $J=200$. Considering their similar final objective value, as shown by Figure \ref{fig:7 sensor convergence 1} and \ref{fig:7 sensor convergence small q_200} (in the Appendix \ref{s:Appendix 5}), a small $J$ appears to be good enough to find a local optimum. Of course, the `ziggy' movement, due to a small $J$, could make the solution diverge from the current valley. To avoid the `ziggy' pattern of small $J$, a small \textcolor{black}{upper-level} learning rate $\rho_m$ is needed. Again, this affects the convergence rate: a large $J=2000$ leads to a smaller \textcolor{black}{lower-level} optimality gap, but the computation of the hypergradient becomes more expensive. Since a smaller \textcolor{black}{lower-level} optimality gap makes the upper bound tighter (as shown in Theorem \ref{theorem2}), there is a trade-off between the upper bound assurance and the computational time affected by $J$. 

\begin{figure}[h!]
     \centering
     \begin{subfigure}[b]{0.47\textwidth}
         \centering
         \includegraphics[width=0.8\textwidth]{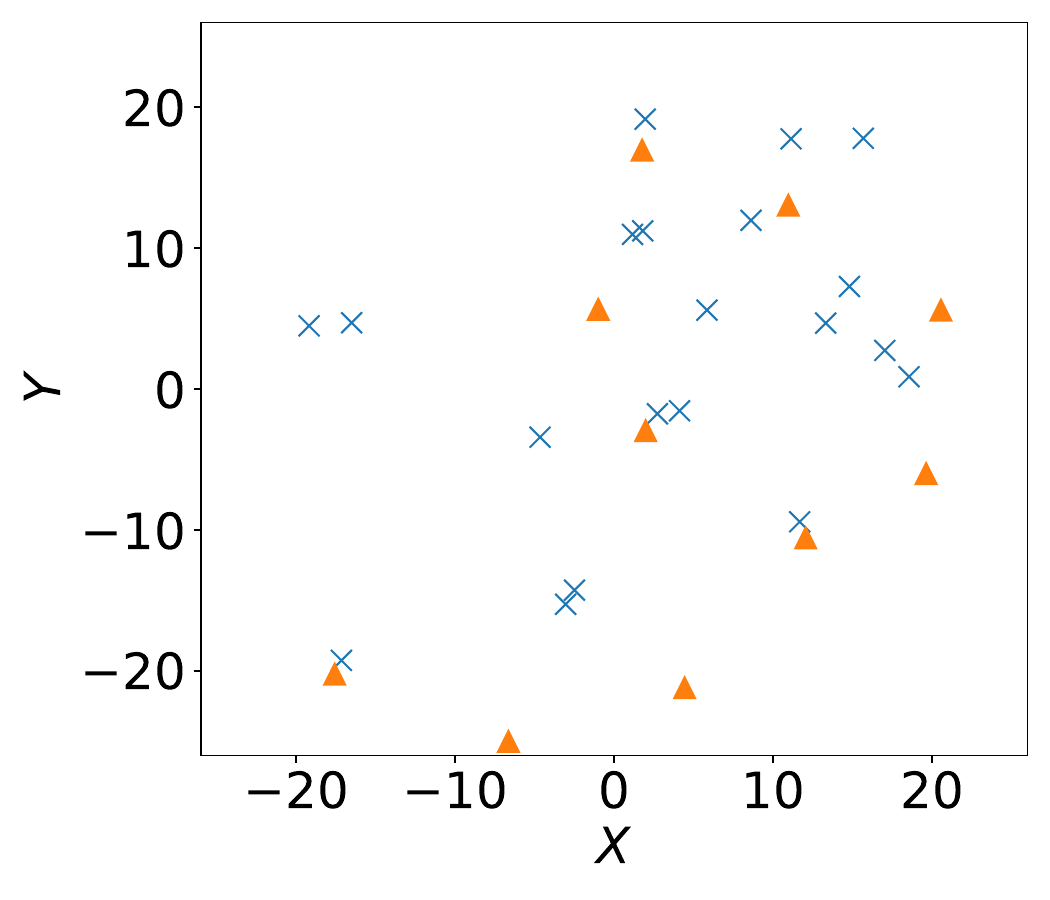}
         \vspace{-8pt}
         \caption{$J=1$, $\rho_m=5\times 10^{-7}$ and $M=3000$}
         \label{fig:20 source 10 sensor placement 1}
     \end{subfigure}
     \hfill
     \vspace{-8pt}
     \begin{subfigure}[b]{0.47\textwidth}
         \centering
         \includegraphics[width=0.8\textwidth]{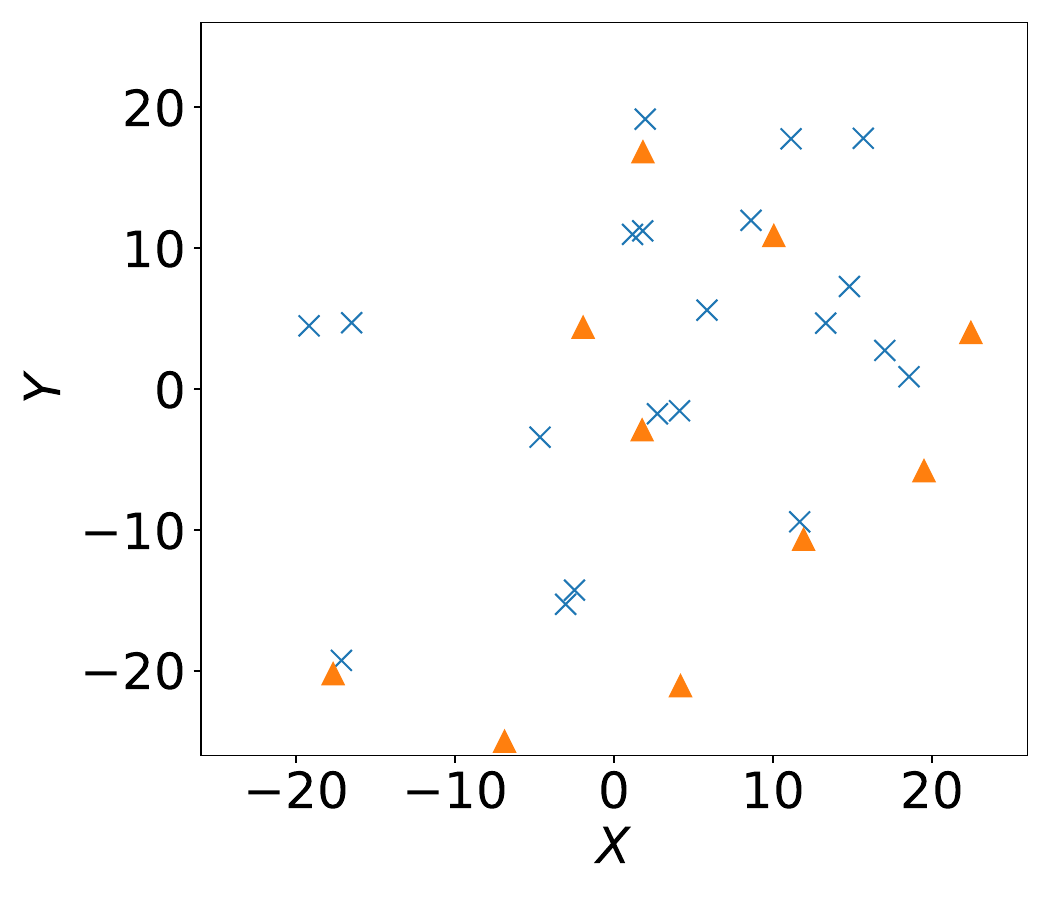}
         \vspace{-8pt}
         \caption{$J=1$, $\rho_m=1\times 10^{-6}$ and $M=3000$}
         \label{fig:20 source 10 sensor placement 2}
     \end{subfigure}
     \vspace{-2pt}
     \caption{Allocation of 10 sensors for 20 emission sources (initial location: $\blacktriangle$; final location: $\bigstar$; sources: $\times$).}
     \label{fig:20 source 10 sensor placement}
\end{figure}

To illustrate the trade-off above, Figure \ref{fig:20 source 10 sensor placement} shows the designs for 20 emission sources whose locations are randomly selected. We compare $\rho_m=5\times 10^{-7}$ and $\rho_m=1\times 10^{-6}$ for $J=1$. 
In this case, $\rho_m=5\times 10^{-7}$ and $\rho_m=1\times 10^{-6}$ lead to similar final designs with the same iteration numbers, so $\rho_m=1\times 10^{-6}$ is better in this case.

\FloatBarrier 

\begin{figure}[h!]
     \centering
     \begin{subfigure}[b]{0.47\textwidth}
         \centering
         \includegraphics[width=0.8\textwidth]{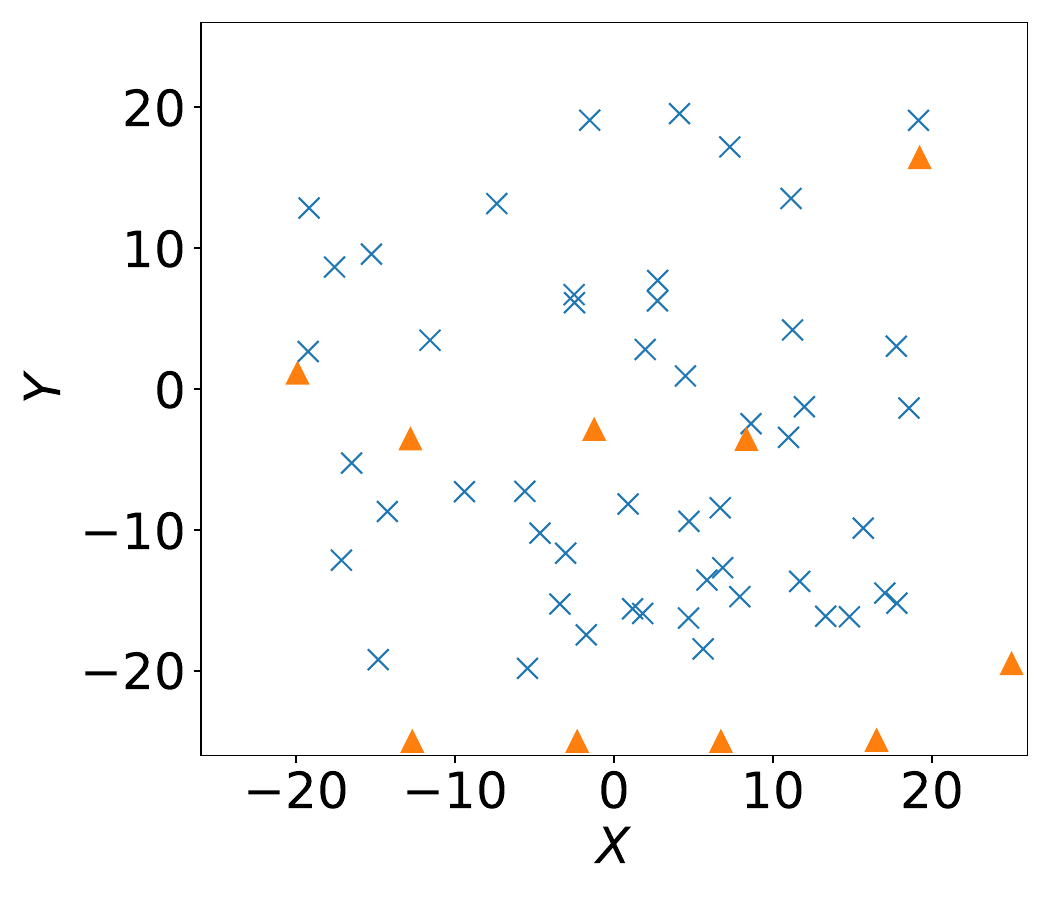}
         \vspace{-8pt}
         \caption{$J=1$, $\rho_m=5\times 10^{-7}$ and $M=3000$}
         \label{fig:20 source 10 sensor placement 2}
     \end{subfigure}
     \hfill
     \vspace{-8pt}
     \begin{subfigure}[b]{0.47\textwidth}
         \centering
         \includegraphics[width=0.8\textwidth]{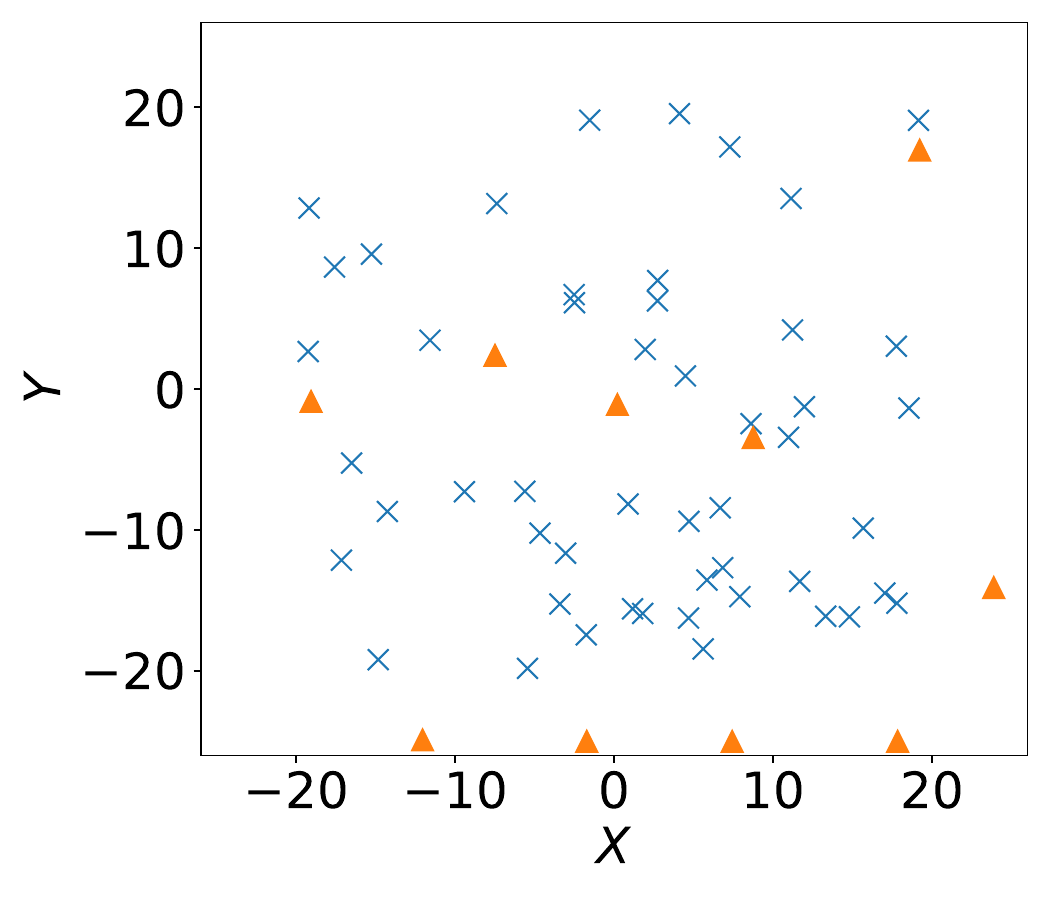}
         \vspace{-8pt}
         \caption{$J=1$, $\rho_m=1\times 10^{-6}$ and $M=300$}
         \label{fig:20 source 10 sensor placement 1}
     \end{subfigure}
     \vspace{-2pt}
     \caption{Allocation of 10 sensors for 50 sources (initial location: $\blacktriangle$; final location: $\bigstar$; sources: $\times$).}
     \label{fig:50 source 10 sensor placement}
\end{figure}


\begin{figure}[h!]
     \centering
     \begin{subfigure}[b]{0.47\textwidth}
         \centering
         \includegraphics[width=0.8\textwidth]{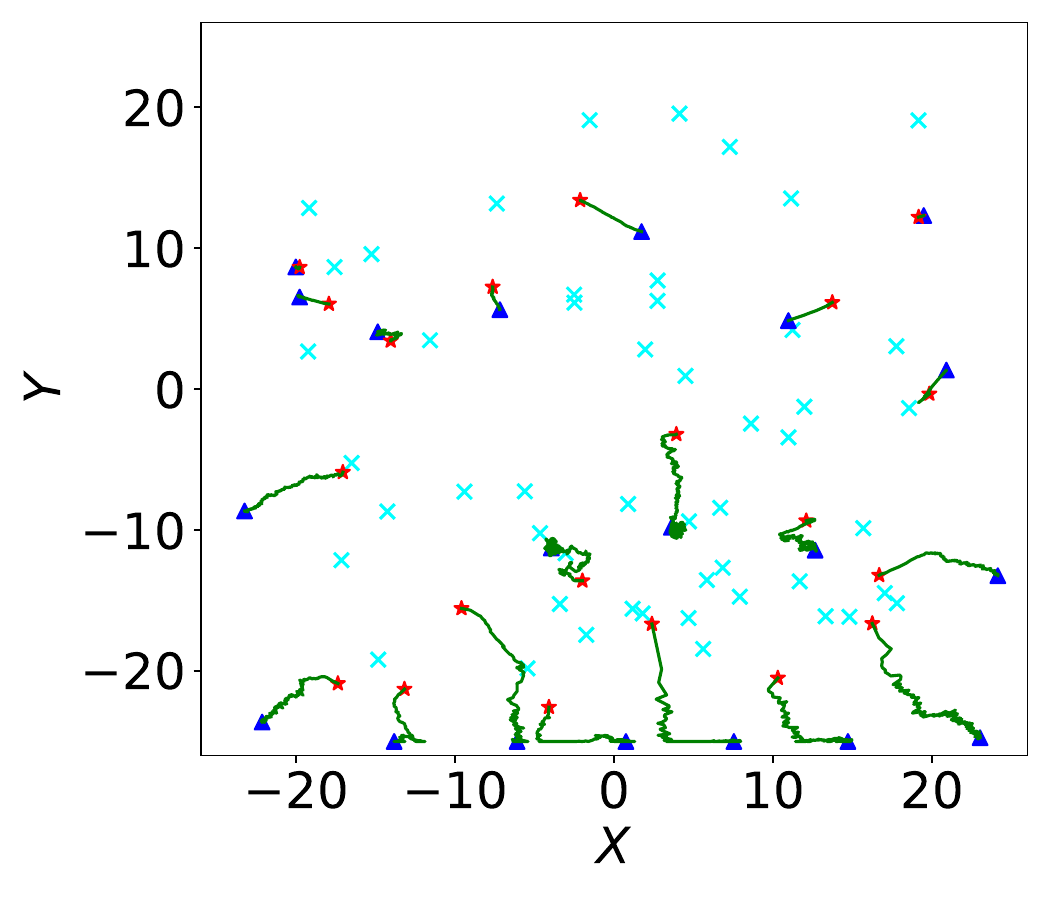}
         \vspace{-8pt}
         \caption{20 sensors, $M$=500}
         \label{fig:50 source 20 sensor 1}
     \end{subfigure}
     \hfill
     \vspace{-4pt}
     \begin{subfigure}[b]{0.47\textwidth}
         \centering
         \includegraphics[width=0.8\textwidth]{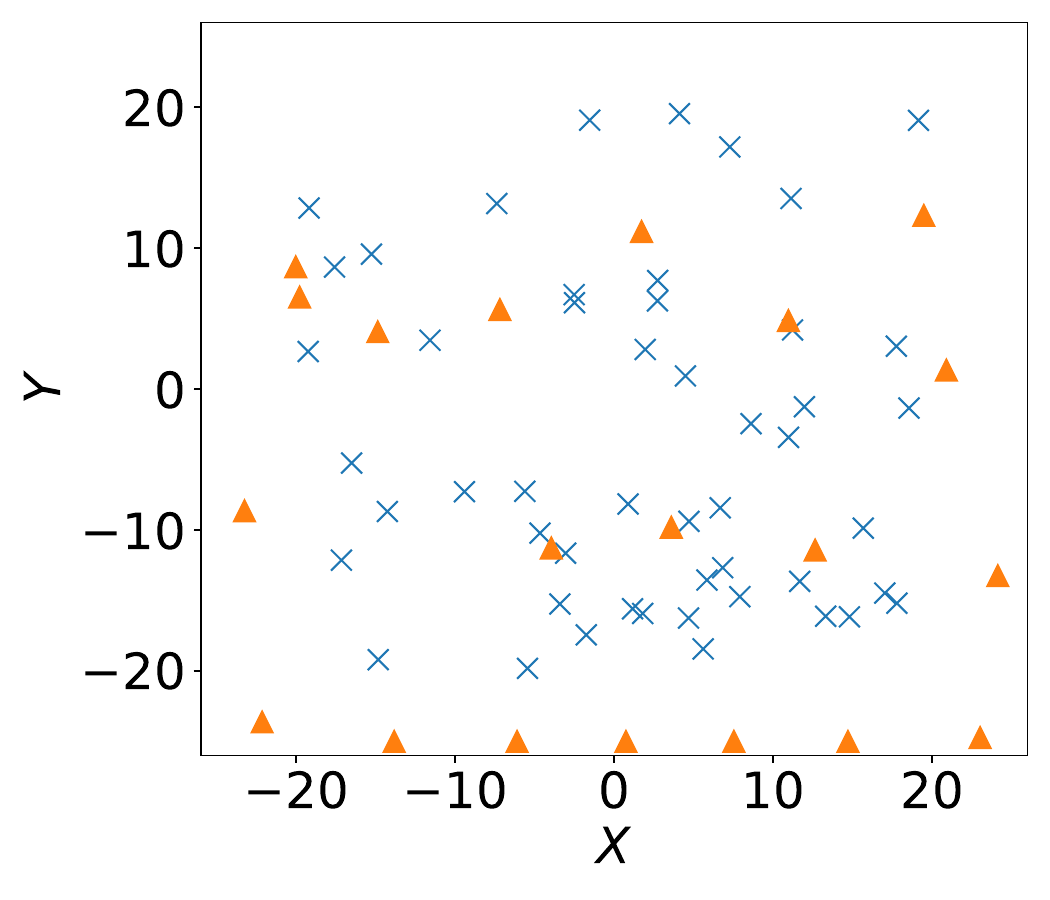}
         \vspace{-8pt}
         \caption{20 sensors, $M$=500}
         \label{fig:50 source 20 sensor 2}
     \end{subfigure}
     \begin{subfigure}[b]{0.47\textwidth}
         \centering
         \includegraphics[width=0.8\textwidth]{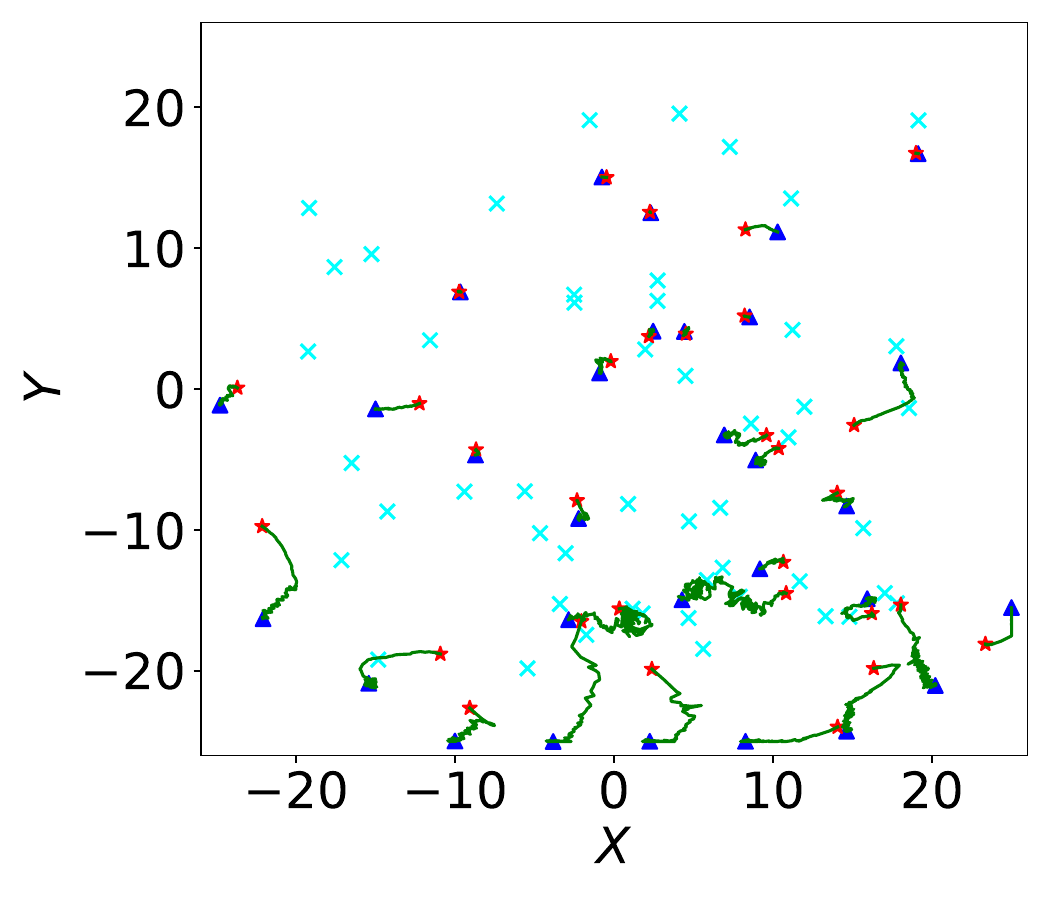}
         \vspace{-8pt}
         \caption{30 sensors, $M$=300}
         \label{fig:50 source 30 sensor 1}
     \end{subfigure}
     \hfill
     \vspace{-8pt}
     \begin{subfigure}[b]{0.47\textwidth}
         \centering
         \includegraphics[width=0.8\textwidth]{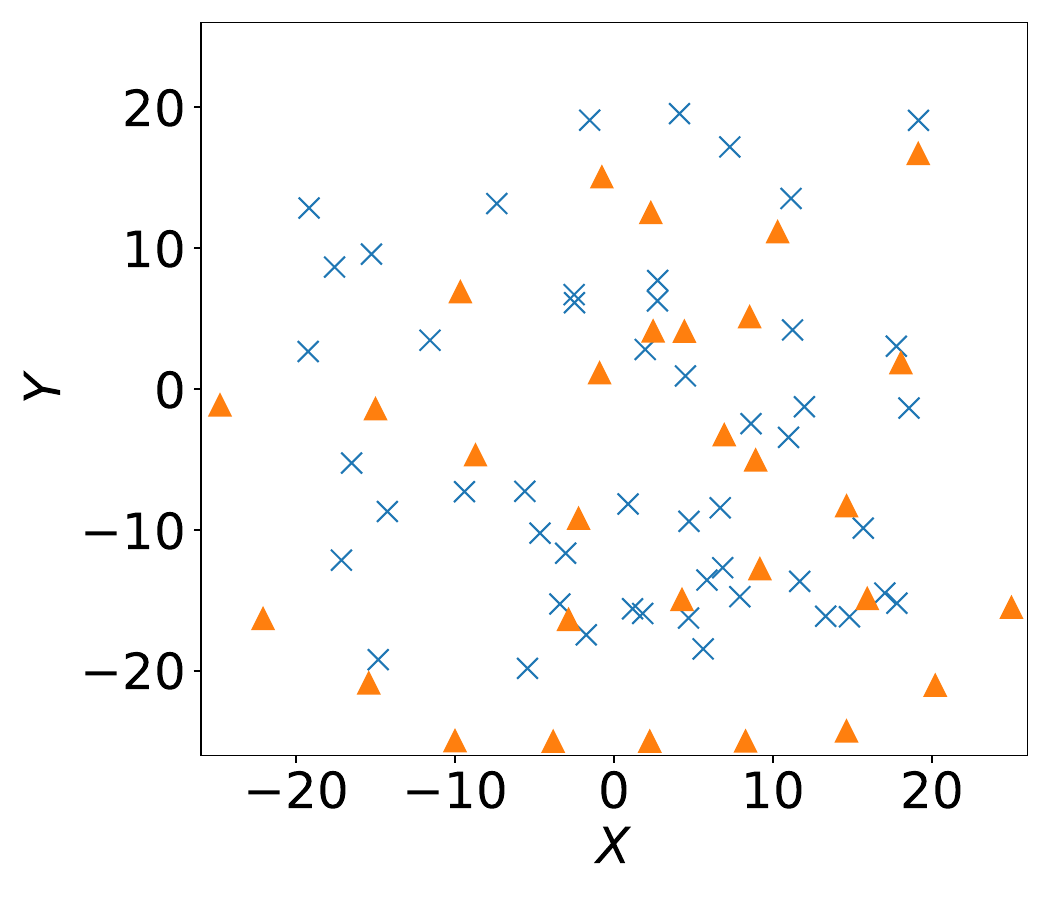}
         \vspace{-8pt}
         \caption{30 sensors, $M$=300}
         \label{fig:50 source 30 sensor 2}
     \end{subfigure}
     \caption{Sensor placement for 50 emission sources (initial location: $\blacktriangle$; final location: $\bigstar$; sources: $\times$).}
     \label{fig:50 source many sensors}
\end{figure}

\begin{figure}[h!]
     \centering
     \begin{subfigure}[b]{0.47\textwidth}
         \centering
         \includegraphics[width=0.8\textwidth]{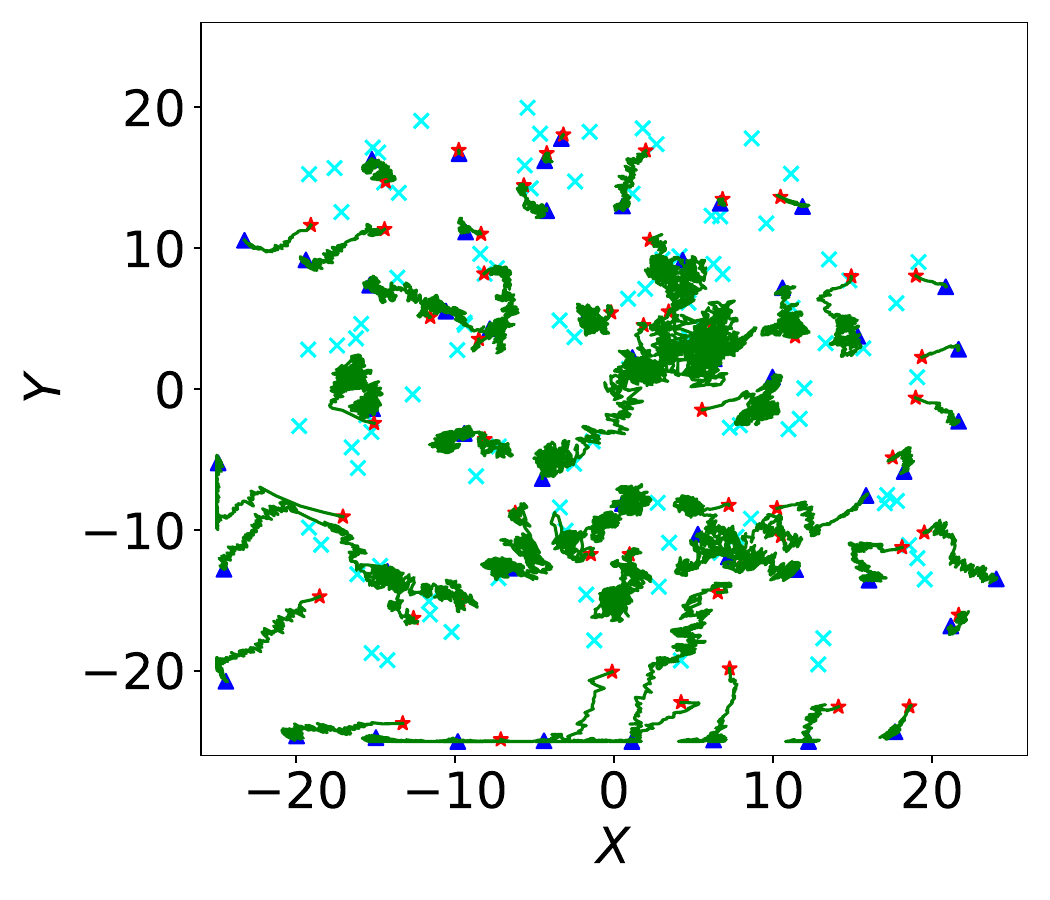}
         \vspace{-8pt}
         \caption{trajectory}
         \label{fig:100 source 50 sensor update}
     \end{subfigure}
     \hfill
     \vspace{-8pt}
     \begin{subfigure}[b]{0.47\textwidth}
         \centering
         \includegraphics[width=0.8\textwidth]{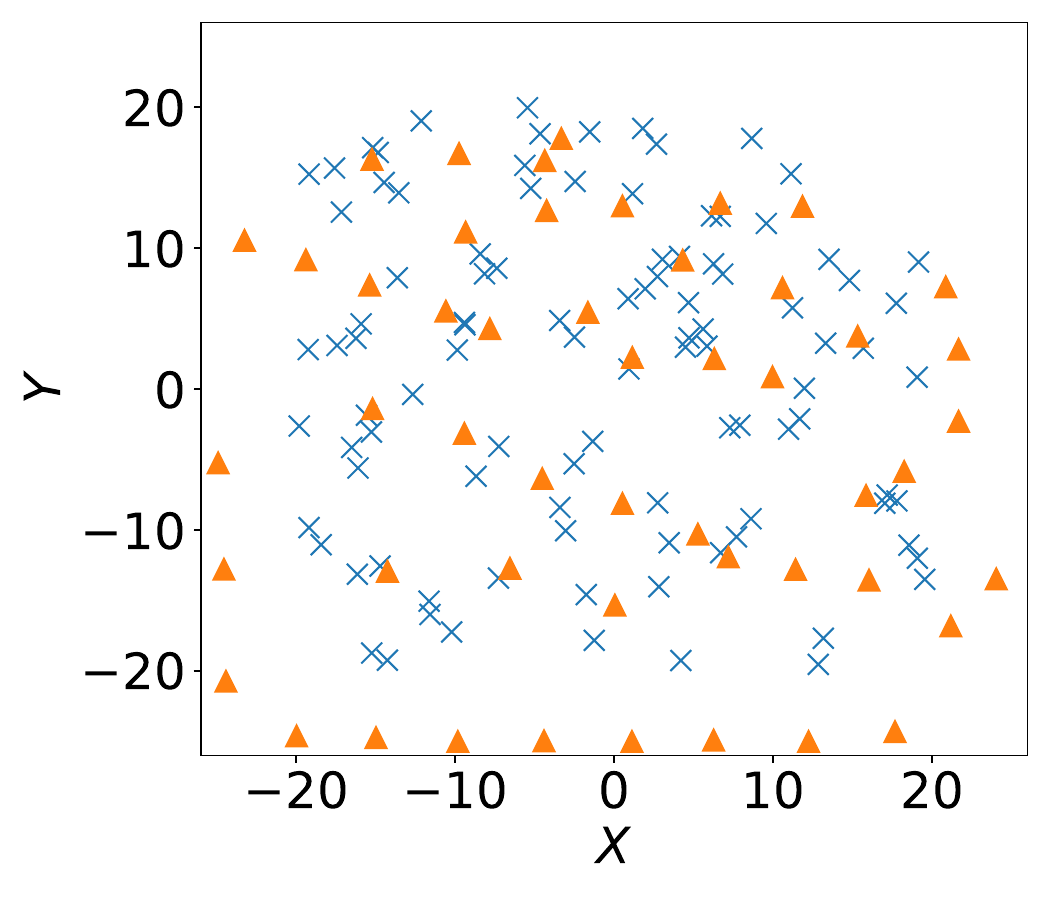}
         \vspace{-8pt}
         \caption{Final locations of sensors}
         \label{fig:100 source 50 sensor location}
     \end{subfigure}
     \caption{Placement of 50 sensors for 100 emission sources with $M=1000$ (initial location: $\blacktriangle$; final location: $\bigstar$; sources: $\times$).}
     \label{fig:100 source many sensors}
\end{figure}

Finally, we place multiple sensors, 10, 20, and 30, for 50 emission sources and place 50 sensors for 100 emission sources. \textcolor{black}{Our optimal sensor allocation method still performs robustly even when the number of sources increases, e.g., 10, 20, 30, 50, and 100 emission sources. It shows scalability that the SBA method can handle larger problem sizes, while maintaining reasonable accuracy by smartly allocating sensors.} When 10 sensors are deployed for 50 sources,  Figure \ref{fig:50 source 10 sensor placement} shows that 4 out of 10 sensors are finally placed on the bottom boundary because of the north-to-south wind direction. The deployment of 20 and 30 sensors are shown in Figure \ref{fig:50 source many sensors}, and the deployment of 50 sensors is shown in Figure \ref{fig:100 source many sensors}. \textcolor{black}{The final designs of sensor locations have a ``space-filling'' pattern that is related to the dispersion processes shown in Figure \ref{fig:comparison of initial guess} or Figure \ref{fig:10 emission scenarios} (in the Appendix \ref{s:Appendix 5}). As the concentration fields of the dispersion processes depend on emission sources, the sensor allocation is determined by the location of the emission sources.} For all of these scenarios, there are always sensors evenly placed on the bottom boundary. \textcolor{black}{Because the wind blows from the north to the south shown in Figure \ref{fig:wind}, there are sensors evenly placed on the bottom (i.e. south). Imagine when the wind blows from the south to the north, one would expect to see sensors being evenly placed on the top (i.e. north). As shown in Figure \ref{fig:100 source many sensors}, some sensors go all the way from north to south and then all the sensors at the bottom become almost evenly distributed. This observation makes perfect sense considering the uncertainty of the dispersion process due to the uncertain wind conditions.}

\FloatBarrier 

\subsection{Validation}\label{s:estimation}
\begin{figure}[h!]
    \centering
    \includegraphics[width=0.6  \linewidth]{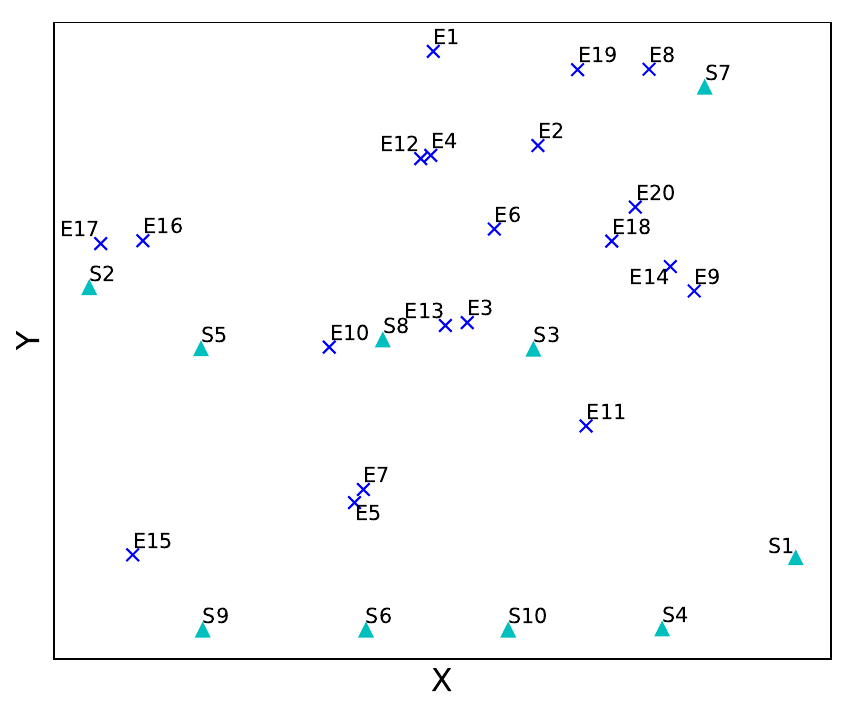}
    \vspace{-12pt}
    \caption{Allocation of 10 sensors (S1-S10) for 20 sources (E1-E20).}
    \label{fig:replot of 10 sensor}
\end{figure}

In this subsection, we further validate the performance of emission estimation based on the sensor allocation obtained above. In particular, we focus on the placement of 10 sensors for 20 sources shown in Figure \ref{fig:replot of 10 sensor}, and compare different designs, emission uncertainties, and observational noise.  

\begin{figure}[h!]
   \centering
   \includegraphics[width=0.65  \linewidth]{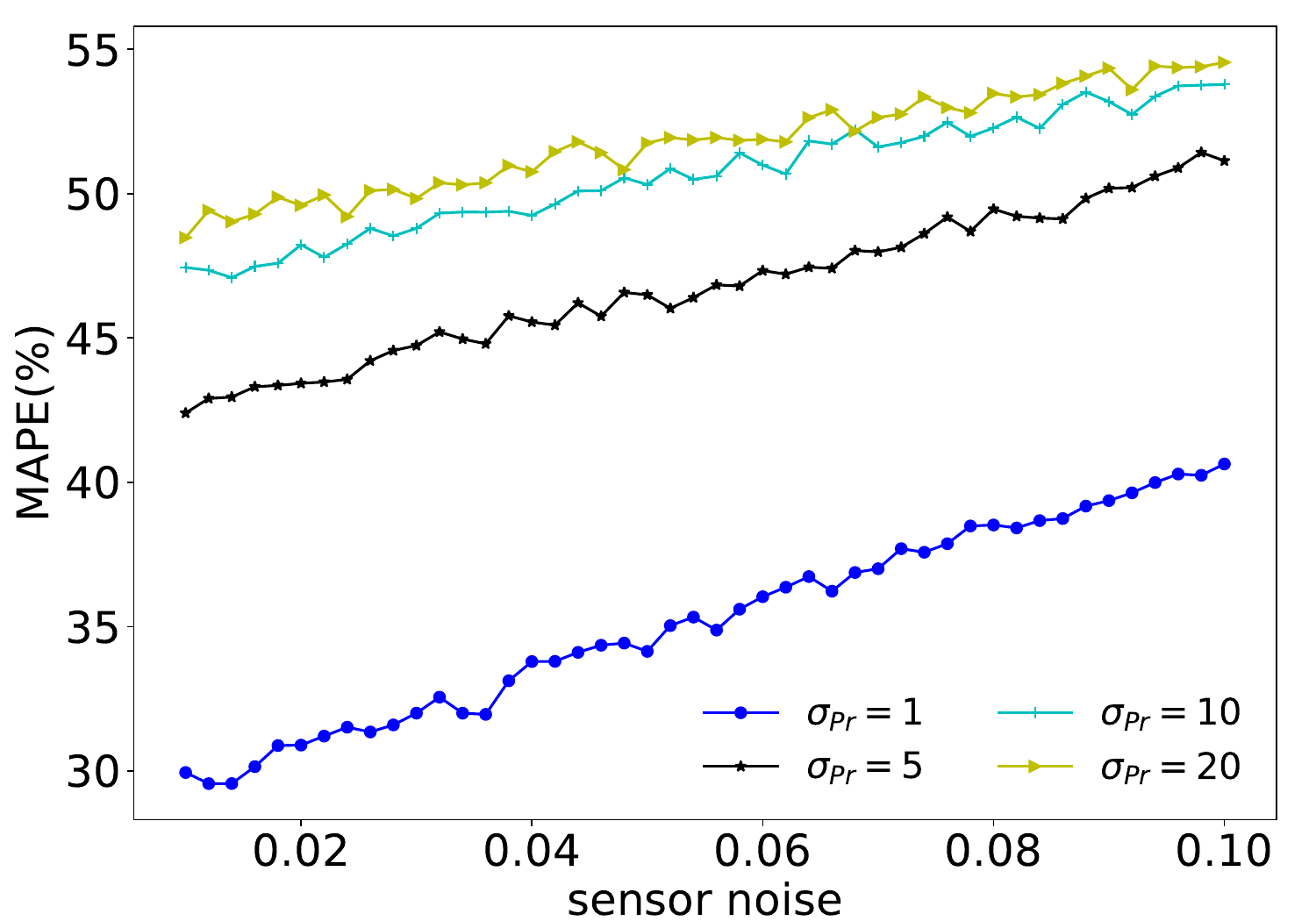}
   \vspace{-8pt}
   \caption{Effect of sensor noise and emission uncertainty on estimation error.}
   \label{fig:MAPE}
\end{figure}

Figure \ref{fig:MAPE} shows the effect of observation noise and emission uncertainty (i.e. $\sigma_{Pr}$) on estimation error. \textcolor{black}{Modern gas-sensing technology can achieve noise levels at or below $2\%$. For example, advanced optical methane detectors (e.g. using infrared spectroscopy) have demonstrated measurement uncertainty on the order of $~1–2\%$ in concentration readings \citep{yang2025improving}. The U.S. EPA’s Method 21 for leak detection requires detectors with noise level within $\pm 2.5\%$ \citep{riddick2023uncertainty}.} It is seen that a larger observation noise increases the estimation error. 
In addition, Figure \ref{fig:MAPE} also shows that the MAPE increases as we increase the uncertainty of emission (i.e., $\sigma_{Pr}$). 

\begin{figure}[h!]
    \begin{subfigure}{\textwidth}
        \includegraphics[width=\textwidth]{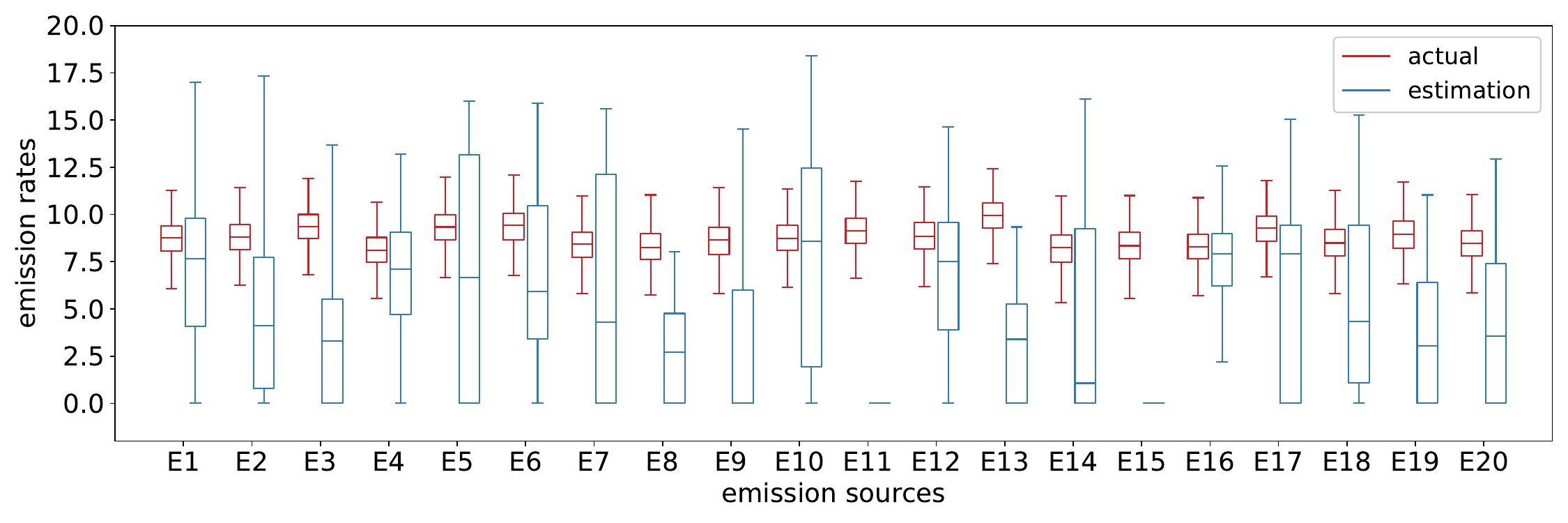}
        \caption{Random design}
        \label{fig:estimation of random design}
    \end{subfigure}
    \begin{subfigure}{\textwidth}
        \includegraphics[width=\textwidth]{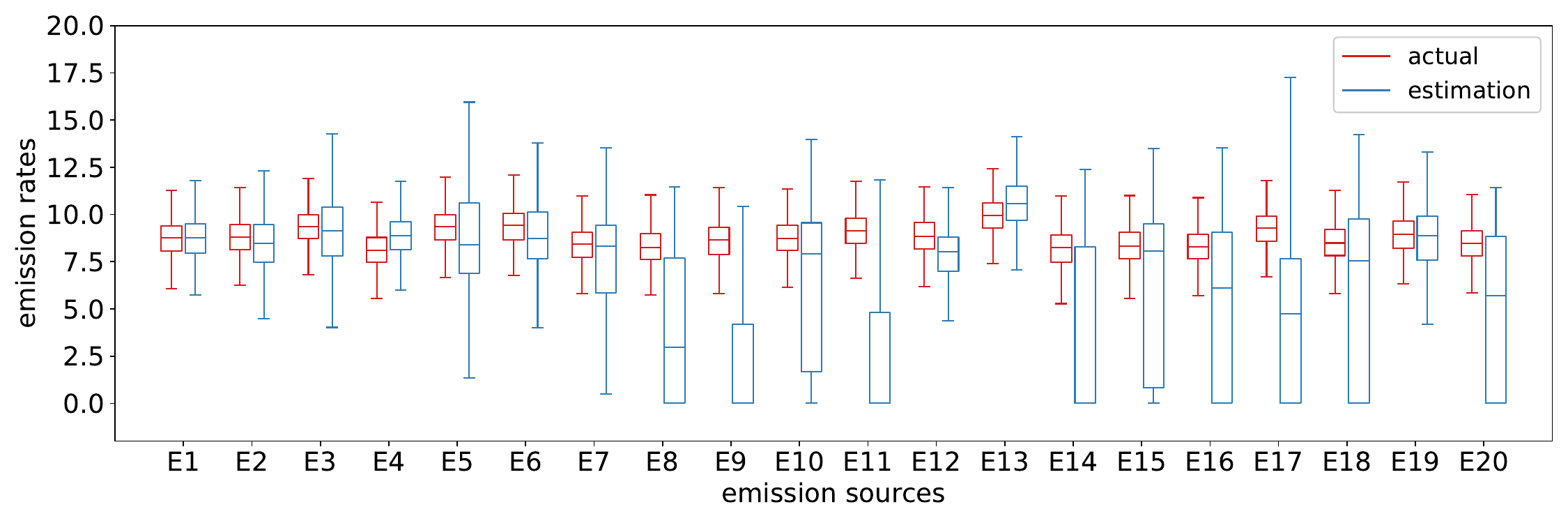}
        \caption{Only using the initial design based on Proposition \ref{p:A2}}
        \label{fig:estimation of A optimal design}
    \end{subfigure}
    \begin{subfigure}{\textwidth}
        \includegraphics[width=\textwidth]{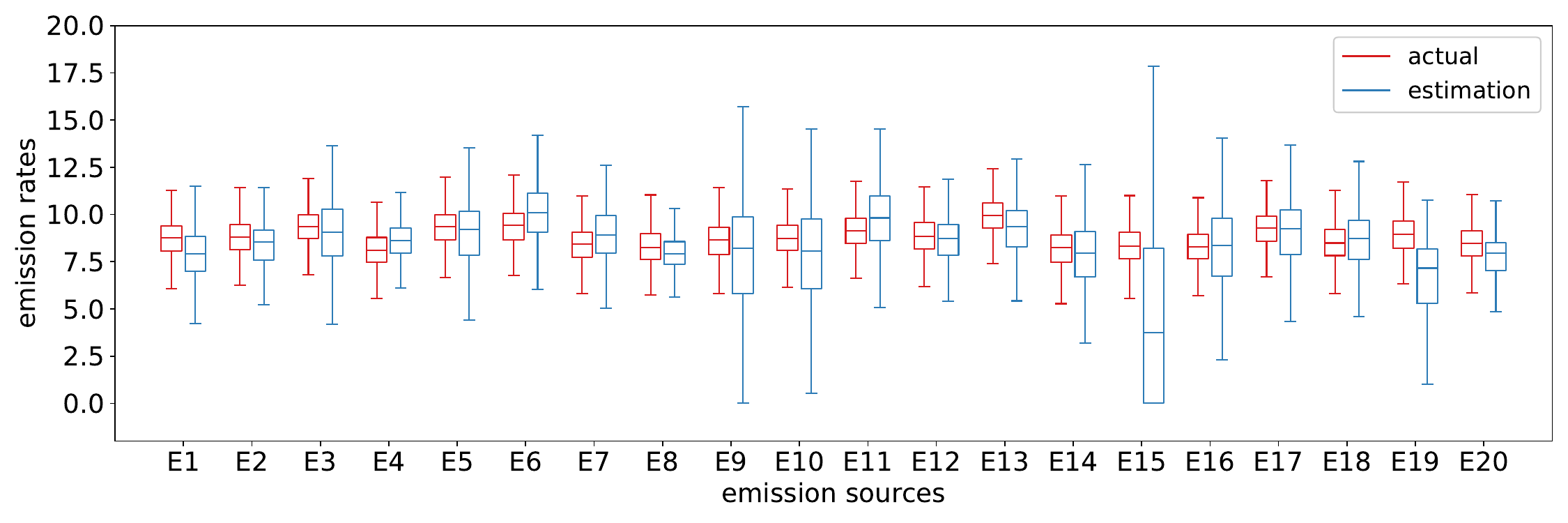}
        \caption{The proposed design by updating the initial design based on Proposition \ref{p:A2}}
        \label{fig:estimation of our design}
    \end{subfigure}
     \caption{Comparison of the estimated emission rates based on different sensor allocations.}
     \label{fig:Estimation of rates}
\end{figure}

Figure \ref{fig:Estimation of rates} shows both the estimated and true emission rates for different emission sources. It is seen that source E15 (at the bottom left corner) is not well covered by the sensor network, and this explains a less accurate estimated emission rate for E15. 
In Figure \ref{fig:Estimation of rates}, we compare the random design (i.e., randomly placed sensors), the initial design based on Proposition \ref{p:A2}, and our design under the same settings. It is seen that the boxplots of actual emission rates are closer to that of the estimated rates based on our design. The MAPE (Mean Absolute Percentage Error) are respectively $69.06\%$, $50.79\%$ and $29.94\%$ for the random design, the initial design based on Proposition \ref{p:A2}, and the optimal design obtained. \textcolor{black}{The U.S. regulatory guidelines for methane leak quantification allow up $\pm 30\%$ uncertainty in emission rates \citep{FederalGuidelines}. As a standard, the American Carbon Registry (ACR) carbon credit methodology, targets $\pm 20\%$ uncertainty. Thus, a $~30\%$ error is at the upper end of acceptable in environmental monitoring. Importantly, our optimized sensor placement achieves this level with a far fewer number of sensors than a naive approach.}
\clearpage

\subsection{Code and Computational Time}
\begin{figure}[h!]
   \centering
   \includegraphics[width=1  \linewidth]{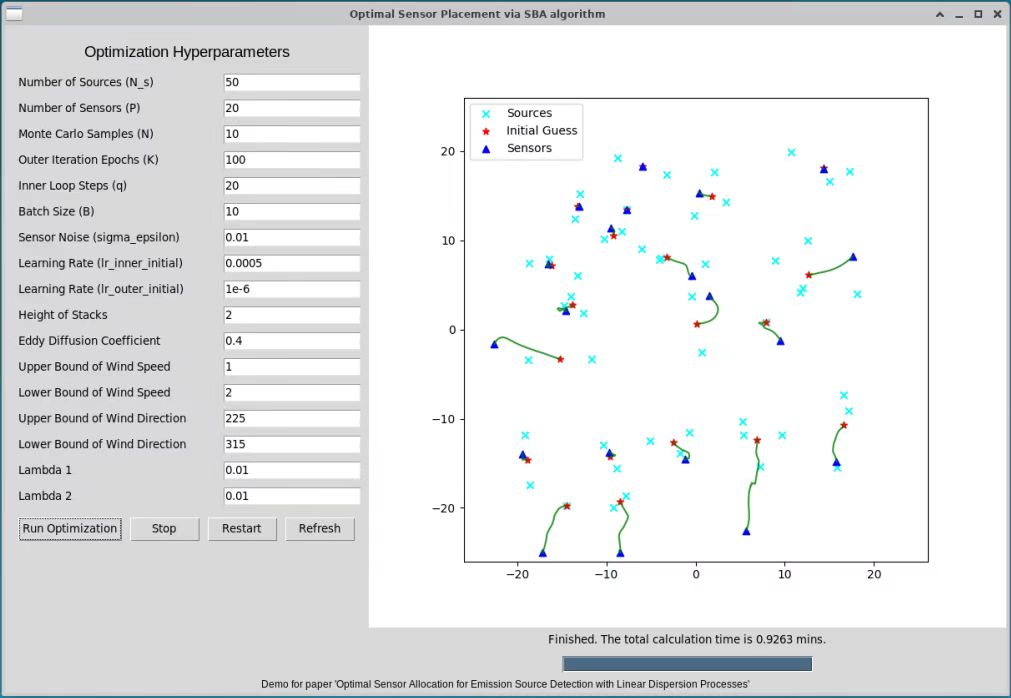}
   \vspace{-8pt}
   \caption{\textcolor{black}{The screenshot of the GUI of the code that implements the proposed approach.}}
   \label{fig:GUI}
\end{figure}

\textcolor{black}{To facilitate the implementation of the approach described in this paper, we provide the code which is available at Github: \textit{https://github.com/lxc95/Optimal-Experimental-Design}. The software GUI is shown in Figure \ref{fig:GUI}. For example, it is seen that the computational time to place 20 sensors for 50 sources is 0.9263 minutes using the hyperparameters shown in  Figure \ref{fig:GUI}. Given the computational complexity of the SBA algorithm, $\mathcal{O}(M\cdot\tilde{N}\cdot(J\cdot N_p+n\cdot N_p^2 + N_p^3))$, the computation time can be dramatically reduced by GPU acceleration (GPU A6000 is used for the computation in our numerical examples). Finally, it is worth noting that the use of the closed-form expression of the gradients provided in Sections \ref{s:update lower-level solution} and \ref{s:update outer solution} makes our approach much faster than the numerical `autograd' function in Pytorch.}

%


\section{Conclusions}\label{s:conclusion}
This paper provided comprehensive investigations, technical details, in-depth discussions and implementation of the optimal sensor placement problem for linear dispersion processes using the framework of bilevel optimization. Compared with the existing linear Gaussian Bayesian inversion framework, the proposed framework provided a more general and realistic solution by relaxing the Gaussian distributional assumption on emission rates, incorporating non-negativity constraints on emission rates as well as parameter uncertainties associated with the forward model. As a consequence, no closed-form solutions are available and the proposed approach must rely on computationally efficient numerical algorithms. 
Therefore, two algorithms, including rSAA and SBA, have been thoroughly investigated for solving the proposed bilevel optimization. Closed-form expressions of the gradients in both the upper- and lower-level problems were obtained that greatly accelerate the algorithms. The convergency results have been established to show the performance guarantee. Comprehensive numerical investigations have been performed, and useful insights have been generated to show how the performance of the algorithms are affected by different model parameter settings. It is shown that the proposed bilevel optimization approach can significantly improve the accuracy of the inverse estimation over some of the existing designs. Finally, code is provided to make it possible to users to adopt our approach. 


An important and also extremely challenging future research is to consider non-linear forward dispersion models.  
When the linear dispersion model is replaced by the nonlinear one, the bilevel optimization framework might still be able to accommodate inverse mapping by approximating the forward models by deep neural networks, such as amortized variational inference (AVI) \citep{ganguly2023amortized} and physics-informed machine learning \citep{daw2022source,wu2023large}. 


\bibliographystyle{chicago}
\spacingset{1}
\bibliography{IISE-Trans}

\begin{thebibliography}{}

\bibitem[\protect\citeauthoryear{Alexanderian, Petra, Stadler, and
  Ghattas}{Alexanderian et~al.}{2014}]{alexanderian2014optimal}
Alexanderian, A., N.~Petra, G.~Stadler, and O.~Ghattas (2014).
\newblock A-optimal design of experiments for infinite-dimensional bayesian
  linear inverse problems with regularized $\ell$\_0-sparsification.
\newblock {\em SIAM Journal on Scientific Computing\/}~{\em 36\/}(5),
  A2122--A2148.

\bibitem[\protect\citeauthoryear{Antil, Di, and Khatri}{Antil
  et~al.}{2020}]{antil2020bilevel}
Antil, H., Z.~W. Di, and R.~Khatri (2020).
\newblock Bilevel optimization, deep learning and fractional laplacian
  regularization with applications in tomography.
\newblock {\em Inverse Problems\/}~{\em 36\/}(6), 064001.

\bibitem[\protect\citeauthoryear{Attia, Leyffer, and Munson}{Attia
  et~al.}{2023}]{attia2023}
Attia, A., S.~Leyffer, and T.~Munson (2023).
\newblock Robust a-optimal experimental design for bayesian inverse problems.
\newblock {\em arXiv preprint arXiv:2305.03855v1\/}.

\bibitem[\protect\citeauthoryear{Brunton, Brunton, Proctor, and Kutz}{Brunton
  et~al.}{2016}]{brunton2016sparse}
Brunton, B.~W., S.~L. Brunton, J.~L. Proctor, and J.~N. Kutz (2016).
\newblock Sparse sensor placement optimization for classification.
\newblock {\em SIAM Journal on Applied Mathematics\/}~{\em 76\/}(5),
  2099--2122.

\bibitem[\protect\citeauthoryear{Chen, Modi, McGaughey, Kimura,
  McDonald-Buller, and Allen}{Chen et~al.}{2022}]{chen2022senet}
Chen, Q., M.~Modi, G.~McGaughey, Y.~Kimura, E.~McDonald-Buller, and D.~T. Allen
  (2022).
\newblock Simulated methane emission detection capabilities of continuous
  monitoring networks in an oil and gas production region.
\newblock {\em Atmosphere\/}~{\em 13\/}(4).

\bibitem[\protect\citeauthoryear{Chepuri and Leus}{Chepuri and
  Leus}{2014}]{chepuri2014continuous}
Chepuri, S.~P. and G.~Leus (2014).
\newblock Continuous sensor placement.
\newblock {\em IEEE signal processing letters\/}~{\em 22\/}(5), 544--548.

\bibitem[\protect\citeauthoryear{Chow, Kosovi{\'c}, and Chan}{Chow
  et~al.}{2008}]{chow2008source}
Chow, F.~K., B.~Kosovi{\'c}, and S.~Chan (2008).
\newblock Source inversion for contaminant plume dispersion in urban
  environments using building-resolving simulations.
\newblock {\em Journal of applied meteorology and climatology\/}~{\em 47\/}(6),
  1553--1572.

\bibitem[\protect\citeauthoryear{Cusworth, Duren, Thorpe, Olson-Duvall,
  Heckler, Chapman, Eastwood, Helmlinger, Green, Asner, et~al.}{Cusworth
  et~al.}{2021}]{cusworth2021intermittency}
Cusworth, D.~H., R.~M. Duren, A.~K. Thorpe, W.~Olson-Duvall, J.~Heckler, J.~W.
  Chapman, M.~L. Eastwood, M.~C. Helmlinger, R.~O. Green, G.~P. Asner, et~al.
  (2021).
\newblock Intermittency of large methane emitters in the permian basin.
\newblock {\em Environmental Science \& Technology Letters\/}~{\em 8\/}(7),
  567--573.

\bibitem[\protect\citeauthoryear{Daw, Karpatne, Yeo, and Klein}{Daw
  et~al.}{2022}]{daw2022source}
Daw, A., A.~Karpatne, K.~Yeo, and L.~Klein (2022).
\newblock Source identification and field reconstruction of advection-diffusion
  process from sparse sensor measurements.
\newblock Conference on Neural Information Processing Systems.

\bibitem[\protect\citeauthoryear{de~Silva, Manohar, Clark, Brunton, Brunton,
  and Kutz}{de~Silva et~al.}{2021}]{de2021pysensors}
de~Silva, B.~M., K.~Manohar, E.~Clark, B.~W. Brunton, S.~L. Brunton, and J.~N.
  Kutz (2021).
\newblock Pysensors: A python package for sparse sensor placement.
\newblock {\em arXiv preprint arXiv:2102.13476\/}.

\bibitem[\protect\citeauthoryear{Ganguly, Jain, and Watchareeruetai}{Ganguly
  et~al.}{2023}]{ganguly2023amortized}
Ganguly, A., S.~Jain, and U.~Watchareeruetai (2023).
\newblock Amortized variational inference: A systematic review.
\newblock {\em Journal of Artificial Intelligence Research\/}~{\em 78},
  167--215.

\bibitem[\protect\citeauthoryear{Giovannelli, Kent, and Vicente}{Giovannelli
  et~al.}{2021}]{giovannelli2021inexact}
Giovannelli, T., G.~Kent, and L.~N. Vicente (2021).
\newblock Inexact bilevel stochastic gradient methods for constrained and
  unconstrained lower-level problems.
\newblock {\em arXiv preprint arXiv:2110.00604\/}.

\bibitem[\protect\citeauthoryear{Golub, Hansen, and O'Leary}{Golub
  et~al.}{1999}]{golub1999tikhonov}
Golub, G.~H., P.~C. Hansen, and D.~P. O'Leary (1999).
\newblock Tikhonov regularization and total least squares.
\newblock {\em SIAM journal on matrix analysis and applications\/}~{\em
  21\/}(1), 185--194.

\bibitem[\protect\citeauthoryear{Guidelines}{Guidelines}{2022}]{FederalGuidelines}
Guidelines, F. (2022).
\newblock Assessing methane emissions from orphaned wells to meet reporting
  requirements of the 2021 infrastructure investment and jobs act (bil):
  Federal program guidelines.
\newblock {\em Available online, https://www.doi.gov/sites/
  doi.gov/files/federal-orphaned-wells-methane-measurement-guidelines-final-for-posting-v2.pdf\/}.

\bibitem[\protect\citeauthoryear{Haber, Horesh, and Tenorio}{Haber
  et~al.}{2009}]{haber2009numerical}
Haber, E., L.~Horesh, and L.~Tenorio (2009).
\newblock Numerical methods for the design of large-scale nonlinear discrete
  ill-posed inverse problems.
\newblock {\em Inverse Problems\/}~{\em 26\/}(2), 025002.

\bibitem[\protect\citeauthoryear{Haber, Magnant, Lucero, and Tenorio}{Haber
  et~al.}{2012}]{haber2012numerical}
Haber, E., Z.~Magnant, C.~Lucero, and L.~Tenorio (2012).
\newblock Numerical methods for a-optimal designs with a sparsity constraint
  for ill-posed inverse problems.
\newblock {\em Computational Optimization and Applications\/}~{\em 52},
  293--314.

\bibitem[\protect\citeauthoryear{Herring, Nagy, and Ruthotto}{Herring
  et~al.}{2018}]{herring2018lap}
Herring, J.~L., J.~G. Nagy, and L.~Ruthotto (2018).
\newblock Lap: a linearize and project method for solving inverse problems with
  coupled variables.
\newblock {\em Sampling Theory in Signal and Image Processing\/}~{\em 17},
  127--151.

\bibitem[\protect\citeauthoryear{Houweling, Kaminski, Dentener, Lelieveld, and
  Heimann}{Houweling et~al.}{1999}]{Houweling99}
Houweling, S., T.~Kaminski, F.~Dentener, J.~Lelieveld, and M.~Heimann (1999).
\newblock Inverse modeling of methane sources and sinks using the adjoint of a
  global transport model.
\newblock {\em Journal of Geophysical Research: Atmospheres\/}~{\em
  104\/}(D21), 26137--26160.

\bibitem[\protect\citeauthoryear{Huan and Marzouk}{Huan and
  Marzouk}{2014}]{huan2014gradient}
Huan, X. and Y.~Marzouk (2014).
\newblock Gradient-based stochastic optimization methods in bayesian
  experimental design.
\newblock {\em International Journal for Uncertainty Quantification\/}~{\em
  4\/}(6).

\bibitem[\protect\citeauthoryear{Huan and Marzouk}{Huan and
  Marzouk}{2013}]{huan2013simulation}
Huan, X. and Y.~M. Marzouk (2013).
\newblock Simulation-based optimal bayesian experimental design for nonlinear
  systems.
\newblock {\em Journal of Computational Physics\/}~{\em 232\/}(1), 288--317.

\bibitem[\protect\citeauthoryear{Hwang, Kim, Chang, Yeo, and Kim}{Hwang
  et~al.}{2019}]{hwang2019bayesian}
Hwang, Y., H.~J. Kim, W.~Chang, K.~Yeo, and Y.~Kim (2019).
\newblock Bayesian pollution source identification via an inverse physics
  model.
\newblock {\em Computational Statistics \& Data Analysis\/}~{\em 134}, 76--92.

\bibitem[\protect\citeauthoryear{Jakkala and Akella}{Jakkala and
  Akella}{2023}]{jakkala2023efficient}
Jakkala, K. and S.~Akella (2023).
\newblock Efficient sensor placement from regression with sparse gaussian
  processes in continuous and discrete spaces.
\newblock {\em arXiv preprint arXiv:2303.00028\/}.

\bibitem[\protect\citeauthoryear{Joshi and Boyd}{Joshi and
  Boyd}{2008}]{joshi2008sensor}
Joshi, S. and S.~Boyd (2008).
\newblock Sensor selection via convex optimization.
\newblock {\em IEEE Transactions on Signal Processing\/}~{\em 57\/}(2),
  451--462.

\bibitem[\protect\citeauthoryear{Khanduri, Tsaknakis, Zhang, Liu, Liu, Zhang,
  and Hong}{Khanduri et~al.}{2023}]{khanduri2023linearly}
Khanduri, P., I.~Tsaknakis, Y.~Zhang, J.~Liu, S.~Liu, J.~Zhang, and M.~Hong
  (2023).
\newblock Linearly constrained bilevel optimization: A smoothed implicit
  gradient approach.

\bibitem[\protect\citeauthoryear{Klein, van Kessel, Nair, Muralidhar, Hamann,
  and Sosa}{Klein et~al.}{2017}]{klein2017monitoring}
Klein, L.~J., T.~van Kessel, D.~Nair, R.~Muralidhar, H.~Hamann, and N.~Sosa
  (2017).
\newblock Monitoring fugitive methane gas emission from natural gas pads.
\newblock In {\em International Electronic Packaging Technical Conference and
  Exhibition}, Volume 58097, pp.\  V001T03A006. American Society of Mechanical
  Engineers.

\bibitem[\protect\citeauthoryear{Klise, Nicholson, and Laird}{Klise
  et~al.}{2017}]{klise2017sensor}
Klise, K.~A., B.~L. Nicholson, and C.~D. Laird (2017).
\newblock Sensor placement optimization using chama.
\newblock Technical report, Sandia National Lab.(SNL-NM), Albuquerque, NM
  (United States).

\bibitem[\protect\citeauthoryear{Krause, Singh, and Guestrin}{Krause
  et~al.}{2008}]{krause2008near}
Krause, A., A.~Singh, and C.~Guestrin (2008).
\newblock Near-optimal sensor placements in gaussian processes: Theory,
  efficient algorithms and empirical studies.
\newblock {\em Journal of Machine Learning Research\/}~{\em 9\/}(2).

\bibitem[\protect\citeauthoryear{Liu and Vicente}{Liu and
  Vicente}{2021}]{liu2021stochastic}
Liu, S. and L.~N. Vicente (2021).
\newblock The stochastic multi-gradient algorithm for multi-objective
  optimization and its application to supervised machine learning.
\newblock {\em Annals of Operations Research\/}, 1--30.

\bibitem[\protect\citeauthoryear{Liu and Yeo}{Liu and
  Yeo}{2023}]{liu2023inverse}
Liu, X. and K.~Yeo (2023).
\newblock Inverse models for estimating the initial condition of
  spatio-temporal advection-diffusion processes.
\newblock {\em Technometrics\/}, 1--14.

\bibitem[\protect\citeauthoryear{Liu, Yeo, Klein, Hwang, Phan, and Liu}{Liu
  et~al.}{2022}]{liu2022optimal}
Liu, X., K.~Yeo, L.~Klein, Y.~Hwang, D.~Phan, and X.~Liu (2022).
\newblock Optimal sensor placement for atmospheric inverse modelling.
\newblock In {\em 2022 IEEE International Conference on Big Data (Big Data)},
  pp.\  4848--4853. IEEE.

\bibitem[\protect\citeauthoryear{Mak and Joseph}{Mak and
  Joseph}{2018}]{mak2018support}
Mak, S. and V.~R. Joseph (2018).
\newblock Support points.
\newblock {\em The Annals of Statistics\/}~{\em 46\/}(6A), 2562--2592.

\bibitem[\protect\citeauthoryear{Manohar, Brunton, Kutz, and Brunton}{Manohar
  et~al.}{2018}]{manohar2018data}
Manohar, K., B.~W. Brunton, J.~N. Kutz, and S.~L. Brunton (2018).
\newblock Data-driven sparse sensor placement for reconstruction: Demonstrating
  the benefits of exploiting known patterns.
\newblock {\em IEEE Control Systems Magazine\/}~{\em 38\/}(3), 63--86.

\bibitem[\protect\citeauthoryear{Manohar, Kutz, and Brunton}{Manohar
  et~al.}{2021}]{manohar2021optimal}
Manohar, K., J.~N. Kutz, and S.~L. Brunton (2021).
\newblock Optimal sensor and actuator selection using balanced model reduction.
\newblock {\em IEEE Transactions on Automatic Control\/}~{\em 67\/}(4),
  2108--2115.

\bibitem[\protect\citeauthoryear{Meng and Li}{Meng and Li}{2020}]{meng2020aug}
Meng, M. and X.~Li (2020).
\newblock Aug-pdg: Linear convergence of convex optimization with inequality
  constraints.
\newblock {\em arXiv preprint arXiv:2011.08569\/}.

\bibitem[\protect\citeauthoryear{Narayanan, Patel, Agnihotri, and
  Batra}{Narayanan et~al.}{2020}]{narayanan2020toolkit}
Narayanan, S.~D., Z.~B. Patel, A.~Agnihotri, and N.~Batra (2020).
\newblock A toolkit for spatial interpolation and sensor placement.
\newblock In {\em Proceedings of the 18th Conference on Embedded Networked
  Sensor Systems}, pp.\  653--654.

\bibitem[\protect\citeauthoryear{Nemirovski, Juditsky, Lan, and
  Shapiro}{Nemirovski et~al.}{2009}]{nemirovski2009robust}
Nemirovski, A., A.~Juditsky, G.~Lan, and A.~Shapiro (2009).
\newblock Robust stochastic approximation approach to stochastic programming.
\newblock {\em SIAM Journal on optimization\/}~{\em 19\/}(4), 1574--1609.

\bibitem[\protect\citeauthoryear{Parise and Ozdaglar}{Parise and
  Ozdaglar}{2017}]{parise2017sensitivity}
Parise, F. and A.~Ozdaglar (2017).
\newblock Sensitivity analysis for network aggregative games.
\newblock In {\em 2017 IEEE 56th Annual Conference on Decision and Control
  (CDC)}, pp.\  3200--3205. IEEE.

\bibitem[\protect\citeauthoryear{Ranieri, Chebira, and Vetterli}{Ranieri
  et~al.}{2014}]{ranieri2014near}
Ranieri, J., A.~Chebira, and M.~Vetterli (2014).
\newblock Near-optimal sensor placement for linear inverse problems.
\newblock {\em IEEE Transactions on signal processing\/}~{\em 62\/}(5),
  1135--1146.

\bibitem[\protect\citeauthoryear{Riddick, Mbua, Riddick, Houlihan, Hodshire,
  and Zimmerle}{Riddick et~al.}{2023}]{riddick2023uncertainty}
Riddick, S.~N., M.~Mbua, J.~C. Riddick, C.~Houlihan, A.~L. Hodshire, and D.~J.
  Zimmerle (2023).
\newblock Uncertainty quantification of methods used to measure methane
  emissions of 1 g ch4 h- 1.
\newblock {\em Sensors\/}~{\em 23\/}(22), 9246.

\bibitem[\protect\citeauthoryear{Ruthotto, Chung, and Chung}{Ruthotto
  et~al.}{2018}]{ruthotto2018optimal}
Ruthotto, L., J.~Chung, and M.~Chung (2018).
\newblock Optimal experimental design for inverse problems with state
  constraints.
\newblock {\em SIAM Journal on Scientific Computing\/}~{\em 40\/}(4),
  B1080--B1100.

\bibitem[\protect\citeauthoryear{Shapiro and Philpott}{Shapiro and
  Philpott}{2007}]{shapiro2007tutorial}
Shapiro, A. and A.~Philpott (2007).
\newblock A tutorial on stochastic programming.
\newblock {\em Manuscript. Available at www2. isye. gatech.
  edu/ashapiro/publications. html\/}~{\em 17}.

\bibitem[\protect\citeauthoryear{Sharrock and Kantas}{Sharrock and
  Kantas}{2022}]{sharrock2022joint}
Sharrock, L. and N.~Kantas (2022).
\newblock Joint online parameter estimation and optimal sensor placement for
  the partially observed stochastic advection-diffusion equation.
\newblock {\em SIAM/ASA Journal on Uncertainty Quantification\/}~{\em 10\/}(1),
  55--95.

\bibitem[\protect\citeauthoryear{Shen and Chan}{Shen and
  Chan}{2002}]{shen2002mathematical}
Shen, J. and T.~F. Chan (2002).
\newblock Mathematical models for local nontexture inpaintings.
\newblock {\em SIAM Journal on Applied Mathematics\/}~{\em 62\/}(3),
  1019--1043.

\bibitem[\protect\citeauthoryear{Sinsbeck and Nowak}{Sinsbeck and
  Nowak}{2017}]{sinsbeck2017sequential}
Sinsbeck, M. and W.~Nowak (2017).
\newblock Sequential design of computer experiments for the solution of
  bayesian inverse problems.
\newblock {\em SIAM/ASA Journal on Uncertainty Quantification\/}~{\em 5\/}(1),
  640--664.

\bibitem[\protect\citeauthoryear{Spantini, Cui, Willcox, Tenorio, and
  Marzouk}{Spantini et~al.}{2017}]{spantini2017goal}
Spantini, A., T.~Cui, K.~Willcox, L.~Tenorio, and Y.~Marzouk (2017).
\newblock Goal-oriented optimal approximations of bayesian linear inverse
  problems.
\newblock {\em SIAM Journal on Scientific Computing\/}~{\em 39\/}(5),
  S167--S196.

\bibitem[\protect\citeauthoryear{Stockie}{Stockie}{2011}]{stockie2011mathematics}
Stockie, J.~M. (2011).
\newblock The mathematics of atmospheric dispersion modeling.
\newblock {\em Siam Review\/}~{\em 53\/}(2), 349--372.

\bibitem[\protect\citeauthoryear{Tarantola}{Tarantola}{2005}]{tarantola2005inverse}
Tarantola, A. (2005).
\newblock {\em Inverse problem theory and methods for model parameter
  estimation}.
\newblock SIAM.

\bibitem[\protect\citeauthoryear{Tsaknakis, Khanduri, and Hong}{Tsaknakis
  et~al.}{2022}]{tsaknakis2022implicit}
Tsaknakis, I., P.~Khanduri, and M.~Hong (2022).
\newblock An implicit gradient-type method for linearly constrained bilevel
  problems.
\newblock In {\em ICASSP 2022-2022 IEEE International Conference on Acoustics,
  Speech and Signal Processing (ICASSP)}, pp.\  5438--5442. IEEE.

\bibitem[\protect\citeauthoryear{Wang, Bardsley, Solonen, Cui, and
  Marzouk}{Wang et~al.}{2017}]{wang2017bayesian}
Wang, Z., J.~M. Bardsley, A.~Solonen, T.~Cui, and Y.~M. Marzouk (2017).
\newblock Bayesian inverse problems with l\_1 priors: a randomize-then-optimize
  approach.
\newblock {\em SIAM Journal on Scientific Computing\/}~{\em 39\/}(5),
  S140--S166.

\bibitem[\protect\citeauthoryear{Willoughby}{Willoughby}{1979}]{willoughby1979solutions}
Willoughby, R.~A. (1979).
\newblock Solutions of ill-posed problems (an tikhonov and vy arsenin).
\newblock {\em SIAM Review\/}~{\em 21\/}(2), 266.

\bibitem[\protect\citeauthoryear{Wu, Chen, and Ghattas}{Wu
  et~al.}{2023}]{wu2023offline}
Wu, K., P.~Chen, and O.~Ghattas (2023).
\newblock An offline-online decomposition method for efficient linear bayesian
  goal-oriented optimal experimental design: Application to optimal sensor
  placement.
\newblock {\em SIAM Journal on Scientific Computing\/}~{\em 45\/}(1), B57--B77.

\bibitem[\protect\citeauthoryear{Wu, O’Leary-Roseberry, Chen, and Ghattas}{Wu
  et~al.}{2023}]{wu2023large}
Wu, K., T.~O’Leary-Roseberry, P.~Chen, and O.~Ghattas (2023).
\newblock Large-scale bayesian optimal experimental design with
  derivative-informed projected neural network.
\newblock {\em Journal of Scientific Computing\/}~{\em 95\/}(1), 30.

\bibitem[\protect\citeauthoryear{Yang, Wen, Chen, Li, Huang, Song, and Li}{Yang
  et~al.}{2025}]{yang2025improving}
Yang, C., M.~Wen, C.~Chen, C.~Li, J.~Huang, L.~Song, and Y.~Li (2025).
\newblock Improving the accuracy of methane sensor with dual measurement modes
  based on off-axis integrated cavity output spectroscopy using white noise
  perturbation.
\newblock {\em Applied Sciences\/}~{\em 15\/}(10), 5562.

\bibitem[\protect\citeauthoryear{Yeo, Hwang, Liu, and Kalagnanam}{Yeo
  et~al.}{2019}]{yeo2019development}
Yeo, K., Y.~Hwang, X.~Liu, and J.~Kalagnanam (2019).
\newblock Development of hp-inverse model by using generalized polynomial
  chaos.
\newblock {\em Computer Methods in Applied Mechanics and Engineering\/}~{\em
  347}, 1--20.

\bibitem[\protect\citeauthoryear{Yu, Zavala, and Anitescu}{Yu
  et~al.}{2018}]{yu2018scalable}
Yu, J., V.~M. Zavala, and M.~Anitescu (2018).
\newblock A scalable design of experiments framework for optimal sensor
  placement.
\newblock {\em Journal of Process Control\/}~{\em 67}, 44--55.

\bibitem[\protect\citeauthoryear{Zou and Hastie}{Zou and
  Hastie}{2005}]{zou2005regularization}
Zou, H. and T.~Hastie (2005).
\newblock Regularization and variable selection via the elastic net.
\newblock {\em Journal of the Royal Statistical Society Series B: Statistical
  Methodology\/}~{\em 67\/}(2), 301--320.

\end{thebibliography}

\vspace{20pt}
\LARGE{\textbf{Supplemental Online Material}}
\appendix
\spacingset{1}
\section{Appendix I}\label{s:Appendix 1}
\subsection{Proof of Proposition 2:}
\textcolor{black}{
We show how (\ref{implicit gradient}) is derived. 
Following the idea of  \cite{parise2017sensitivity,tsaknakis2022implicit}, the Lagrangian function of the \textcolor{black}{lower-level} QP problem is written as
\begin{equation}
    h(\bm{s}, \bm{\theta}, \bm{\eta}) = \frac{1}{2}{\bm{\theta}}^T\bm{C}^{(i)} \bm{\theta}+(\textcolor{black}{\bm{d}^{(i)}})^T \bm{\theta} - \bm{\eta}^T\bm{\theta}.
\end{equation}
Consider a KKT point $(\bm{\theta}, \bm{\eta})$ for some fixed $\bm{s}\in \Omega^{\bm{s}}$, we have
\begin{equation*}
    \begin{split}
&\nabla_{\bm{\theta}}h(\bm{s}, \bm{\theta}, \bm{\eta})=\bm{C}^{(i)} \bm{\theta}+\textcolor{black}{\bm{d}^{(i)}} -\bm{\eta}=\bm{0},\\[2mm]
&\bm{\eta}\bm{\theta}=\bm{0}, \bm{\eta}\ge \bm{0}, \bm{\theta}\ge \bm{0}.
    \end{split}
\end{equation*}
By considering only the active constraints at $(\bm{\theta}, \bm{\eta})$, the KKT conditions can be equivalently written as
\begin{equation*}
    \bm{C}^{(i)} \bm{\theta}+\textcolor{black}{\bm{d}^{(i)}} -\bm{\bar{I}}^T\bm{\bar{\eta}}=\bm{0},\, \bm{\bar{I}}\bm{\theta}=\bm{0},\, \bm{\bar{\eta}}>\bm{0},
\end{equation*}
Then, computing the gradient of the KKT conditions w.r.t. $\bm{s}$, we obtain
\begin{equation}
    \nabla_{\bm{s}}(\bm{C}^{(i)})\bm{\theta}+\nabla_{\bm{s}}\textcolor{black}{\bm{d}^{(i)}} + \bm{C}^{(i)} \nabla_{\bm{s}}\bm{\theta} - \bm{\bar{I}}^T\nabla_{\bm{s}}\bm{\bar{\eta}} = \bm{0},
    \label{chain rule KKT 1}
\end{equation}
\begin{equation}
    \bm{\bar{I}}\nabla_{\bm{s}}\bm{\theta} = \bm{0}.
    \label{chain rule KKT 2}
\end{equation}
Re-arranging (\ref{chain rule KKT 1}) yields the first line of (\ref{implicit gradient}), and substituting the first line (\ref{implicit gradient}) into (\ref{chain rule KKT 2}) yields the second line of (\ref{implicit gradient}). 
}

\subsection{Proof of Proposition 3:}\label{A:proof A-optimal}
The derivation of Proposition 3 is obtained following \cite{ruthotto2018optimal}. Consider an observation model as follows,
\begin{equation}
\bm{\Phi}(\bm{\beta},\bm{s})=\mathcal{F}(\bm{\beta},\bm{s})\bm{\theta}+\bm{\epsilon},\ \bm{\epsilon}\sim \mathcal{N}(\bm{0},\bm{\Gamma}_{\bm{\epsilon}})
\end{equation}
where $\bm{\epsilon}$ is the additive Gaussian noise, and $\mathcal{F}:\mathbb{R}^{N_p}\mapsto \mathbb{R}^{d}$ is a linear parameter-to-observation mapping. Let $\bm{\theta}\sim \mathcal{N}(\bm{\mu}_{\text{pr}},\bm{\Gamma}_{\text{pr}})$ be the prior distribution of $\bm{\theta}$, we obtain the posterior distribution $\bm{\theta}_{\text{post}}\sim \mathcal{N}(\bm{\mu}_{\text{post}},\bm{\Gamma}_{\text{post}})$, and
\begin{equation}
    \begin{split}
        \bm{\mu}_{\text{post}}&=\bm{\Gamma}_{\text{post}}(\mathcal{F}^*(\bm{\beta},\bm{s})\bm{\Gamma}^{-1}_{\bm{\epsilon}}(\bm{s})\bm{\Phi}(\bm{\beta},\bm{s})+\bm{\Gamma}^{-1}_{\text{pr}}\bm{\mu}_{\text{pr}})\\
        \bm{\Gamma}_{\text{post}}&=(\mathcal{F}^*(\bm{\beta},\bm{s})\bm{\Gamma}^{-1}_{\bm{\epsilon}}(\bm{s})\mathcal{F}(\bm{\beta},\bm{s})+\bm{\Gamma}^{-1}_{pr})^{-1}
    \end{split}
\end{equation}
where $F^*(\bm{\beta},\bm{s})$ is the adjoint of $F$, e.g., by solving the adjoint PDE model. It is noted that $F^*(\bm{\beta},\bm{s})=F^T(\bm{\beta},\bm{s})$ because of its linear operator property. 

Then, the Bayesian risk is defined as,
\begin{equation}
    \begin{split}
        \Psi_{\text{risk, linear, Gaussian}}(\bm{s})&=\mathbb{E}_{\bm{\theta},\bm{\beta}}\Bigg\{\mathbb{E}_{\bm{\Phi}|\bm{\theta},\bm{\beta}}\Big\{\Big\|\hat{\bm{\theta}}_{MAP}(\bm{\Phi},\bm{\beta},\bm{s})-\bm{\theta}\Big\|^2_2\Big\}\Bigg\}\\
        &=\mathbb{E}_{\bm{\beta}}\Bigg\{\mathbb{E}_{\bm{\theta}|\bm{\beta}}\Bigg\{\mathbb{E}_{\bm{\Phi}|\bm{\theta},\bm{\beta}}\Big\{\Big\|\hat{\bm{\theta}}_{MAP}(\bm{\Phi},\bm{\beta},\bm{s})-\bm{\theta}\Big\|^2_2\Big\}\Bigg\}\Bigg\}
    \end{split}
    \label{risk 3 expectation}
\end{equation}

For convenience, we respectively denote $\mathcal{F}(\bm{\beta},\bm{s})$, $\mathcal{F}^*(\bm{\beta},\bm{s})$, $\bm{\Phi}(\bm{\beta},\bm{s})$, $\bm{\Gamma}_{\text{post}}$ and $\bm{\Gamma}^{-1}_{\bm{\epsilon}}(\bm{s})$ by $\mathcal{F}$, $\mathcal{F}^*$, $\bm{\Phi}$, $\bm{\Gamma}_{\text{post}}$, and $\bm{\Gamma}^{-1}_{\bm{\epsilon}}$. Then, 
we expand the $L^2$ loss function as
\begin{equation}
    \begin{split}
        \Big\|\hat{\bm{\theta}}_{MAP}(\bm{\Phi},\bm{\beta},\bm{s})-\bm{\theta}\Big\|^2_2=\Big\|(\bm{\Gamma}_{\text{post}}\mathcal{F}^*\bm{\Gamma}^{-1}_{\bm{\epsilon}}\mathcal{F}-\bm{I})\bm{\theta} + \bm{\Gamma}_{\text{post}}(\mathcal{F}^*\bm{\Gamma}^{-1}_{\bm{\epsilon}}\bm{\epsilon}+\bm{L}^T\bm{L}\bm{\mu}_{\text{pr}})\Big\|^2_2
    \end{split}
    \label{expand loss}
\end{equation}
where $\bm{L}^T\bm{L}=\bm{\Gamma}^{-1}_{pr}$. 

Denote $\bm{M}(\bm{s})=\bm{\Gamma}_{\text{post}}\mathcal{F}^*\bm{\Gamma}^{-1}_{\bm{\epsilon}}\mathcal{F}-\bm{I}$, we can further obtain
\begin{equation}
    \begin{split}
        \Big\|\hat{\bm{\theta}}_{MAP}(\bm{\Phi},\bm{\beta},\bm{s})-\bm{\theta}\Big\|^2_2=\Big\|\bm{M}(\bm{s})\bm{\theta} + \bm{\Gamma}_{\text{post}}(\mathcal{F}^*\bm{\Gamma}^{-1}_{\bm{\epsilon}}\bm{\epsilon}+\bm{L}^T\bm{L}\bm{\mu}_{\text{pr}})\Big\|^2_2
    \end{split}
    \label{expand loss 2}
\end{equation}

Then, plugging (\ref{expand loss 2}) into the expectation over $\bm{\theta}|\bm{\beta}$ yields
\begin{equation}
    \begin{split}
        &\ \ \ \mathbb{E}_{\bm{\theta}|\bm{\beta}}\Bigg\{\mathbb{E}_{\bm{\Phi}|\bm{\theta},\bm{\beta}}\Big\{\Big\|\hat{\bm{\theta}}_{MAP}(\bm{\Phi},\bm{\beta},\bm{s})-\bm{\theta}\Big\|^2_2\Big\}\Bigg\}\\
        &=\mathbb{E}_{\bm{\theta}|\bm{\beta}}\Bigg\{\mathbb{E}_{\bm{\Phi}|\bm{\theta},\bm{\beta}}\Big\{\bm{\theta}^T\bm{M}^T(\bm{s})\bm{M}(\bm{s})\bm{\theta}\Big\}\Bigg\}+\mathbb{E}_{\bm{\theta}|\bm{\beta}}\Bigg\{\mathbb{E}_{\bm{\Phi}|\bm{\theta},\bm{\beta}}\Big\{2\bm{\theta}^T\bm{M}^T(\bm{s})\bm{\Gamma}_{\text{post}}(\mathcal{F}^*\bm{\Gamma}^{-1}_{\bm{\epsilon}}\bm{\epsilon}+\bm{L}^T\bm{L}\bm{\mu}_{\text{pr}})\Big\}\Bigg\}\\
        &\ \ \ +\mathbb{E}_{\bm{\theta}|\bm{\beta}}\Bigg\{\mathbb{E}_{\bm{\Phi}|\bm{\theta},\bm{\beta}}\Big\{(\mathcal{F}^*\bm{\Gamma}^{-1}_{\bm{\epsilon}}\bm{\epsilon}+\bm{L}^T\bm{L}\bm{\mu}_{\text{pr}})^T\bm{\Gamma}^T_{\text{post}}\bm{\Gamma}_{\text{post}}(\mathcal{F}^*\bm{\Gamma}^{-1}_{\bm{\epsilon}}\bm{\epsilon}+\bm{L}^T\bm{L}\bm{\mu}_{\text{pr}})\Big\}\Bigg\}
    \end{split}
    \label{expand loss 3}
\end{equation}

Recall that $\bm{\epsilon}\sim \mathcal{N}(\bm{0},\bm{\Gamma}_{\bm{\epsilon}}(\bm{s}))$, $\bm{\theta}\sim \mathcal{N}(\bm{\mu}_{\text{pr}},\bm{\Gamma}_{\text{pr}})$, and $\bm{\theta}\sim \mathcal{N}(\bm{\mu}_{\text{pr}},\bm{\Gamma}_{\text{pr}})$, we obtain
\begin{equation}
    \begin{split}
        &\ \ \ \mathbb{E}_{\bm{\theta}|\bm{\beta}}\Bigg\{\mathbb{E}_{\bm{\Phi}|\bm{\theta},\bm{\beta}}\Big\{\Big\|\hat{\bm{\theta}}_{MAP}(\bm{\Phi},\bm{\beta},\bm{s})-\bm{\theta}\Big\|^2_2\Big\}\Bigg\}\\
        &=\mathbb{E}_{\bm{\theta}|\bm{\beta}}\Bigg\{\mathbb{E}_{\bm{\Phi}|\bm{\theta},\bm{\beta}}\Big\{\bm{\theta}^T\bm{M}^T(\bm{s})\bm{M}(\bm{s})\bm{\theta}\Big\}\Bigg\}+
        2\bm{\mu}_{\text{pr}}^T\bm{M}^T(\bm{s})\bm{\Gamma}_{\text{post}}\bm{L}^T\bm{L}\bm{\mu}_{\text{pr}}\\
        &\ \ \ +\mathbb{E}_{\bm{\theta}|\bm{\beta}}\Bigg\{\mathbb{E}_{\bm{\Phi}|\bm{\theta},\bm{\beta}}\Big\{\bm{\epsilon}^T\bm{\Gamma}^{-T}_{\bm{\epsilon}}\mathcal{F}\bm{\Gamma}^T_{\text{post}}\bm{\Gamma}_{\text{post}}\mathcal{F}^*\bm{\Gamma}^{-1}_{\bm{\epsilon}}\bm{\epsilon}\Big\}\Bigg\} \bm{\mu}^T_{\text{pr}}\bm{L}^T\bm{L}\bm{\Gamma}^T_{\text{post}}\bm{\Gamma}_{\text{post}}\bm{L}^T\bm{L}\bm{\mu}_{\text{pr}}.
    \end{split}
    \label{expand loss 4}
\end{equation}

Because $\mathbb{E}(\bm{\delta}^T\bm{\Lambda}\bm{\delta})=\bm{\mu}^T_{\bm{\delta}}\bm{\Lambda}\bm{\mu}_{\bm{\delta}}+\text{tr}(\bm{\Lambda}\bm{\Gamma}_{\bm{\delta}})$, where $\bm{\delta}\sim \mathcal{N}(\bm{\mu}_{\bm{\delta}},\bm{\Gamma}_{\bm{\delta}})$, it follows from (\ref{expand loss 4}) that 
\begin{equation}
    \begin{split}
        &\ \ \ \mathbb{E}_{\bm{\theta}|\bm{\beta}}\Bigg\{\mathbb{E}_{\bm{\Phi}|\bm{\theta},\bm{\beta}}\Big\{\Big\|\hat{\bm{\theta}}_{MAP}(\bm{\Phi},\bm{\beta},\bm{s})-\bm{\theta}\Big\|^2_2\Big\}\Bigg\}\\
        &=\bm{\mu}^T_{\text{pr}}\bm{M}^T(\bm{s})\bm{M}(\bm{s})\bm{\mu}_{\text{pr}}+ \text{tr}(\bm{M}^T(\bm{s})\bm{M}(\bm{s})\bm{\Gamma}_{\text{pr}})
        +
        2\bm{\mu}_{\text{pr}}^T\bm{M}^T(\bm{s})\bm{\Gamma}_{\text{post}}\bm{L}^T\bm{L}\bm{\mu}_{\text{pr}}+\text{tr}(\bm{\Gamma}^{-T}_{\bm{\epsilon}}\mathcal{F}\bm{\Gamma}^T_{\text{post}}\bm{\Gamma}_{\text{post}}\mathcal{F}^*)\\
        &\ \ \ +\bm{\mu}^T_{\text{pr}}\bm{L}^T\bm{L}\bm{\Gamma}^T_{\text{post}}\bm{\Gamma}_{\text{post}}\bm{L}^T\bm{L}\bm{\mu}_{\text{pr}}
    \end{split}
    \label{expand loss 5}
\end{equation}
where the first, third and fifth terms on the right hand side can be written as $\|\bm{M}(\bm{s})\bm{\mu}_{\text{pr}}+\bm{\Gamma}_{\text{post}}\bm{L}^T\bm{L}\bm{\mu}_{\text{pr}}\|^2_2$,  which turns out to be zero as follows
\begin{equation}
    \begin{split}
        &\|\bm{M}(\bm{s})\bm{\mu}_{\text{pr}}+\bm{\Gamma}_{\text{post}}\bm{L}^T\bm{L}\bm{\mu}_{\text{pr}}\|^2_2\\
        &=\|(\bm{M}(\bm{s})+\bm{\Gamma}_{\text{post}}\bm{L}^T\bm{L})\bm{\mu}_{\text{pr}}\|^2_2\\
        &=\|(\bm{\Gamma}_{\text{post}}\mathcal{F}^*\bm{\Gamma}^{-1}_{\bm{\epsilon}}\mathcal{F}-\bm{I}+\bm{\Gamma}_{\text{post}}\bm{L}^T\bm{L})\bm{\mu}_{\text{pr}}\|^2_2\\
        &=\|(\bm{\Gamma}_{\text{post}}(\mathcal{F}^*\bm{\Gamma}^{-1}_{\bm{\epsilon}}\mathcal{F}+\bm{L}^T\bm{L})-\bm{I})\bm{\mu}_{\text{pr}}\|^2_2\\
        &=\bm{0}
    \end{split}
    \label{proof of 0}
\end{equation}

Then we can rewrite (\ref{expand loss 5}) as
\begin{equation}
    \begin{split}
        \mathbb{E}_{\bm{\theta}|\bm{\beta}}\Bigg\{\mathbb{E}_{\bm{\Phi}|\bm{\theta},\bm{\beta}}\Big\{\Big\|\hat{\bm{\theta}}_{MAP}(\bm{\Phi},\bm{\beta},\bm{s})-\bm{\theta}\Big\|^2_2\Big\}\Bigg\}= \text{tr}(\bm{M}^T(\bm{s})\bm{M}(\bm{s})\bm{\Gamma}_{\text{pr}}) + 
        \text{tr}(\bm{\Gamma}^{-T}_{\bm{\epsilon}}\mathcal{F}\bm{\Gamma}^T_{\text{post}}\bm{\Gamma}_{\text{post}}\mathcal{F}^*) 
    \end{split}
    \label{expand loss 6}
\end{equation}
where the first term on the right hand side can be further transformed according to $\bm{M}(\bm{s})\bm{L}^{-1}=-\bm{\Gamma}_{\text{post}}\bm{L}^T$ given by (\ref{proof of 0}), 
\begin{equation}
    \begin{split}
        \text{tr}(\bm{M}^T(\bm{s})\bm{M}(\bm{s})\bm{\Gamma}_{\text{pr}})=\Big\|\bm{\Gamma}_{\text{post}}\bm{L}^T\Big\|^2_F
    \end{split}
    \label{transform first term}
\end{equation}

For the second term on the right hand side of (\ref{expand loss 6}), we further transform it by defining $\bm{\Gamma}^{-1}_{\bm{\epsilon}}=\bm{U}^T\bm{U}$ as follows
\begin{equation}
    \begin{split}
        \text{tr}(\bm{\Gamma}^{-T}_{\bm{\epsilon}}\mathcal{F}\bm{\Gamma}^T_{\text{post}}\bm{\Gamma}_{\text{post}}\mathcal{F}^*)=\Big\|\bm{\Gamma}_{\text{post}}\mathcal{F}^*\bm{U}^T\Big\|^2_F
    \end{split}
    \label{transform second term}
\end{equation}

After plugging (\ref{transform first term})(\ref{transform second term}) into (\ref{expand loss 6}), we achieve
\begin{equation}
    \begin{split}
        \mathbb{E}_{\bm{\theta}|\bm{\beta}}\Bigg\{\mathbb{E}_{\bm{\Phi}|\bm{\theta},\bm{\beta}}\Big\{\Big\|\hat{\bm{\theta}}_{MAP}(\bm{\Phi},\bm{\beta},\bm{s})-\bm{\theta}\Big\|^2_2\Big\}\Bigg\}=\Big\|\bm{\Gamma}_{\text{post}}\bm{L}^T\Big\|^2_F + \Big\|\bm{\Gamma}_{\text{post}}\mathcal{F}^*\bm{U}^T\Big\|^2_F
    \end{split}
    \label{expand loss 7}
\end{equation}

Finally, we plug (\ref{expand loss 7}) into (\ref{risk 3 expectation}) to obtain the closed-form optimization objective
\begin{equation}
    \begin{split}
         \hat{\Psi}_{\text{risk, linear, Gaussian}}(\bm{s})=\mathbb{E}_{\bm{\beta}}\Bigg\{\Big\|\bm{\Gamma}_{\text{post}}\bm{L}^T\Big\|^2_F + \Big\|\bm{\Gamma}_{\text{post}}\mathcal{F}^*\bm{U}^T\Big\|^2_F\Bigg\}
    \end{split}.
\end{equation}

\section{Appendix II}\label{s:Appendix 2}
Here we derive the closed-form gradients for the linear dispersion process. To compute $\nabla_{\bm{s}}\big(\bm{C}(\bm{\beta}^{(i)}, \bm{s})\bm{\theta}^{(i)}+\bm{d}^T(\bm{\beta}^{(i)},\bm{\Phi}^{(i)}, \bm{s})\big)$, we need the gradients
\begin{equation}
    \frac{\partial A_mA_n}{\partial s_{i,1}}, \ \frac{\partial A_mA_n}{\partial s_{i,2}}, \ 
    \frac{\partial A_m}{\partial s_{i,1}}, \ \frac{\partial A_m}{\partial s_{i,2}}
\end{equation}
Below are the derivations of these gradients:

Given the Gaussian Plume kernel \citep{stockie2011mathematics}
\begin{equation}
    A_j(\bm{s}_i)=\frac{1}{2\pi K \|(\bm{s}_i-\bm{x}_j)\cdot\bm{\beta}^{\parallel}\|}\text{exp}\Big(-\frac{u\big(\|(\bm{s}_i-\bm{x}_j)\cdot\bm{\beta}^{\perp}\|^2 + H^2\big)}{4K\|(\bm{s}_i-\bm{x}_j)\cdot\bm{\beta}^{\parallel}\|}\Big),
\end{equation}
let $\bm{r}_{\parallel}^{(j)} = (\bm{s}_i-\bm{x}_j)\cdot\bm{\beta}^{\parallel}$ and $\bm{r}_{\perp}^{(j)} = (\bm{s}_i-\bm{x}_j)\cdot\bm{\beta}^{\perp}$ for simplicity, we denote
\begin{equation}
    A_j=\frac{1}{2\pi K |\bm{r}_{\parallel}^{(j)}|}\text{exp}\Big(-\frac{u\big(|\bm{r}_{\perp}^{(j)}|^2 + H^2\big)}{4K|\bm{r}_{\parallel}^{(j)}|}\Big).
\end{equation}
Then, we can get
\begin{equation}
    A_mA_n=\frac{1}{4\pi^2 K^2 |\bm{r}_{\parallel}^{(m)}| \cdot|\bm{r}_{\parallel}^{(n)}|}\text{exp}\Big(\frac{-u\big(|\bm{r}_{\perp}^{(m)}|^2 + H^2\big)}{4K|\bm{r}_{\parallel}^{(m)}|} + \frac{-u\big(|\bm{r}_{\perp}^{(n)}|^2 + H^2\big)}{4K|\bm{r}_{\parallel}^{(n)}|}\Big).
\end{equation}

By denoting $\bm{w}=(w_1,w_2)=\frac{\bm{\beta}}{|\bm{\beta}|}$, $\bm{s}_i=(s_{i,1}, s_{i,2})$, $\bm{x}_j = (x_{j,1}, x_{j,2})$, and $\bm{r}^{(j)}=(s_{i,1}-x_{j,1}, s_{i,2}-x_{j,2})$, we can derive $\bm{r}_{\parallel}^{(j)}$, $|\bm{r}_{\parallel}^{(j)}|$, $\bm{r}_{\perp}^{(j)}$, $|\bm{r}_{\perp}^{(j)}|$ as,
\begin{equation}
    \begin{split}
        &\bm{r}_{\parallel}^{(j)}=(w_1(s_{i,1} - x_{j,1}) + w_2(s_{i,2} - x_{j,2}))\cdot (w_1,w_2)\\
        &|\bm{r}_{\parallel}^{(j)}| = w_1(s_{i,1} - x_{j,1}) + w_2(s_{i,2} - x_{j,2})\\
        &\bm{r}_{\perp}^{(j)} = \bm{r}^{(j)}-\bm{r}_{\parallel}^{(j)}\\
        &=\Big(s_{i,1}-x_{j,1}-w_1[w_1(s_{i,1} - x_{j,1}) + w_2(s_{i,2} - x_{j,2})],s_{i,2}-x_{j,2}-w_2[w_1(s_{i,1} - x_{j,1}) + w_2(s_{i,2} - x_{j,2})]\Big)\\
        &=\Big((1-w_1^2)(s_{i,1}-x_{j,1})-w_1w_2(s_{i,2}-x_{j,2}),-w_1w_2(s_{i,1}-x_{j,1})+(1-w_2^2)(s_{i,2}-x_{j,2})\Big)\\
        &|\bm{r}_{\perp}^{(j)}|=\sqrt{\Big(\big[(1-w_1^2)(s_{i,1}-x_{j,1})-w_1w_2(s_{i,2}-x_{j,2})\big]^2+\big[-w_1w_2(s_{i,1}-x_{j,1})+(1-w_2^2)(s_{i,2}-x_{j,2})\big]^2\Big)}
    \end{split}
    \nonumber
\end{equation}

Then we can derive the gradients of $A_mA_n$ w.r.t $s_{i,1}$ and $s_{i,2}$,
\begin{equation}
\begin{split}
        \frac{\partial A_mA_n}{\partial s_{i,1}}=&\frac{-1}{\big[4\pi^2K^2|\bm{r}_{\parallel}^{(m)}|\cdot|\bm{r}_{\parallel}^{(n)}|\big]^2}4\pi^2K^2w_1\Big[|\bm{r}_{\parallel}^m|+|\bm{r}_{\parallel}^n|\Big]\cdot \text{exp}\Big(\frac{-u\big(|\bm{r}_{\perp}^{(m)}|^2 + H^2\big)}{4K|\bm{r}_{\parallel}^{(m)}|} + \frac{-u\big(|\bm{r}_{\perp}^{(n)}|^2 + H^2\big)}{4K|\bm{r}_{\parallel}^{(n)}|}\Big)+\\
        &\frac{1}{4\pi^2 K^2 |\bm{r}_{\parallel}^{(m)}| \cdot|\bm{r}_{\parallel}^{(n)}|}\cdot\text{exp}\Big(\frac{-u\big(|\bm{r}_{\perp}^{(m)}|^2 + H^2\big)}{4K|\bm{r}_{\parallel}^{(m)}|} + \frac{-u\big(|\bm{r}_{\perp}^{(n)}|^2 + H^2\big)}{4K|\bm{r}_{\parallel}^{(n)}|}\Big)\cdot \Big(\frac{\partial\normalsize{\textcircled{\scriptsize{1}}}\normalsize}{\partial s_{i,1}} + \frac{\partial\normalsize{\textcircled{\scriptsize{2}}}\normalsize}{\partial s_{i,1}}\Big)
\end{split}
\end{equation}
and similarly, we can obtain the gradients of $A_m$ w.r.t $s_{i,1}$,
\begin{equation}
    \frac{\partial A_m}{\partial s_{i,1}} = \frac{-1\cdot 2\pi K w_1}{(2\pi K|\bm{r}_{\parallel}^{(m)}|)^2} \text{exp}\Big(\frac{-u\big(|\bm{r}_{\perp}^{(m)}|^2 + H^2\big)}{4K|\bm{r}_{\parallel}^{(m)}|}\Big) + \frac{1}{2\pi K|\bm{r}_{\parallel}^{(m)}|}\text{exp}\Big(\frac{-u\big(|\bm{r}_{\perp}^{(m)}|^2 + H^2\big)}{4K|\bm{r}_{\parallel}^{(m)}|}\Big)\cdot \frac{\partial\normalsize{\textcircled{\scriptsize{1}}}\normalsize}{\partial s_{i,1}}
\end{equation}
where
\begin{equation}
\begin{split}
&\frac{\partial\normalsize{\textcircled{\scriptsize{1}}}\normalsize}{\partial s_{i,1}}=\frac{-u\Big[2\cdot \normalsize{\textcircled{\scriptsize{3}}}\normalsize\cdot (1-w_1^2) + 2\cdot \normalsize{\textcircled{\scriptsize{4}}}\normalsize\cdot (-w_1w_2)\Big]4K|\bm{r}_{\parallel}^{(m)}|-\Big(-u\big(|\bm{r}_{\perp}^{(m)}|^2 + H^2\big)4Kw_1\Big)}{(4K|\bm{r}_{\parallel}^{(m)}|)^2}\\
&\frac{\partial\normalsize{\textcircled{\scriptsize{2}}}\normalsize}{\partial s_{i,1}}=\frac{-u\Big[2\cdot \normalsize{\textcircled{\scriptsize{5}}}\normalsize\cdot (1-w_1^2) + 2\cdot \normalsize{\textcircled{\scriptsize{6}}}\normalsize\cdot (-w_1w_2)\Big]4K|\bm{r}_{\parallel}^{(n)}|-\Big(-u\big(|\bm{r}_{\perp}^{(n)}|^2 + H^2\big)4Kw_1\Big)}{(4K|\bm{r}_{\parallel}^{(n)}|)^2}
\end{split}
    \nonumber
\end{equation}
with $\normalsize{\textcircled{\scriptsize{3}}}\normalsize=[(1-w_1^2)(s_{i,1}-x_{m,1})-w_1w_2(s_{i,2}-x_{m,2})]$, $\normalsize{\textcircled{\scriptsize{4}}}\normalsize=[-w_1w_2(s_{i,1}-x_{m,1})+(1-w_2^2)(s_{i,2}-x_{m,2})]$, $\normalsize{\textcircled{\scriptsize{5}}}\normalsize=[(1-w_1^2)(s_{i,1}-x_{n,1})-w_1w_2(s_{i,2}-x_{n,2})]$ and $\normalsize{\textcircled{\scriptsize{6}}}\normalsize=[-w_1w_2(s_{i,1}-x_{n,1})+(1-w_2^2)(s_{i,2}-x_{n,2})]$.

Next, 
\begin{equation}
\begin{split}
        \frac{\partial A_mA_n}{\partial s_{i,2}}=&\frac{-1}{\big[4\pi^2K^2|\bm{r}_{\parallel}^{(m)}|\cdot|\bm{r}_{\parallel}^{(n)}|\big]^2}4\pi^2K^2w_2\Big[|\bm{r}_{\parallel}^m|+|\bm{r}_{\parallel}^n|\Big]\cdot \text{exp}\Big(\frac{-u\big(|\bm{r}_{\perp}^{(m)}|^2 + H^2\big)}{4K|\bm{r}_{\parallel}^{(m)}|} + \frac{-u\big(|\bm{r}_{\perp}^{(n)}|^2 + H^2\big)}{4K|\bm{r}_{\parallel}^{(n)}|}\Big)+\\
        &\frac{1}{4\pi^2 K^2 |\bm{r}_{\parallel}^{(m)}| \cdot|\bm{r}_{\parallel}^{(n)}|}\cdot\text{exp}\Big(\frac{-u\big(|\bm{r}_{\perp}^{(m)}|^2 + H^2\big)}{4K|\bm{r}_{\parallel}^{(m)}|} + \frac{-u\big(|\bm{r}_{\perp}^{(n)}|^2 + H^2\big)}{4K|\bm{r}_{\parallel}^{(n)}|}\Big)\cdot \Big(\frac{\partial\normalsize{\textcircled{\scriptsize{1}}}\normalsize}{\partial s_{i,2}} + \frac{\partial\normalsize{\textcircled{\scriptsize{2}}}\normalsize}{\partial s_{i,2}}\Big)
\end{split}
\end{equation}
and similarly, we obtain the gradients of $A_m$ w.r.t $s_{i,2}$,
\begin{equation}
    \frac{\partial A_m}{\partial s_{i,2}} = \frac{-1\cdot 2\pi K w_2}{(2\pi K|\bm{r}_{\parallel}^{(m)}|)^2} \text{exp}\Big(\frac{-u\big(|\bm{r}_{\perp}^{(m)}|^2 + H^2\big)}{4K|\bm{r}_{\parallel}^{(m)}|}\Big) + \frac{1}{2\pi K|\bm{r}_{\parallel}^{(m)}|}\text{exp}\Big(\frac{-u\big(|\bm{r}_{\perp}^{(m)}|^2 + H^2\big)}{4K|\bm{r}_{\parallel}^{(m)}|}\Big)\cdot \frac{\partial\normalsize{\textcircled{\scriptsize{1}}}\normalsize}{\partial s_{i,2}}
\end{equation}
where
\begin{equation}
\begin{split}
&\frac{\partial\normalsize{\textcircled{\scriptsize{1}}}\normalsize}{\partial s_{i,2}}=\frac{-u\Big[2\cdot \normalsize{\textcircled{\scriptsize{3}}}\normalsize\cdot (1-w_2^2) + 2\cdot \normalsize{\textcircled{\scriptsize{4}}}\normalsize\cdot (-w_1w_2)\Big]4K|\bm{r}_{\parallel}^{(m)}|-\Big(-u\big(|\bm{r}_{\perp}^{(m)}|^2 + H^2\big)4Kw_2\Big)}{(4K|\bm{r}_{\parallel}^{(m)}|)^2}\\
&\frac{\partial\normalsize{\textcircled{\scriptsize{2}}}\normalsize}{\partial s_{i,2}}=\frac{-u\Big[2\cdot \normalsize{\textcircled{\scriptsize{5}}}\normalsize\cdot (1-w_2^2) + 2\cdot \normalsize{\textcircled{\scriptsize{6}}}\normalsize\cdot (-w_1w_2)\Big]4K|\bm{r}_{\parallel}^{(n)}|-\Big(-u\big(|\bm{r}_{\perp}^{(n)}|^2 + H^2\big)4Kw_2\Big)}{(4K|\bm{r}_{\parallel}^{(n)}|)^2}
\end{split}
    \nonumber
\end{equation}
with$\normalsize{\textcircled{\scriptsize{3}}}\normalsize=[(1-w_1^2)(s_{i,1}-x_{m,1})-w_1w_2(s_{i,2}-x_{m,2})]$, $\normalsize{\textcircled{\scriptsize{4}}}\normalsize=[-w_1w_2(s_{i,1}-x_{m,1})+(1-w_2^2)(s_{i,2}-x_{m,2})]$, $\normalsize{\textcircled{\scriptsize{5}}}\normalsize=[(1-w_1^2)(s_{i,1}-x_{n,1})-w_1w_2(s_{i,2}-x_{n,2})]$ and $\normalsize{\textcircled{\scriptsize{6}}}\normalsize=[-w_1w_2(s_{i,1}-x_{n,1})+(1-w_2^2)(s_{i,2}-x_{n,2})]$.

\section{Appendix III}\label{s:Appendix 4}

\subsection{Proof of Lemma 1(a) and 1(b)} \label{s:Appendix 4 lemma 1 bc}

We first introduce the following assumption,
\begin{assumption}
For any $i$-th sample, we assume the following bounds for different gradients,
\begin{equation}
    \begin{split}
        \|\bm{\theta}-\bm{\theta}^{(i)}\|&\leq\mathcal{C}_{\bm{\theta}},\\
        \|\nabla_{\bm{s}}\bm{\theta}\|&\leq\mathcal{C}_{\nabla\bm{\theta}},\\
        \|(\bm{C}^{(i)})^{-1}\|&\leq \mathcal{C}_{\nabla_{\bm{\theta}\bm{\theta}}L},\\
        \|\nabla_{\bm{s}}(\bm{C}^{(i)})\bm{\theta}+\nabla_{\bm{s}}\textcolor{black}{\bm{d}^{(i)}}\|&\leq\mathcal{C}_{\nabla_{\bm{\theta}\bm{s}}L},\\
        \|\nabla_{\bm{s}}(\bm{C}^{(i)})\|&\leq\mathcal{C}_{\nabla_{\bm{s}}C}
    \end{split}
\end{equation}
where $\mathcal{C}_{\bm{\theta}}$, $\mathcal{C}_{\nabla\bm{\theta}}$, $\mathcal{C}_{\nabla_{\bm{\theta}\bm{\theta}}L}$, $\mathcal{C}_{\nabla_{\bm{\theta}\bm{s}}L}$ and $\mathcal{C}_{\nabla_{\bm{s}}C}$ are some constants; $\|\bm{\theta}-\bm{\theta}^{(i)}\|=\frac{1}{2}\nabla_{\bm{\theta}}\hat{\Psi}^{(i)}$; $\bm{C}^{(i)}=\nabla_{\bm{\theta}\bm{\theta}}L$; $\nabla_{\bm{s}}(\bm{C}^{(i)})\bm{\theta}+\nabla_{\bm{s}}\textcolor{black}{\bm{d}^{(i)}}=\nabla_{\bm{\theta}\bm{s}}L$.
\label{A:bounds 1}
\end{assumption}

\begin{assumption}
Following the similar idea by \cite{khanduri2023linearly}, we assume
\begin{equation}
    \big\|\bar{\bm{I}}^T\big(\bar{\bm{I}}(\bm{C}^{(i)})^{-1}\bar{\bm{I}}^T\big)^{-1}\bar{\bm{I}}-\bar{\bm{I}}^{*T}\big(\bar{\bm{I}}^*(\bm{C}^{(i)})^{-1}\bar{\bm{I}}^{*T}\big)^{-1}\bar{\bm{I}}^*\big\|\leq \mathcal{L}_{C}\cdot\delta
\end{equation}
where $\bar{\bm{I}}^*$ denotes the active rows of the identity matrix for true solutions; $\mathcal{L}_{C}$ is a constant.
\label{A:lagrangian bound}
\end{assumption}

We define 
$\nabla_{\bm{s}} \hat{\Psi}(\bm{s}; \{\hat{\bm{\theta}}^{(i)}\}_{i=1}^{\tilde{N}})=\frac{2}{\tilde{N}}\sum_{i=1}^{\tilde{N}}(\nabla_{\bm{s}}\hat{\bm{\theta}}^{(i)})^T(\hat{\bm{\theta}}^{(i)}-\bm{\theta}^{(i)})$, and
$\nabla_{\bm{s}} \hat{\Psi}(\bm{s}; \{\hat{\bm{\theta}}^{*(i)}\}_{i=1}^{\tilde{N}})=\frac{2}{\tilde{N}}\sum_{i=1}^{\tilde{N}}(\nabla_{\bm{s}}\hat{\bm{\theta}}^{*(i)})^T(\hat{\bm{\theta}}^{*(i)}-\bm{\theta}^{(i)})$.
For simplicity, we denote $\nabla_{\bm{s}} \hat{\Psi}(\bm{s}; \{\hat{\bm{\theta}}^{(i)}\}_{i=1}^{\tilde{N}})$ and $\nabla_{\bm{s}} \hat{\Psi}(\bm{s}; \{\hat{\bm{\theta}}^{*(i)}\}_{i=1}^{\tilde{N}})$ as $\nabla_{\bm{s}}\hat{\Psi}_{\tilde{N}}(\bm{s})$ and $\nabla_{\bm{s}}\hat{\Psi}_{\tilde{N}}^{*}(\bm{s})$ respectively.

Then, we have
\begin{equation}
    \begin{split}
        \|\nabla_{\bm{s}}\hat{\Psi}_{\tilde{N}}(\bm{s})-\nabla_{\bm{s}}\hat{\Psi}_{\tilde{N}}^{*}(\bm{s})\|&=\big\|\frac{2}{\tilde{N}}\sum_{i=1}^{\tilde{N}}\big((\nabla_{\bm{s}}\hat{\bm{\theta}}^{(i)})^T(\hat{\bm{\theta}}^{(i)}-\bm{\theta}^{(i)})-(\nabla_{\bm{s}}\hat{\bm{\theta}}^{*(i)})^T(\hat{\bm{\theta}}^{*(i)}-\bm{\theta}^{(i)})\big)\big\|\\
        &\leq \frac{2}{\tilde{N}}\sum_{i=1}^{\tilde{N}}\big\|(\nabla_{\bm{s}}\hat{\bm{\theta}}^{(i)})^T(\hat{\bm{\theta}}^{(i)}-\bm{\theta}^{(i)})-(\nabla_{\bm{s}}\hat{\bm{\theta}}^{*(i)})^T(\hat{\bm{\theta}}^{*(i)}-\bm{\theta}^{(i)})\big\|\\
        &\leq \frac{2}{\tilde{N}}\sum_{i=1}^{\tilde{N}}\big\|(\nabla_{\bm{s}}\hat{\bm{\theta}}^{(i)})^T(\hat{\bm{\theta}}^{(i)}-\bm{\theta}^{(i)})-(\nabla_{\bm{s}}\hat{\bm{\theta}}^{*(i)})^T(\hat{\bm{\theta}}^{(i)}-\bm{\theta}^{(i)})\big\| \\
        &\ \ + \frac{2}{\tilde{N}}\sum_{i=1}^{\tilde{N}}\big\|(\nabla_{\bm{s}}\hat{\bm{\theta}}^{*(i)})^T(\hat{\bm{\theta}}^{(i)}-\bm{\theta}^{(i)})-(\nabla_{\bm{s}}\hat{\bm{\theta}}^{*(i)})^T(\hat{\bm{\theta}}^{*(i)}-\bm{\theta}^{(i)})\big\|\\
        &\leq \frac{2}{\tilde{N}}\sum_{i=1}^{\tilde{N}}\|\nabla_{\bm{s}}\hat{\bm{\theta}}^{(i)}-\nabla_{\bm{s}}\hat{\bm{\theta}}^{*(i)}\|\|\hat{\bm{\theta}}^{(i)}-\bm{\theta}^{(i)}\| + \frac{2}{\tilde{N}}\sum_{i=1}^{\tilde{N}}\|\nabla_{\bm{s}}\hat{\bm{\theta}}^{*(i)}\|\|\hat{\bm{\theta}}^{(i)}-\hat{\bm{\theta}}^{*(i)}\|\\
        &\leq \frac{2}{\tilde{N}}\sum_{i=1}^{\tilde{N}}\|\nabla_{\bm{s}}\hat{\bm{\theta}}^{(i)}-\nabla_{\bm{s}}\hat{\bm{\theta}}^{*(i)}\|\mathcal{C}_{\bm{\theta}} + \mathcal{C}_{\nabla\bm{\theta}}\delta
    \end{split}
    \label{Psi error}
\end{equation}
where the upper bound of $\|\hat{\bm{\theta}}^{(i)}-\hat{\bm{\theta}}^{*(i)}\|$ is shown in Assumption 3; $\|\hat{\bm{\theta}}^{(i)}-\bm{\theta}^{(i)}\|$ and $\|\nabla_{\bm{s}}\hat{\bm{\theta}}^{*(i)}\|$ have bounds defined in assumption 4. The upper bound of $\|\nabla_{\bm{s}}\hat{\bm{\theta}}^{(i)}-\nabla_{\bm{s}}\hat{\bm{\theta}}^{*(i)}\|$ is derived as follows
\begin{equation}
\begin{split}
        \|\nabla_{\bm{s}}\hat{\bm{\theta}}^{(i)}-\nabla_{\bm{s}}\hat{\bm{\theta}}^{*(i)}\|&=\big\|(\bm{C}^{(i)})^{-1}\big(-\nabla_{\bm{s}}(\bm{C}^{(i)})\hat{\bm{\theta}}^{(i)}-\nabla_{\bm{s}}\textcolor{black}{\bm{d}^{(i)}}+\bar{\bm{I}}^T\nabla_{\bm{s}}\bar{\bm{\eta}}^{(i)}\big)\\
    &\ \ -(\bm{C}^{(i)})^{-1}\big(-\nabla_{\bm{s}}(\bm{C}^{(i)})\hat{\bm{\theta}^{*(i)}}-\nabla_{\bm{s}}(\textcolor{black}{\bm{d}^{(i)}})^T+\bar{\bm{I}}^{*T}\nabla_{\bm{s}}\bar{\bm{\eta}}^{*(i)}\big)
    \big\|\\
    &=\|(\bm{C}^{(i)})^{-1}\big(-\nabla_{\bm{s}}(\bm{C}^{(i)})(\hat{\bm{\theta}}^{(i)}-\hat{\bm{\theta}}^{*(i)})+(\bar{\bm{I}}^T\nabla_{\bm{s}}\bar{\bm{\eta}}^{(i)}-\bar{\bm{I}}^{*T}\nabla_{\bm{s}}\bar{\bm{\eta}}^{*(i)})\big)\|\\
    &\leq\|(\bm{C}^{(i)})^{-1}\nabla_{\bm{s}}(\bm{C}^{(i)})(\hat{\bm{\theta}}^{(i)}-\hat{\bm{\theta}}^{*(i)})\|+\|(\bm{C}^{(i)})^{-1}(\bar{\bm{I}}^T\nabla_{\bm{s}}\bar{\bm{\eta}}^{(i)}-\bar{\bm{I}}^{*T}\nabla_{\bm{s}}\bar{\bm{\eta}}^{*(i)})\|\\
    &\leq\|(\bm{C}^{(i)})^{-1}\|\|\nabla_{\bm{s}}(\bm{C}^{(i)})\|\|\hat{\bm{\theta}}^{(i)}-\hat{\bm{\theta}}^{*(i)}\|+\|(\bm{C}^{(i)})^{-1}\|\|\bar{\bm{I}}^T\nabla_{\bm{s}}\bar{\bm{\eta}}^{(i)}-\bar{\bm{I}}^{*T}\nabla_{\bm{s}}\bar{\bm{\eta}}^{*(i)}\|\\
    &\leq \mathcal{C}_{\nabla_{\bm{\theta}\bm{\theta}}L}\mathcal{C}_{\nabla_{\bm{s}}C}\delta+\mathcal{C}_{\nabla_{\bm{\theta}\bm{\theta}}L}\|\bar{\bm{I}}^T\nabla_{\bm{s}}\bar{\bm{\eta}}^{(i)}-\bar{\bm{I}}^{*T}\nabla_{\bm{s}}\bar{\bm{\eta}}^{*(i)}\|
\end{split}
\label{theta error}
\end{equation}
where the last inequality is based on Assumption 4. 

The upper bound of $\|\bar{\bm{I}}^T\nabla_{\bm{s}}\bar{\bm{\eta}}^{(i)}-\bar{\bm{I}}^{*T}\nabla_{\bm{s}}\bar{\bm{\eta}}^{*(i)}\|$ is derived as
\begin{equation}
    \begin{split}
        \|\bar{\bm{I}}^T\nabla_{\bm{s}}\bar{\bm{\eta}}^{(i)}-\bar{\bm{I}}^{*T}\nabla_{\bm{s}}\bar{\bm{\eta}}^{*(i)}\|&=\big\|\bar{\bm{I}}^T\big(\bar{\bm{I}}(\bm{C}^{(i)})^{-1}\bar{\bm{I}}^T\big)^{-1}\bar{\bm{I}}(\bm{C}^{(i)})^{-1}\big(\nabla_{\bm{s}}(\bm{C}^{(i)})\hat{\bm{\theta}}^{(i)}+\nabla_{\bm{s}}\textcolor{black}{\bm{d}^{(i)}}\big)\\
        &\ \ -\bar{\bm{I}}^{*T}\big(\bar{\bm{I}}^*(\bm{C}^{(i)})^{-1}\bar{\bm{I}}^{*T}\big)^{-1}\bar{\bm{I}}^*(\bm{C}^{(i)})^{-1}\big(\nabla_{\bm{s}}(\bm{C}^{(i)})\hat{\bm{\theta}}^{*(i)}+\nabla_{\bm{s}}\textcolor{black}{\bm{d}^{(i)}}\big)
        \big\|\\
        &=\big\|\bar{\bm{I}}^T\big(\bar{\bm{I}}(\bm{C}^{(i)})^{-1}\bar{\bm{I}}^T\big)^{-1}\bar{\bm{I}}(\bm{C}^{(i)})^{-1}\big(\nabla_{\bm{s}}(\bm{C}^{(i)})\hat{\bm{\theta}}^{(i)}+\nabla_{\bm{s}}\textcolor{black}{\bm{d}^{(i)}}\big)\\
        &\ \ -\bar{\bm{I}}^{T}\big(\bar{\bm{I}}(\bm{C}^{(i)})^{-1}\bar{\bm{I}}^T\big)^{-1}\bar{\bm{I}}(\bm{C}^{(i)})^{-1}\big(\nabla_{\bm{s}}(\bm{C}^{(i)})\hat{\bm{\theta}}^{*(i)}+\nabla_{\bm{s}}\textcolor{black}{\bm{d}^{(i)}}\big)\\
        &\ \ +\bar{\bm{I}}^T\big(\bar{\bm{I}}(\bm{C}^{(i)})^{-1}\bar{\bm{I}}^T\big)^{-1}\bar{\bm{I}}(\bm{C}^{(i)})^{-1}\big(\nabla_{\bm{s}}(\bm{C}^{(i)})\hat{\bm{\theta}}^{*(i)}+\nabla_{\bm{s}}\textcolor{black}{\bm{d}^{(i)}}\big)\\
        &\ \ -\bar{\bm{I}}^{*T}\big(\bar{\bm{I}}^*(\bm{C}^{(i)})^{-1}\bar{\bm{I}}^{*T}\big)^{-1}\bar{\bm{I}}^*(\bm{C}^{(i)})^{-1}\big(\nabla_{\bm{s}}(\bm{C}^{(i)})\hat{\bm{\theta}}^{*(i)}+\nabla_{\bm{s}}\textcolor{black}{\bm{d}^{(i)}}\big)\big\|\\
        &\leq \big\| \bar{\bm{I}}^T\big(\bar{\bm{I}}(\bm{C}^{(i)})^{-1}\bar{\bm{I}}^T\big)^{-1}\bar{\bm{I}}(\bm{C}^{(i)})^{-1}\nabla_{\bm{s}}(\bm{C}^{(i)})(\hat{\bm{\theta}}^{(i)}-\hat{\bm{\theta}}^{*(i)})\\
        +&\big(\bar{\bm{I}}^T\big(\bar{\bm{I}}(\bm{C}^{(i)})^{-1}\bar{\bm{I}}^T\big)^{-1}\bar{\bm{I}}-\bar{\bm{I}}^{*T}\big(\bar{\bm{I}}^*(\bm{C}^{(i)})^{-1}\bar{\bm{I}}^{*T}\big)^{-1}\bar{\bm{I}}^*\big)\big(\nabla_{\bm{s}}(\bm{C}^{(i)})\hat{\bm{\theta}}^{*(i)}+\nabla_{\bm{s}}\textcolor{black}{\bm{d}^{(i)}}\big)\big\|\\
        &\leq \big\| \bar{\bm{I}}^T\big(\bar{\bm{I}}(\bm{C}^{(i)})^{-1}\bar{\bm{I}}^T\big)^{-1}\bar{\bm{I}}\big\|\big\|(\bm{C}^{(i)})^{-1}\big\|\big\|\nabla_{\bm{s}}(\bm{C}^{(i)})\big\|\big\|\hat{\bm{\theta}}^{(i)}-\hat{\bm{\theta}}^{*(i)}\big\|\\
        +&\big\|\bar{\bm{I}}^T\big(\bar{\bm{I}}(\bm{C}^{(i)})^{-1}\bar{\bm{I}}^T\big)^{-1}\bar{\bm{I}}-\bar{\bm{I}}^{*T}\big(\bar{\bm{I}}^*(\bm{C}^{(i)})^{-1}\bar{\bm{I}}^{*T}\big)^{-1}\bar{\bm{I}}^*\big\|\big\|\nabla_{\bm{s}}(\bm{C}^{(i)})\hat{\bm{\theta}}^{*(i)}+\nabla_{\bm{s}}\textcolor{black}{\bm{d}^{(i)}}\big\|\\
        &\leq \mathcal{C}_{\nabla_{\bm{\theta}\bm{\theta}}L}^2\mathcal{C}_{\nabla_{\bm{s}}C}\delta+\mathcal{L}_{C}\delta\mathcal{C}_{\nabla_{\bm{\theta}\bm{s}}L}
    \end{split}
    \label{eta error}
\end{equation}
where the last inequality is based on Assumptions 4 and 5. Plugging inequality (\ref{eta error}) into (\ref{theta error}) yields
\begin{equation}
    \|\nabla_{\bm{s}}\hat{\bm{\theta}}^{(i)}-\nabla_{\bm{s}}\hat{\bm{\theta}}^{*(i)}\| \leq \mathcal{C}_{\nabla_{\bm{\theta}\bm{\theta}}L}\mathcal{C}_{\nabla_{\bm{s}}C}\delta+\mathcal{C}_{\nabla_{\bm{\theta}\bm{\theta}}L}(\mathcal{C}_{\nabla_{\bm{\theta}\bm{\theta}}L}^2\mathcal{C}_{\nabla_{\bm{s}}C}\delta+\mathcal{L}_{C}\delta\mathcal{C}_{\nabla_{\bm{\theta}\bm{s}}L})
    \label{theta error 2}
\end{equation}
Plugging inequality (\ref{theta error 2}) into inequality (\ref{Psi error}), we obtain
\begin{equation}
    \|\nabla_{\bm{s}}\hat{\Psi}_{\tilde{N}}(\bm{s})-\nabla_{\bm{s}}\hat{\Psi}_{\tilde{N}}^{*}(\bm{s})\|\leq \big(\mathcal{C}_{\nabla_{\bm{\theta}\bm{\theta}}L}\mathcal{C}_{\nabla_{\bm{s}}C}\delta+\mathcal{C}_{\nabla_{\bm{\theta}\bm{\theta}}L}(\mathcal{C}_{\nabla_{\bm{\theta}\bm{\theta}}L}^2\mathcal{C}_{\nabla_{\bm{s}}C}\delta+\mathcal{L}_{C}\delta\mathcal{C}_{\nabla_{\bm{\theta}\bm{s}}L})\big)\mathcal{C}_{\bm{\theta}} + \mathcal{C}_{\nabla\bm{\theta}}\delta
    \label{e:lemma a}
\end{equation}
where the RHS can be rewritten as $\big(\mathcal{C}_{\nabla_{\bm{\theta}\bm{\theta}}L}(\mathcal{C}_{\nabla_{\bm{s}}C}+\mathcal{C}_{\nabla_{\bm{\theta}\bm{\theta}}L}^2\mathcal{C}_{\nabla_{\bm{s}}C}+\mathcal{L}_{C}\mathcal{C}_{\nabla_{\bm{\theta}\bm{s}}L})\mathcal{C}_{\bm{\theta}}+\mathcal{C}_{\nabla\bm{\theta}}\big)\delta$. By letting $\mathcal{L}_{\Psi} :=  \mathcal{C}_{\nabla_{\bm{\theta}\bm{\theta}}L}(\mathcal{C}_{\nabla_{\bm{s}}C}+\mathcal{C}_{\nabla_{\bm{\theta}\bm{\theta}}L}^2\mathcal{C}_{\nabla_{\bm{s}}C}+\mathcal{L}_{C}\mathcal{C}_{\nabla_{\bm{\theta}\bm{s}}L})\mathcal{C}_{\bm{\theta}}+\mathcal{C}_{\nabla\bm{\theta}}$.


We introduce the assumption following the idea in \cite{giovannelli2021inexact},
\begin{assumption}
\begin{equation}
    \big\|\nabla_{\bm{s}}\hat{\Psi}(\bm{s};\hat{\bm{\theta}}(\bm{\xi}),\bm{\xi})-\nabla_{\bm{s}}\hat{\Psi}(\bm{s};\hat{\bm{\theta}})\big\|\leq \mathcal{L}_{D}\big\|D(\bm{s},\hat{\bm{\theta}}(\bm{\xi}), \hat{\bm{\eta}}(\bm{\xi}))-D(\bm{s},\hat{\bm{\theta}}, \hat{\bm{\eta}})\big\|
\end{equation}
where there is a difference from \cite{giovannelli2021inexact} that we are not approximating the calculation of any gradients, Hessians and Jacobians; $\bm{\xi}$ denotes the combination of random samples of uncertain parameters and $\hat{\bm{\theta}}(\bm{\xi})$ denotes the inversion estimates $\hat{\bm{\theta}}$ for the corresponding samples; $\mathcal{L}_{D}$ is a constant; $D(\cdot)$ denotes the data used to evaluate $\nabla_{\bm{s}}\hat{\Psi}(\cdot)$; we assume $D(\bm{s},\hat{\bm{\theta}}(\bm{\xi}^{(i)}), \hat{\bm{\eta}}(\bm{\xi}^{(i)}))\in \mathbb{R}^{n_{cov}}$ is normally distributed with mean $D(\bm{s},\hat{\bm{\theta}}, \hat{\bm{\eta}})$ and covariance $\sigma^2\bm{I}_{n_{cov}}$, where $\{\bm{\xi}^{(i)}\}_{i=0}^{\tilde{N}-1}$ are realizations of $\bm{\xi}$ and $D(\bm{s},\hat{\bm{\theta}}(\bm{\xi}), \hat{\bm{\eta}}(\bm{\xi}))=\frac{1}{\tilde{N}}\sum_{i=1}^{\tilde{N}}D(\bm{s},\hat{\bm{\theta}}(\bm{\xi}^{(i)}), \hat{\bm{\eta}}(\bm{\xi}^{(i)}))$ for each \textcolor{black}{upper-level} iteration step in the SGD algorithm. According to \cite{giovannelli2021inexact,liu2021stochastic}, we have
\begin{equation}
    \begin{split}
        \mathbb{E}\big(\|D(\bm{s},\hat{\bm{\theta}}(\bm{\xi}), \hat{\bm{\eta}}(\bm{\xi}))-D(\bm{s},\hat{\bm{\theta}}, \hat{\bm{\eta}})\|^2\big)\leq \frac{\sigma^2}{\tilde{N}}\\
        \mathbb{E}\big(\|D(\bm{s},\hat{\bm{\theta}}(\bm{\xi}), \hat{\bm{\eta}}(\bm{\xi}))-D(\bm{s},\hat{\bm{\theta}}, \hat{\bm{\eta}})\|\big)\leq \frac{\sigma\sqrt{n_{cov}}}{\sqrt{\tilde{N}}}
    \end{split}
    \label{data inequality}
\end{equation}
\label{A:data}
\end{assumption}
   For Lemma 1(a), we have
\begin{equation}
\begin{split}
         &\mathbb{E}\big(\|\nabla_{\bm{s}} \hat{\Psi}(\bm{s}; \hat{\bm{\theta}}(\bm{\xi}),\bm{\xi}) - \nabla_{\bm{s}} \hat{\Psi}(\bm{s};\hat{\bm{\theta}}^*)\|\big)\\
         &\leq  \underbrace{\mathbb{E}\big(\|\nabla_{\bm{s}} \hat{\Psi}(\bm{s}; \hat{\bm{\theta}}(\bm{\xi}),\bm{\xi}) - \nabla_{\bm{s}} \hat{\Psi}(\bm{s};\hat{\bm{\theta}})\|\big)}_{\normalsize{\textcircled{\scriptsize{1}}}}+\underbrace{\mathbb{E}\big(\|\nabla_{\bm{s}} \hat{\Psi}(\bm{s}; \hat{\bm{\theta}}) - \nabla_{\bm{s}} \hat{\Psi}(\bm{s};\hat{\bm{\theta}}^*)\|\big)}_{\normalsize{\textcircled{\scriptsize{2}}}}
\end{split}
\end{equation}
where $\normalsize{\textcircled{\scriptsize{1}}}\leq \mathcal{L}_{D}\frac{\sigma\sqrt{n_{cov}}}{\sqrt{\tilde{N}}}$ according to Assumption 6 and the  inequality (\ref{data inequality}), and $\normalsize{\textcircled{\scriptsize{2}}}\leq \mathcal{L}_{\Psi}\delta$ according to (\ref{e:lemma a}) when $\tilde{N}$ goes to infinity. Lemma 1(a) is proved.

For Lemma 1(b), we have
\begin{equation}
\begin{split}
         &\mathbb{E}\big(\|\nabla_{\bm{s}} \hat{\Psi}(\bm{s}; \hat{\bm{\theta}}(\bm{\xi}),\bm{\xi}) - \nabla_{\bm{s}} \hat{\Psi}(\bm{s};\hat{\bm{\theta}}^*)\|^2\big)\\
         &\leq  2\underbrace{\mathbb{E}\big(\|\nabla_{\bm{s}} \hat{\Psi}(\bm{s}; \hat{\bm{\theta}}(\bm{\xi}),\bm{\xi}) - \nabla_{\bm{s}} \hat{\Psi}(\bm{s};\hat{\bm{\theta}})\|^2\big)}_{\normalsize{\textcircled{\scriptsize{3}}}}+2\underbrace{\mathbb{E}\big(\|\nabla_{\bm{s}} \hat{\Psi}(\bm{s}; \hat{\bm{\theta}}) - \nabla_{\bm{s}} \hat{\Psi}(\bm{s};\hat{\bm{\theta}}^*)\|^2\big)}_{\normalsize{\textcircled{\scriptsize{4}}}}
\end{split}
\end{equation}
where $\normalsize{\textcircled{\scriptsize{3}}}\leq \mathcal{L}_{D}^2\frac{\sigma^2}{\tilde{N}}$ according to assumption 6 and inequality (\ref{data inequality}), and $\normalsize{\textcircled{\scriptsize{4}}}\leq \mathcal{L}_{\Psi}^2\delta^2$ according to (\ref{e:lemma a}) when $\tilde{N}$ goes to infinity. Hence, Lemma 1(b) is proved.

\subsection{Proof of Theorem 1} \label{s:proof of theorem 2}
\begin{assumption}
We assume the bounded gradients,
  \begin{equation}
      \|\nabla_{\bm{s}} \hat{\Psi}(\bm{s}; \hat{\bm{\theta}}^{*})\| \leq \mathcal{C}_{\nabla\Psi}
      \label{e:bounded hypergradient 1}
  \end{equation}
  \begin{equation}
      \|\nabla_{\bm{s}} \hat{\Psi}(\bm{s}; \hat{\bm{\theta}})\| \leq \mathcal{C}_{\nabla\Psi}
      \label{e:bounded hypergradient 2}
  \end{equation}
  \label{A:gradient bound}
 \end{assumption}
According to the smoothness assumption (Assumption 2) and Taylor's formula, we have
    \begin{equation}
    \begin{split}
        \hat{\Psi}(\bm{s}_{m+1};\hat{\bm{\theta}}^{*})-\hat{\Psi}(\bm{s}_{m};\hat{\bm{\theta}}^{*})\leq  \Big[\nabla_{\bm{s}}\hat{\Psi}(\bm{s}_{m};\hat{\bm{\theta}}^{*})\Big]^T(\bm{s}_{m+1}-\bm{s}_m)+\frac{1}{2}\mathcal{L}_{\nabla\Psi}\|\bm{s}_{m+1}-\bm{s}_m\|^2.
    \end{split}
    \label{e:smoothness inequality}
    \end{equation}

    Recall that our algorithm has $\bm{s}_{m+1}=P_{\Omega^s}(\bm{s}_{m}-\rho_m\nabla_{\bm{s}}\hat{\Psi} (\bm{s}_{m};\hat{\bm{\theta}}^{*}))$ and we assume $\Omega^s$ is $\mathbb{R}^{n\times 2}$, we have $\bm{s}_{m+1} - \bm{s}_{m} = -\rho_m\nabla_{\bm{s}}\hat{\Psi} (\bm{s}_{m};\hat{\bm{\theta}}^{*})$, which can be plugged into (\ref{e:smoothness inequality}),
    \begin{equation}
        \begin{split}
        \hat{\Psi}(\bm{s}_{m+1};\hat{\bm{\theta}}^{*})-\hat{\Psi}(\bm{s}_{m};\hat{\bm{\theta}}^{*}) \leq  -\rho_m\Big[\nabla_{\bm{s}}\hat{\Psi}(\bm{s}_{m};\hat{\bm{\theta}}^{*})\Big]^T\nabla_{\bm{s}}\hat{\Psi} (\bm{s}_{m};\hat{\bm{\theta}}^{*})+\frac{1}{2}\rho_m^2\mathcal{L}_{\nabla\Psi}\|\nabla_{\bm{s}}\hat{\Psi} (\bm{s}_{m};\hat{\bm{\theta}}^{*})\|^2.
    \end{split}
    \end{equation}

    Adding and subtracting $\rho_m[\nabla_{\bm{s}}\Psi(\bm{s}_{m};\hat{\bm{\theta}}^{*})]^T\nabla_{\bm{s}}\Psi(\bm{s}_{m};\hat{\bm{\theta}}^{*})$, to prove the first part of Theorem 1, we adopt the smoothness assumption and obtain
    \begin{equation}
    \begin{split}
        \hat{\Psi}(\bm{s}_{m+1};\hat{\bm{\theta}}^*)-\hat{\Psi}(\bm{s}_m;\hat{\bm{\theta}}^*)\leq &\rho_m\Big[\nabla_{\bm{s}}\hat{\Psi}(\bm{s}_m;\hat{\bm{\theta}}^*)\Big]^T\Big(\nabla_{\bm{s}}\hat{\Psi}(\bm{s}_m;\hat{\bm{\theta}}^*)-\nabla_{\bm{s}}\hat{\Psi}(\bm{s}_m;\hat{\bm{\theta}}(\bm{\xi}),\bm{\xi})\Big)\\
        &-\rho_m\|\nabla_{\bm{s}}\hat{\Psi}(\bm{s}_m;\hat{\bm{\theta}}^*)\|^2 + \frac{1}{2}\rho_m^2 \mathcal{L}_{\nabla\Psi}\|\nabla_{\bm{s}}\hat{\Psi}(\bm{s};
        \hat{\bm{\theta}}(\bm{\xi}),\bm{\xi})\|^2.
    \end{split}
    \end{equation}
    Then, according to Cauchy-Schwarz inequality, we have
    \begin{equation}
    \begin{split}
        \hat{\Psi}(\bm{s}_{m+1};\hat{\bm{\theta}}^*)-\hat{\Psi}(\bm{s}_m;\hat{\bm{\theta}}^*)\leq &\rho_m\|\nabla_{\bm{s}}\hat{\Psi}(\bm{s}_m;\hat{\bm{\theta}}^*)\|\cdot\|\nabla_{\bm{s}}\hat{\Psi}(\bm{s}_m;\hat{\bm{\theta}}^*)-\nabla_{\bm{s}}\hat{\Psi}(\bm{s}_m;\hat{\bm{\theta}}(\bm{\xi}),\bm{\xi})\|\\
        &-\rho_m\|\nabla_{\bm{s}}\hat{\Psi}(\bm{s}_m;\hat{\bm{\theta}}^*)\|^2 + \frac{1}{2}\rho_m^2 \mathcal{L}_{\nabla\Psi}\|\nabla_{\bm{s}}\hat{\Psi}(\bm{s};
        \hat{\bm{\theta}}(\bm{\xi}),\bm{\xi})\|^2.
    \end{split}
    \label{e:stochastic inequality part 2}
    \end{equation}
    By expanding the last term on the RHS, we get
    \begin{equation}
    \begin{split}
        \hat{\Psi}(\bm{s}_{m+1};\hat{\bm{\theta}}^*)-&\hat{\Psi}(\bm{s}_m;\hat{\bm{\theta}}^*)\leq(\rho_m+\rho_m^2\mathcal{L}_{\nabla\Psi})\|\nabla_{\bm{s}}\hat{\Psi}(\bm{s}_m;\hat{\bm{\theta}}^*)\|\cdot\|\nabla_{\bm{s}}\hat{\Psi}(\bm{s}_m;\hat{\bm{\theta}}^*)-\nabla_{\bm{s}}\hat{\Psi}(\bm{s}_m;\hat{\bm{\theta}}(\bm{\xi}),\bm{\xi})\|\\
        &-(\rho_m-\frac{1}{2}\rho_m^2\mathcal{L}_{\nabla\Psi})\|\nabla_{\bm{s}}\hat{\Psi}(\bm{s}_m;\hat{\bm{\theta}}^*)\|^2 + \frac{1}{2}\rho_m^2 \mathcal{L}_{\nabla\Psi}\|\nabla_{\bm{s}}\hat{\Psi}(\bm{s}_m;\hat{\bm{\theta}}^*)-\nabla_{\bm{s}}\hat{\Psi}(\bm{s}_m;
        \hat{\bm{\theta}}(\bm{\xi}),\bm{\xi})\|^2.
    \end{split}
    \end{equation}
    Because the distribution of $\bm{\xi}$ is known, we obtain the expectation
    \begin{equation}
    \begin{split}
        E\Big[\hat{\Psi}(\bm{s}_{m+1};\hat{\bm{\theta}}^*)\Big]-&\hat{\Psi}(\bm{s}_m;\hat{\bm{\theta}}^*)\leq(\rho_m+\rho_m^2\mathcal{L}_{\nabla\Psi})\|\nabla_{\bm{s}}\hat{\Psi}(\bm{s}_m;\hat{\bm{\theta}}^*)\|\cdot E\Big[\|\nabla_{\bm{s}}\hat{\Psi}(\bm{s}_m;\hat{\bm{\theta}}^*)-\nabla_{\bm{s}}\hat{\Psi}(\bm{s}_m;\hat{\bm{\theta}}(\bm{\xi}),\bm{\xi})\|\Big]\\
        -(\rho_m-\frac{1}{2}&\rho_m^2\mathcal{L}_{\nabla\Psi})\|\nabla_{\bm{s}}\hat{\Psi}(\bm{s}_m;\hat{\bm{\theta}}^*)\|^2 + \frac{1}{2}\rho_m^2 \mathcal{L}_{\nabla\Psi}\cdot E\Big[\|\nabla_{\bm{s}}\hat{\Psi}(\bm{s}_m;\hat{\bm{\theta}}^*)-\nabla_{\bm{s}}\hat{\Psi}(\bm{s}_m;
        \hat{\bm{\theta}}(\bm{\xi}),\bm{\xi})\|^2\Big]
    \end{split}
    \end{equation}
    According to Lemma 1 and Assumption 7, we have
    \begin{equation}
    \begin{split}
        E\Big[\hat{\Psi}(\bm{s}_{m+1};\hat{\bm{\theta}}^*)\Big]-&\hat{\Psi}(\bm{s}_m;\hat{\bm{\theta}}^*)\leq(\rho_m+\rho_m^2\mathcal{L}_{\nabla\Psi})\mathcal{C}_{\nabla\Psi}(\mathcal{L}_{\Psi}\delta+\mathcal{L}_{D}\frac{\sigma\sqrt{n_{cov}}}{\sqrt{\tilde{N}}})\\
        &-(\rho_m-\frac{1}{2}\rho_m^2\mathcal{L}_{\nabla\Psi})\|\nabla_{\bm{s}}\hat{\Psi}(\bm{s}_m;\hat{\bm{\theta}}^*)\|^2 + \frac{1}{2}\rho_m^2 \mathcal{L}_{\nabla\Psi}(\mathcal{L}_{\Psi}^2\delta^2+\mathcal{L}_{D}^2\frac{\sigma^2}{\tilde{N}}).
    \end{split}
    \end{equation}
    and
    \begin{equation}
    \begin{split}
        E\Big[\hat{\Psi}(\bm{s}_{m+1};\hat{\bm{\theta}}^*)\Big]-&E\Big[\hat{\Psi}(\bm{s}_m;\hat{\bm{\theta}}^*)\Big]\leq(\rho_m+\rho_m^2\mathcal{L}_{\nabla\Psi})\mathcal{C}_{\nabla\Psi}(\mathcal{L}_{\Psi}\delta+\mathcal{L}_{D}\frac{\sigma\sqrt{n_{cov}}}{\sqrt{\tilde{N}}})\\
        &-(\rho_m-\frac{1}{2}\rho_m^2\mathcal{L}_{\nabla\Psi})E\Big[\|\nabla_{\bm{s}}\hat{\Psi}(\bm{s}_m;\hat{\bm{\theta}}^*)\|^2\Big] + \frac{1}{2}\rho_m^2 \mathcal{L}_{\nabla\Psi}(\mathcal{L}_{\Psi}^2\delta^2+\mathcal{L}_{D}^2\frac{\sigma^2}{\tilde{N}})
    \end{split}
    \end{equation}
    By taking the sum of this inequality from $m=0$ to $m=M-1$, we have
    \begin{equation}
    \begin{split}
        &\sum_{m=0}^{M-1}(\rho_m-\frac{1}{2}\rho_m^2\mathcal{L}_{\nabla\Psi})E\Big[\|\nabla_{\bm{s}}\hat{\Psi}(\bm{s}_m;\hat{\bm{\theta}}^*)\|^2\Big] \leq E\Big[\hat{\Psi}(\bm{s}_{0};\hat{\bm{\theta}}^*)\Big]-E\Big[\hat{\Psi}(\bm{s}_M;\hat{\bm{\theta}}^*)\Big]\\
        &+ \mathcal{C}_{\nabla\Psi}(\mathcal{L}_{\Psi}\delta+\mathcal{L}_{D}\frac{\sigma\sqrt{n_{cov}}}{\sqrt{\tilde{N}}})\sum_{m=0}^{M-1}(\rho_m+\rho_m^2\mathcal{L}_{\nabla\Psi})+ \frac{1}{2} \mathcal{L}_{\nabla\Psi}(\mathcal{L}_{\Psi}^2\delta^2+\mathcal{L}_{D}^2\frac{\sigma^2}{\tilde{N}})\sum_{m=0}^{M-1}\rho_m^2.
    \end{split}
    \end{equation}
    If $\rho_m$ is a constant, i.e., $\rho_m = \rho$, $0<\rho<\frac{2}{\mathcal{L}_{\nabla\Psi}}$, and according to the fact that $E[\hat{\Psi}(\cdot)]$ is always positive, we can achieve the final inequality after divide both sides with $M$.

 Next, we prove the second part of Theorem 1.
     We start from (\ref{e:stochastic inequality part 2}). According to Lemma 1, we have
    \begin{equation}
    \begin{split}
        &\sum_{m=0}^{M-1}E\Big[\|\nabla_{\bm{s}}\hat{\Psi}(\bm{s}_m;\hat{\bm{\theta}}^*)\|^2\Big] \leq E\Big[\hat{\Psi}(\bm{s}_{0};\hat{\bm{\theta}}^*)\Big]-E\Big[\hat{\Psi}(\bm{s}_M;\hat{\bm{\theta}}^*)\Big]\\
        &+ \mathcal{C}_{\nabla\Psi}(\mathcal{L}_{\Psi}\delta+\mathcal{L}_{D}\frac{\sigma\sqrt{n_{cov}}}{\sqrt{ \tilde{N}}})\sum_{m=0}^{M-1}\rho_m+ \frac{1}{2} \mathcal{L}_{\nabla\Psi}\mathcal{C}_{\nabla\Psi}^2\sum_{m=0}^{M-1}\rho_m^2.
    \end{split}
    \end{equation}
    Define $A_M=\sum_{m=0}^{M-1}\frac{1}{m+1}$ and divide the both sides with $A_M$,
    \begin{equation}
    \begin{split}
        &\frac{1}{A_M}\sum_{m=0}^{M-1}E\Big[\|\nabla_{\bm{s}}\hat{\Psi}(\bm{s}_m;\hat{\bm{\theta}}^*)\|^2\Big] \leq \frac{E\Big[\hat{\Psi}(\bm{s}_{0};\hat{\bm{\theta}}^*)\Big]-E\Big[\hat{\Psi}(\bm{s}_M;\hat{\bm{\theta}}^*)\Big]}{A_M}\\
        &+ \mathcal{C}_{\nabla\Psi}(\mathcal{L}_{\Psi}\delta+\mathcal{L}_{D}\frac{\sigma\sqrt{n_{cov}}}{\sqrt{ \tilde{N}}})\frac{\sum_{m=0}^{M-1}\rho_m}{A_M}+ \frac{1}{2} \mathcal{L}_{\nabla\Psi}\mathcal{C}_{\nabla\Psi}^2\frac{\sum_{m=0}^{M-1}\rho_m^2}{A_M}.
    \end{split}
    \end{equation}
    Then, it is easy to see that
    \begin{equation}
    \begin{split}
        \lim_{M\rightarrow\infty}\Bigg[\frac{1}{A_M}\sum_{m=0}^{M-1}E\Big[\|\nabla_{\bm{s}}\hat{\Psi}(\bm{s}_m;\hat{\bm{\theta}}^*)\|^2\Big]\Bigg] \leq \mathcal{C}_{\nabla\Psi}(\mathcal{L}_{\Psi}\delta+\mathcal{L}_{D}\frac{\sigma\sqrt{n_{cov}}}{\sqrt{ \tilde{N}}})\rho_0
    \end{split}
    \end{equation}
   Let $\bm{s}_M=\bm{s}_m$ with probability $\frac{1}{A_M(m+1)}$, the second part of Theorem 1 is proved.

\section{Appendix IV}\label{s:Appendix 5}

Following the investigations in Example II,  we present additional results on different scenarios of the number of sensors, number of emission sources, initial sensor locations, and \textcolor{black}{lower-level} problem iteration limit $J$. 

\begin{figure}[h!]
     \centering
     \begin{subfigure}[b]{0.49\textwidth}
         \centering
         \includegraphics[width=0.8\textwidth]{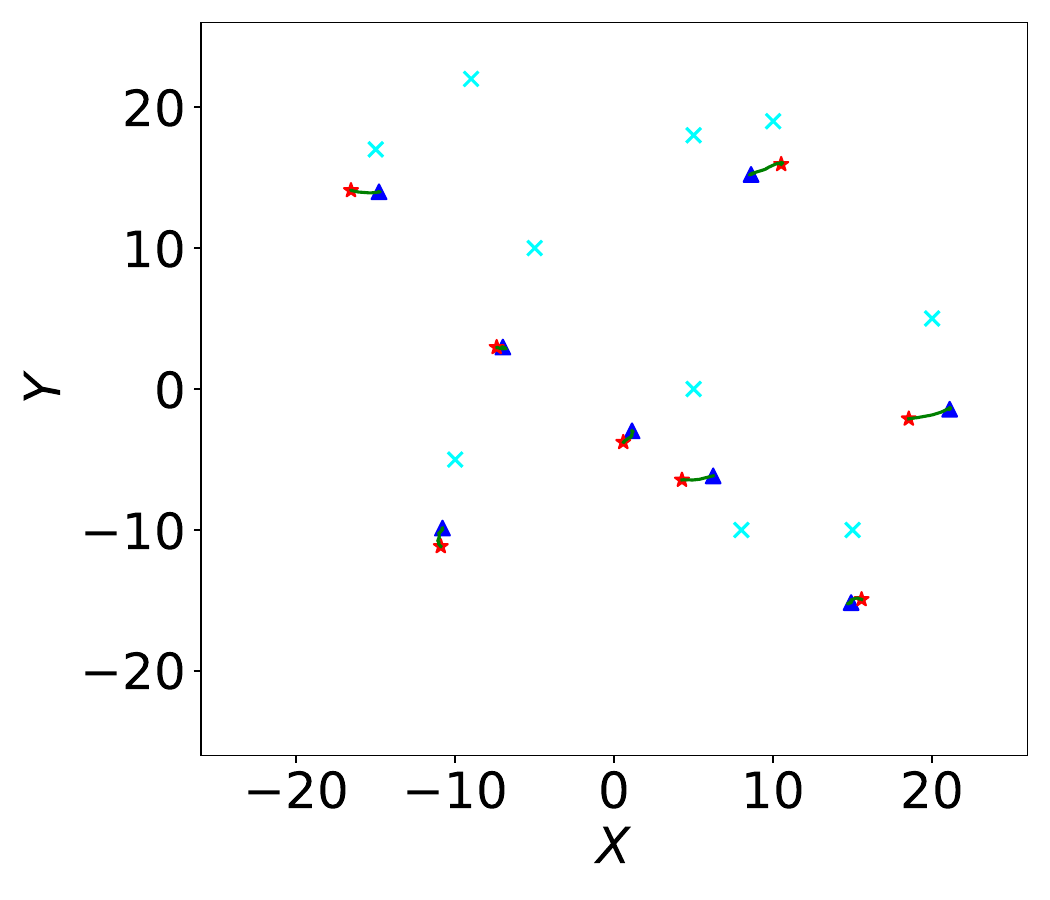}
         \caption{update of sensor locations, $J=2000$}
         \label{fig:8 sensor update}
     \end{subfigure}
     \hfill
     \begin{subfigure}[b]{0.49\textwidth}
         \centering
         \includegraphics[width=0.8\textwidth]{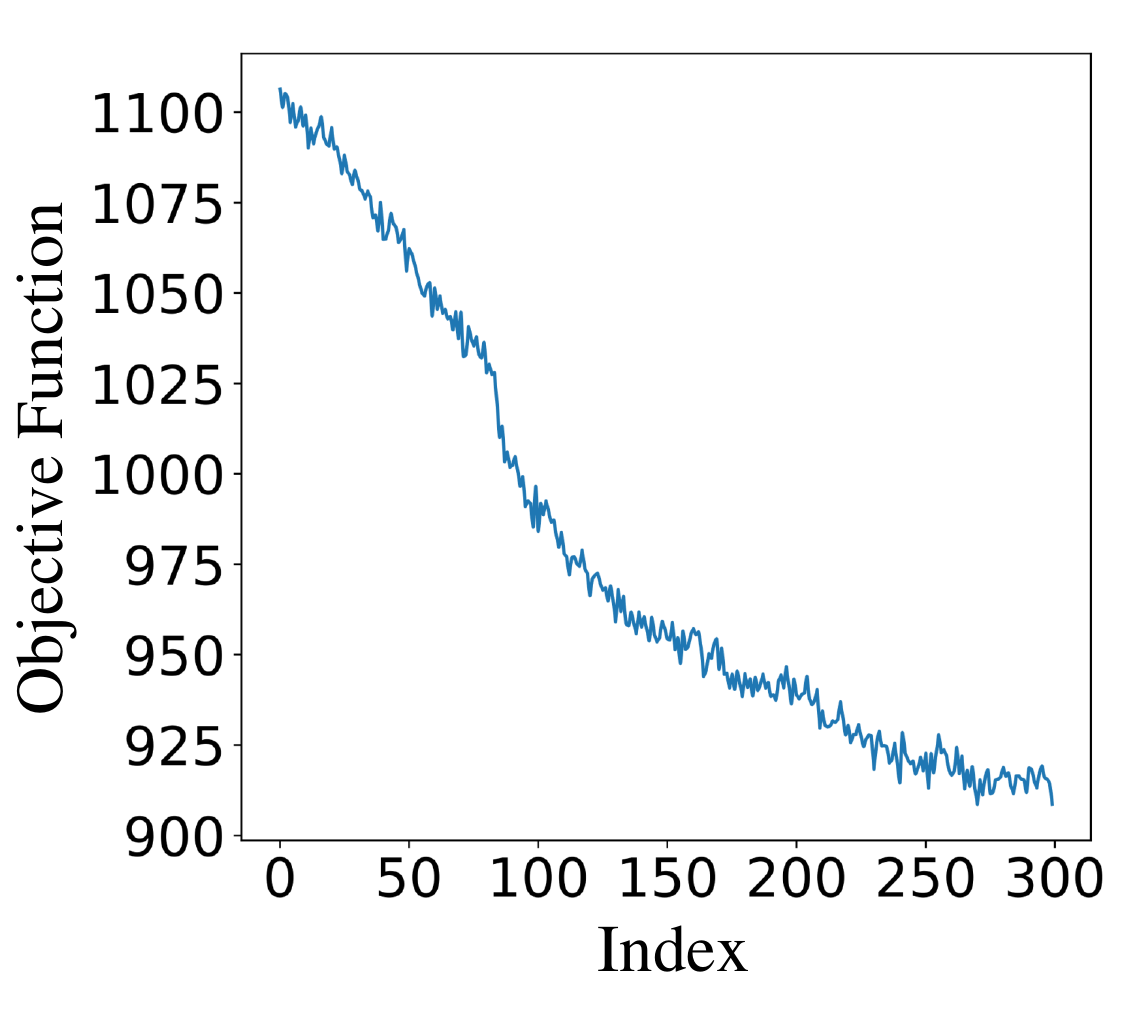}
         \caption{objective value along iterations, $J=2000$}
         \label{fig:8 sensor convergence}
     \end{subfigure}
     \caption{Allocation of 8 sensors for 10 emission sources ($\rho_m=0.00005$)}
     \label{fig:8 sensors}
\end{figure}

\begin{figure}[h!]
     \centering
     \begin{subfigure}[b]{0.49\textwidth}
         \centering
         \includegraphics[width=0.8\textwidth]{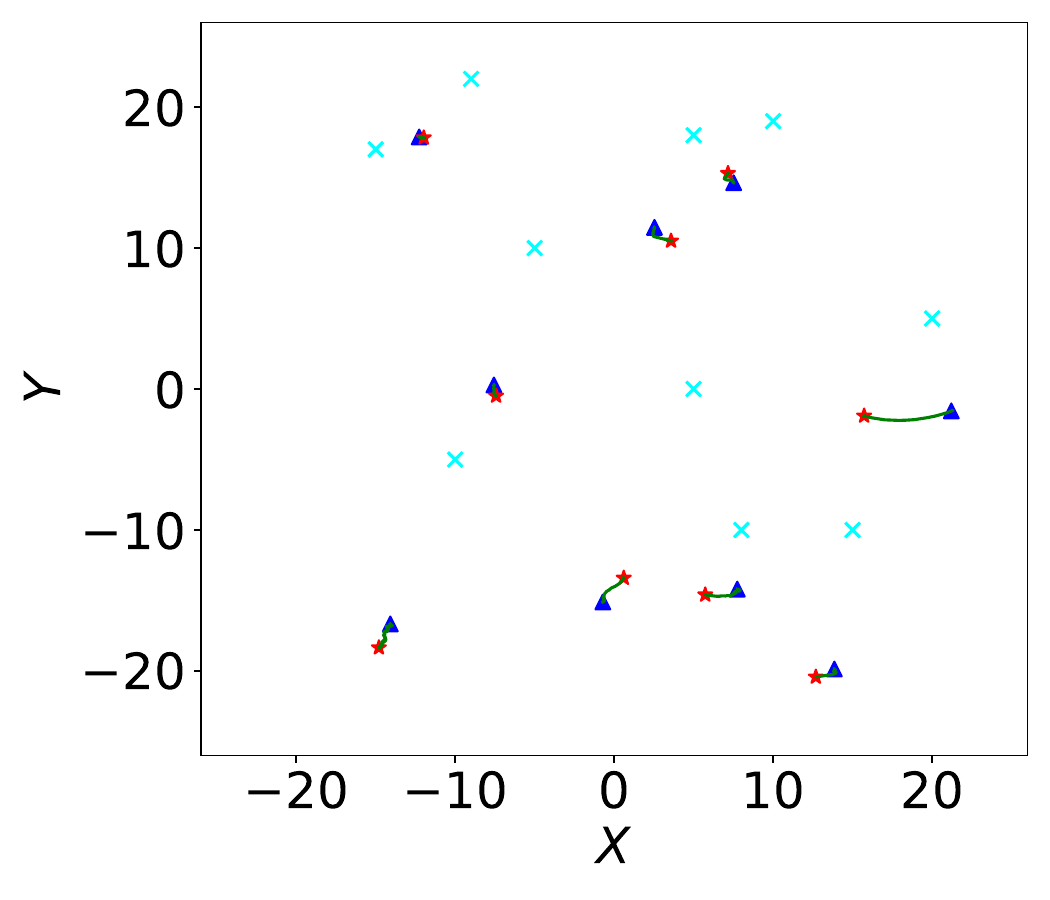}
         \caption{update of sensor locations, $J=2000$}
         \label{fig:9 sensor update}
     \end{subfigure}
     \hfill
     \begin{subfigure}[b]{0.49\textwidth}
         \centering
         \includegraphics[width=0.8\textwidth]{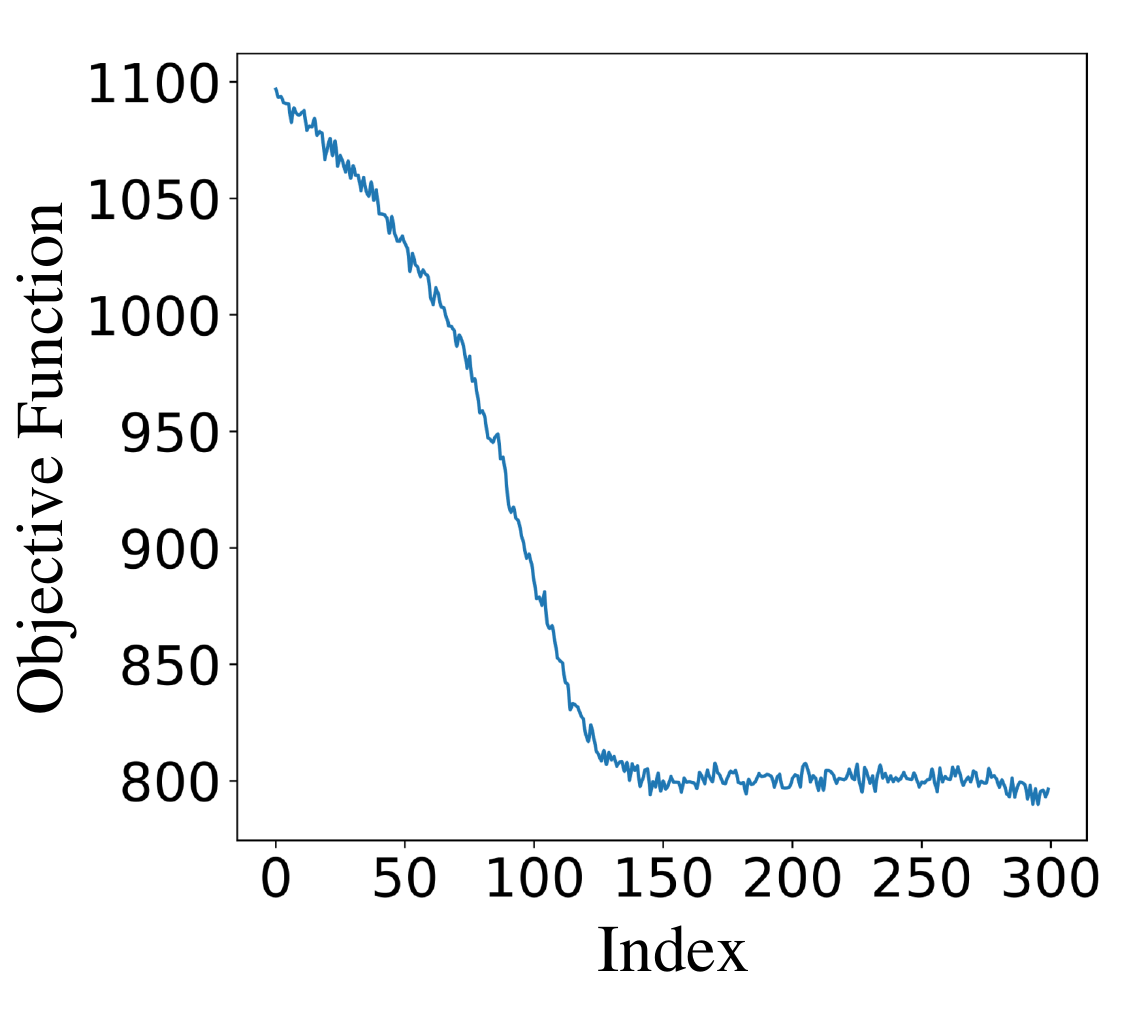}
         \caption{objective value along iterations, $J=2000$}
         \label{fig:9 sensor convergence}
     \end{subfigure}
     \caption{Allocation of 9 sensors for 10 emission sources ($\rho_m=0.00005$)}
     \label{fig:9 sensors}
\end{figure}

\begin{figure}[h!]
     \centering
     \begin{subfigure}[b]{0.49\textwidth}
         \centering
         \includegraphics[width=0.8\textwidth]{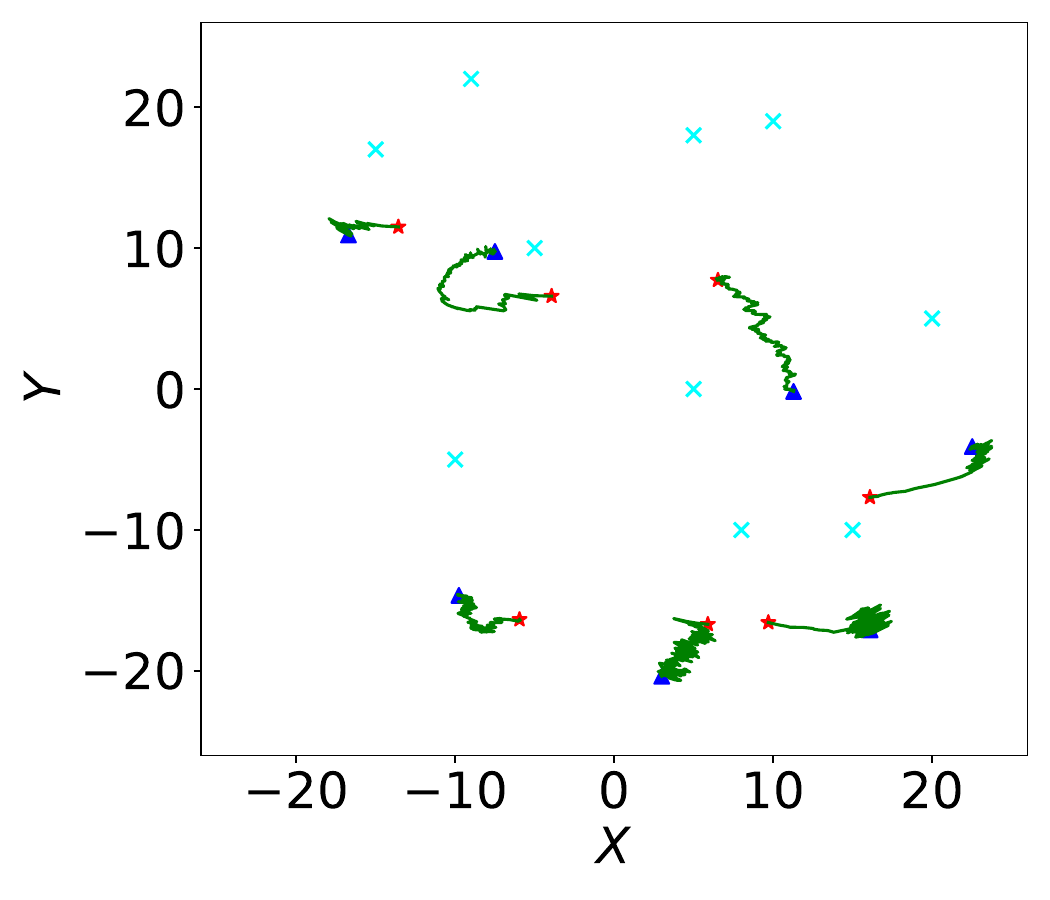}
         \caption{update of sensor locations, $J=200$}
         \label{fig:7 sensor update small q_200}
     \end{subfigure}
     \hfill
     \begin{subfigure}[b]{0.49\textwidth}
         \centering
         \includegraphics[width=0.8\textwidth]{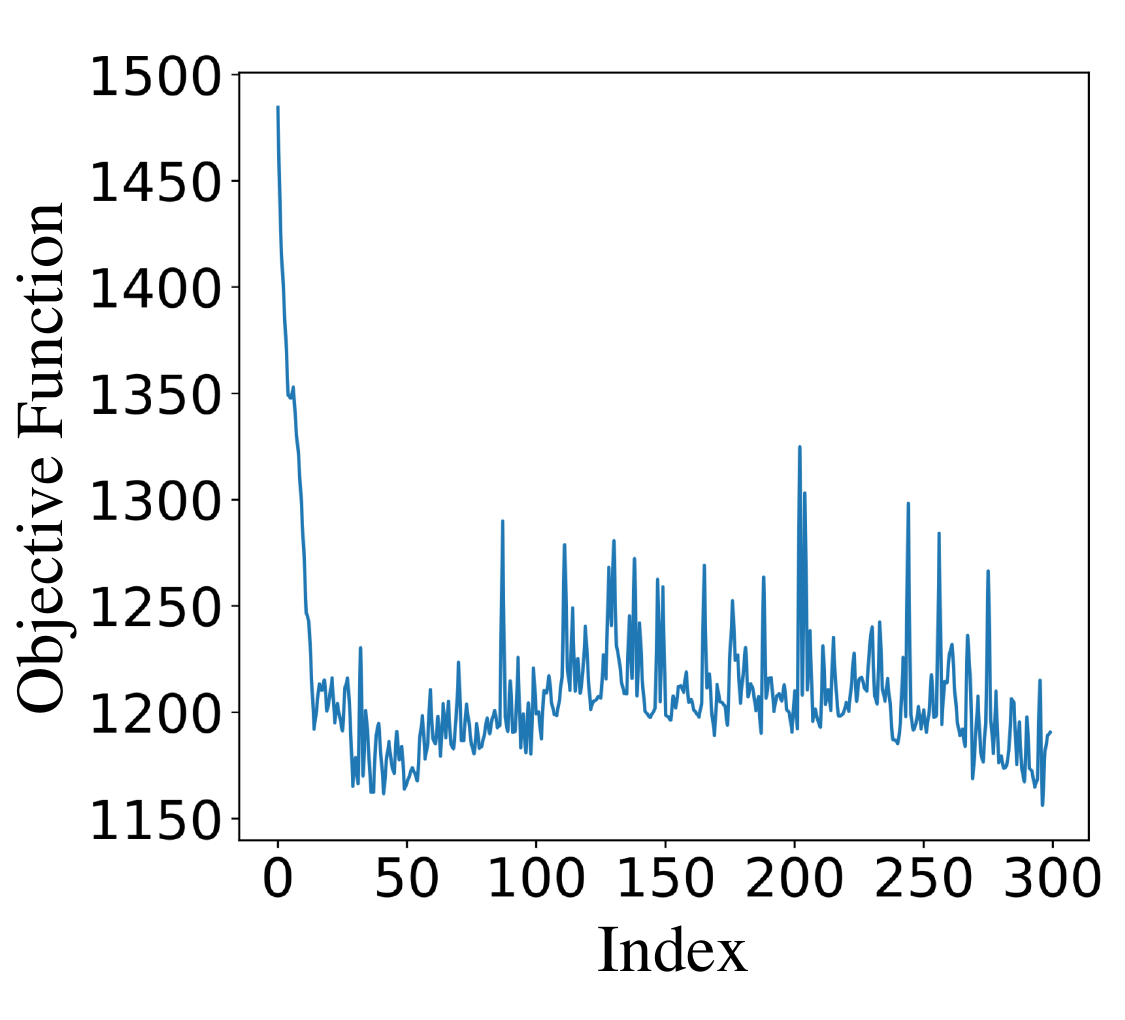}
         \caption{objective value along iterations, $J=200$}
         \label{fig:7 sensor convergence small q_200}
     \end{subfigure}
     \caption{Allocation of 7 sensors for 10 emission sources ($\rho_m=0.00005$).}
     \label{fig:7 sensors q_200}
\end{figure}

\begin{figure}[h!]
     \centering
     \begin{subfigure}[b]{0.49\textwidth}
         \centering
         \includegraphics[width=0.8\textwidth]{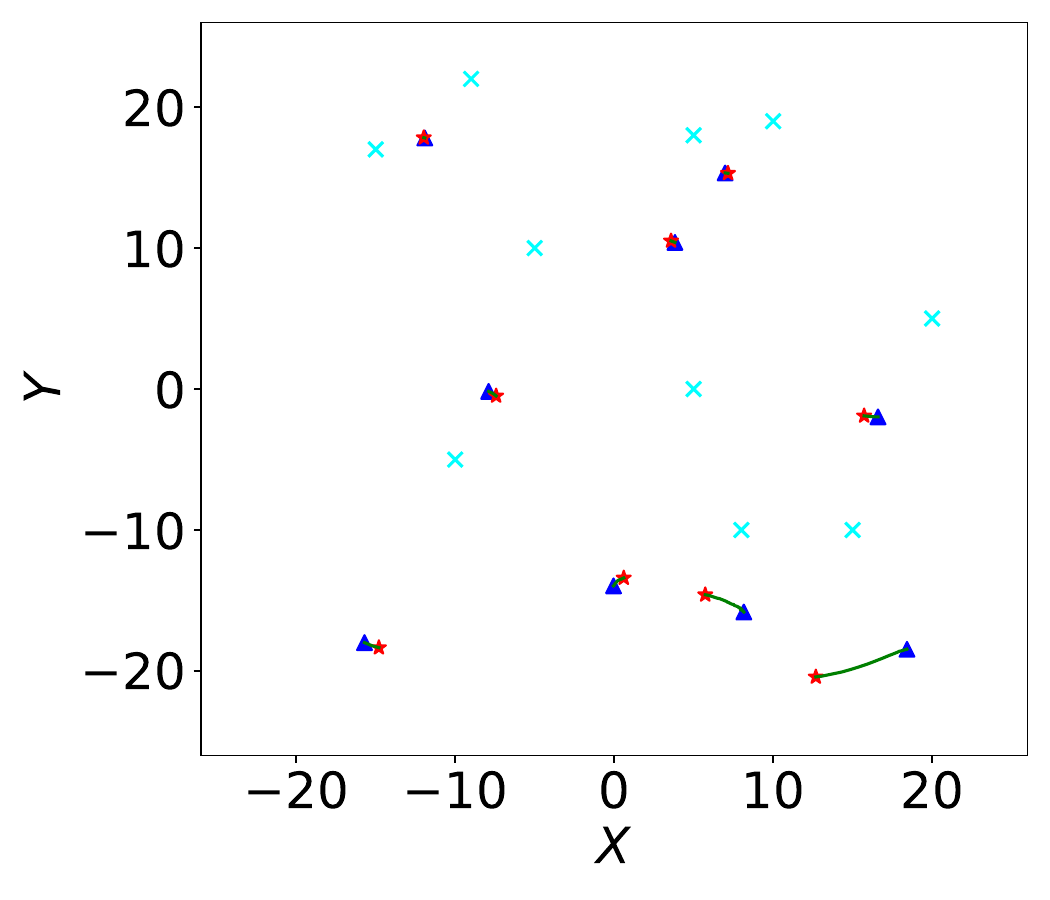}
         \caption{update of sensor locations, $J=1$}
         \label{fig:9 sensor update small q}
     \end{subfigure}
     \hfill
     \begin{subfigure}[b]{0.49\textwidth}
         \centering
         \includegraphics[width=0.8\textwidth]{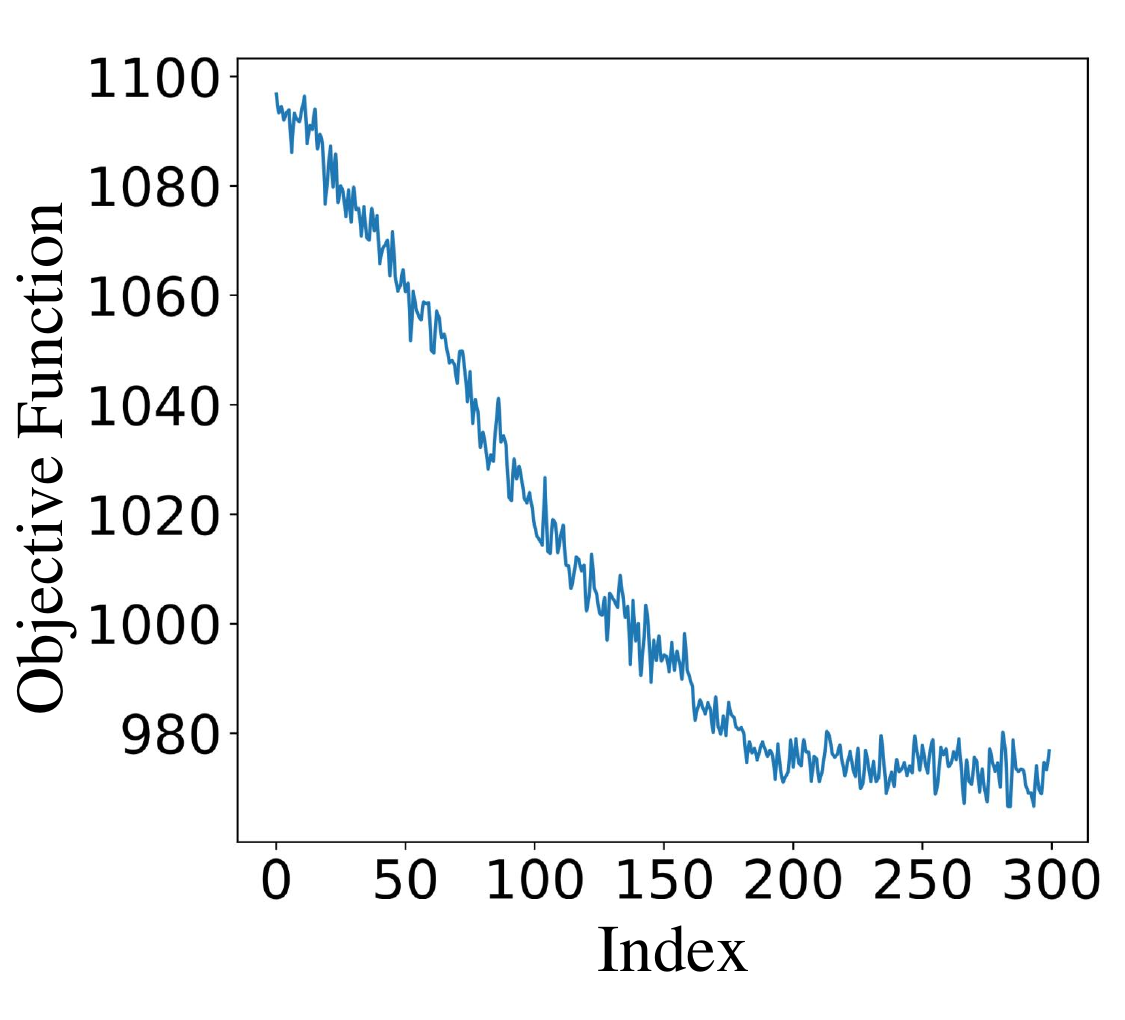}
         \caption{objective value along iterations, $J=1$}
         \label{fig:9 sensor convergence small q}
     \end{subfigure}
     \caption{Allocation of 9 sensors for 10 emission sources ($\rho_m=0.000001$).}
     \label{fig:9 sensors}
\end{figure}

\begin{figure}[h!]
     \centering
     \begin{subfigure}[b]{0.49\textwidth}
         \centering
         \includegraphics[width=0.8\textwidth]{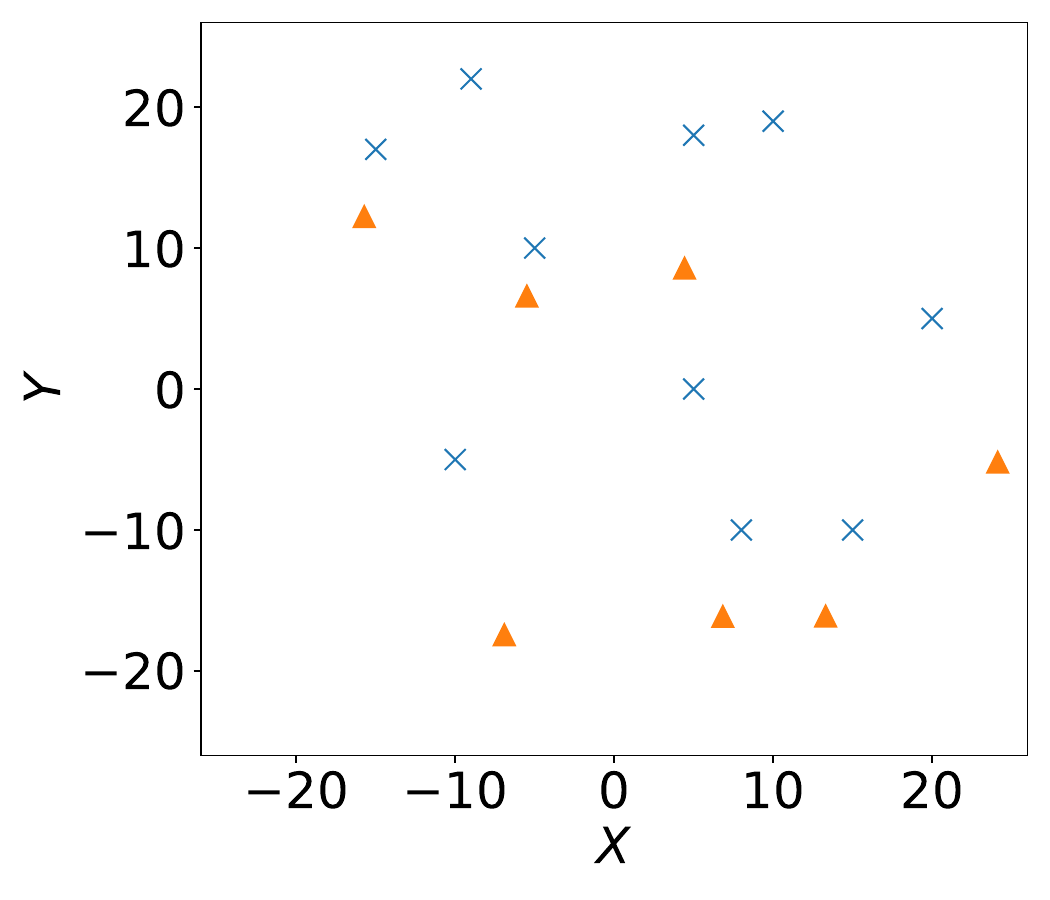}
         \caption{$J=2000$, $\rho_m=0.00005$, $M=300$}
         \label{fig:7 sensor final 1}
     \end{subfigure}
     \hfill
     \begin{subfigure}[b]{0.49\textwidth}
         \centering
         \includegraphics[width=0.8\textwidth]{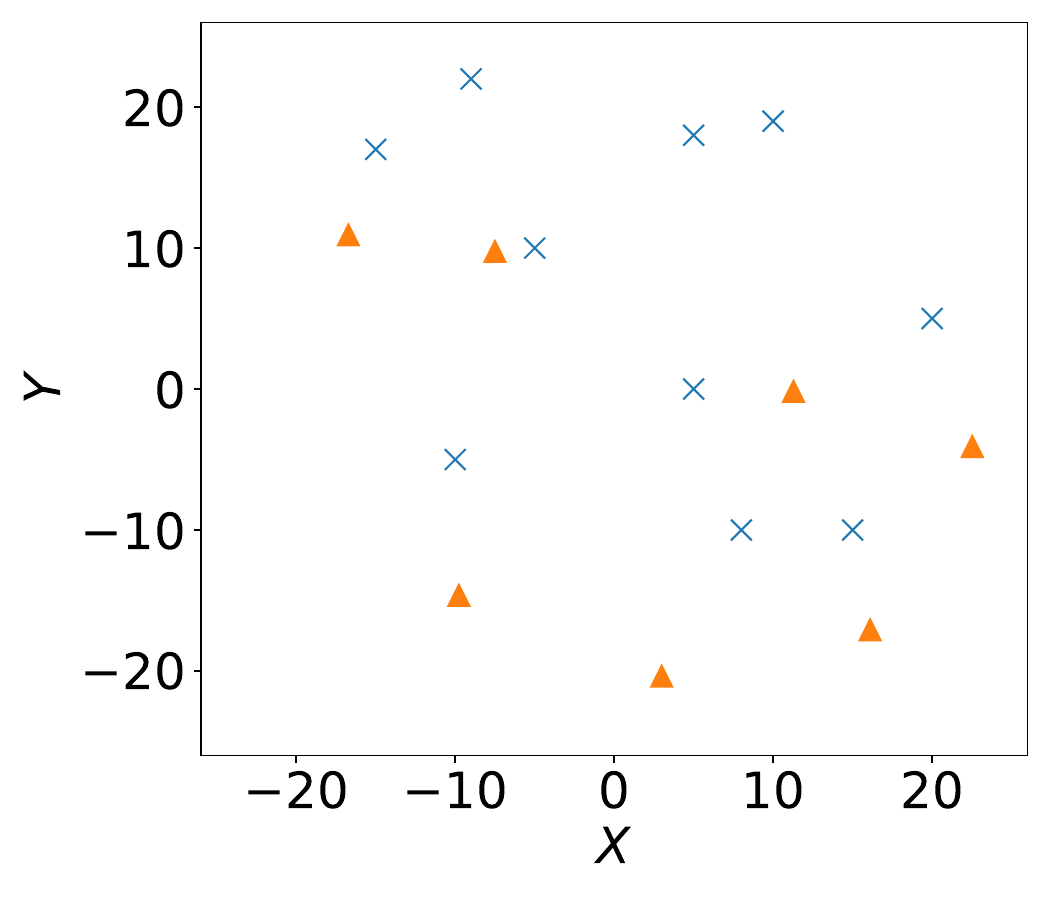}
         \caption{$J=200$, $\rho_m=0.00005$, $M=300$}
         \label{fig:7 sensor final 2}
     \end{subfigure}
     \begin{subfigure}[b]{0.49\textwidth}
         \centering
         \includegraphics[width=0.8\textwidth]{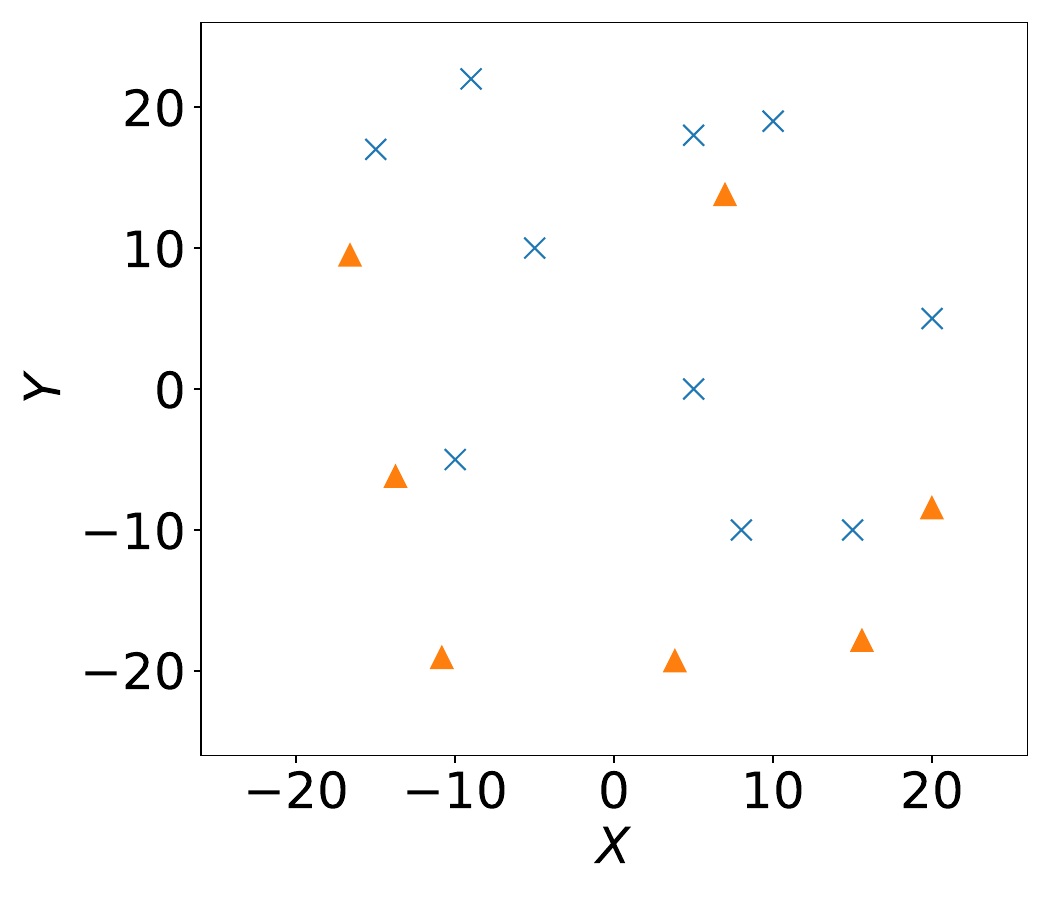}
         \caption{$J=1$, $\rho_m=0.0000005$, $M=2000$}
         \label{fig:7 sensor final small q}
     \end{subfigure}
     \begin{subfigure}[b]{0.49\textwidth}
         \centering
         \includegraphics[width=0.8\textwidth]{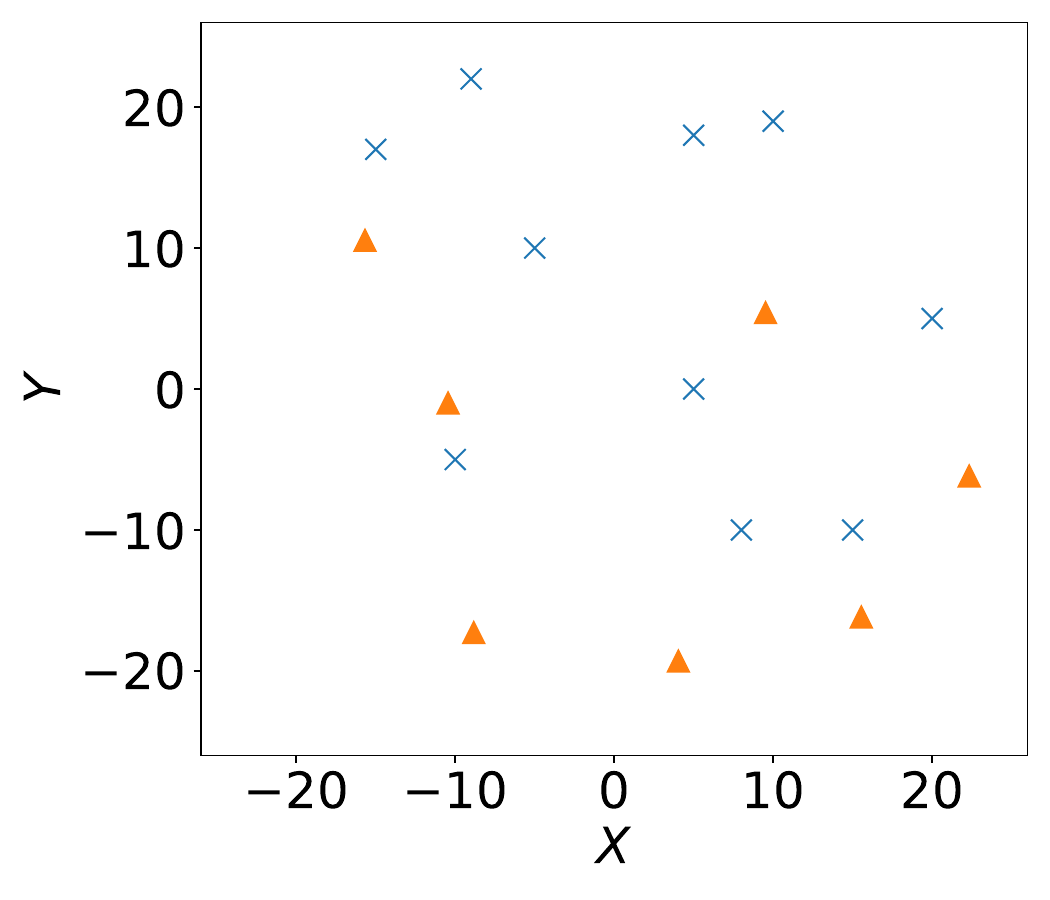}
         \caption{$J=1$, $\rho_m=0.000001$, $M=3000$}
         \label{fig:7 sensor final small q}
     \end{subfigure}
     \caption{Comparison of final designs between different hyperparameters (10 emission sources and 7 sensors, $N'=20$)}
     \label{fig:different hyperparameters}
\end{figure}

\begin{figure}[h!]
     \centering
     \begin{subfigure}[b]{0.49\textwidth}
         \centering
         \includegraphics[width=0.8\textwidth]{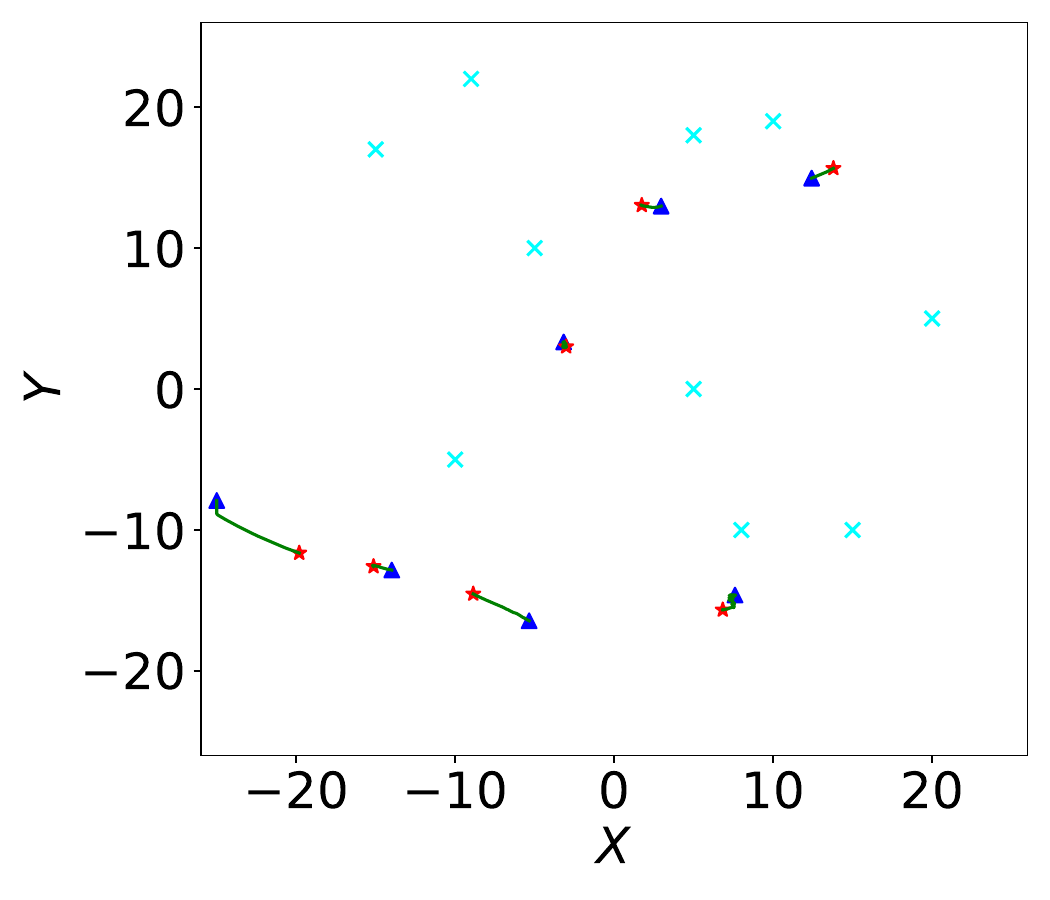}
         \caption{update of sensor locations, $J=2000$}
         \label{fig:random guess 1}
     \end{subfigure}
     \hfill
     \begin{subfigure}[b]{0.49\textwidth}
         \centering
         \includegraphics[width=0.8\textwidth]{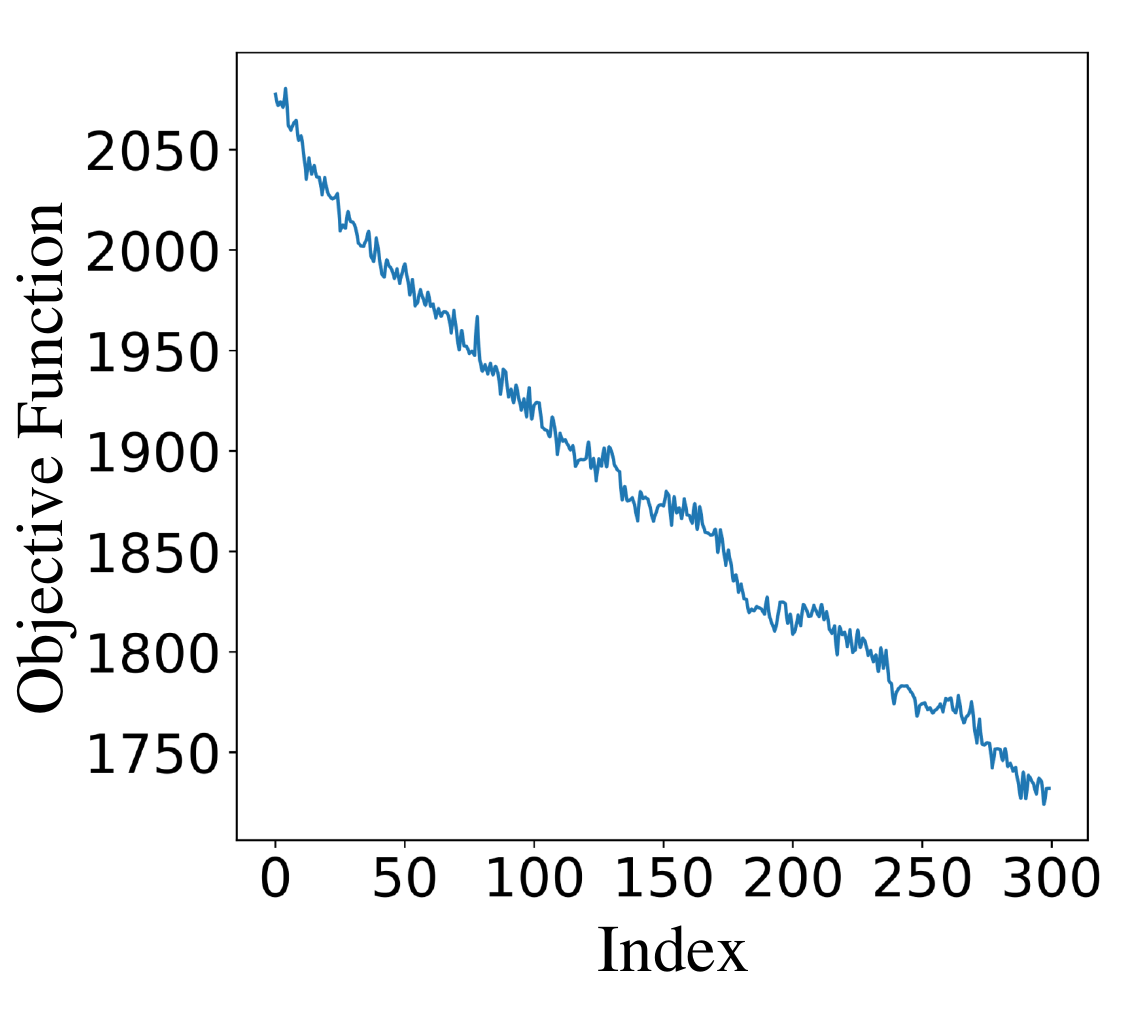}
         \caption{objective value along iterations, $J=2000$}
         \label{fig:random guess 2}
     \end{subfigure}
     \begin{subfigure}[b]{0.49\textwidth}
         \centering
         \includegraphics[width=0.8\textwidth]{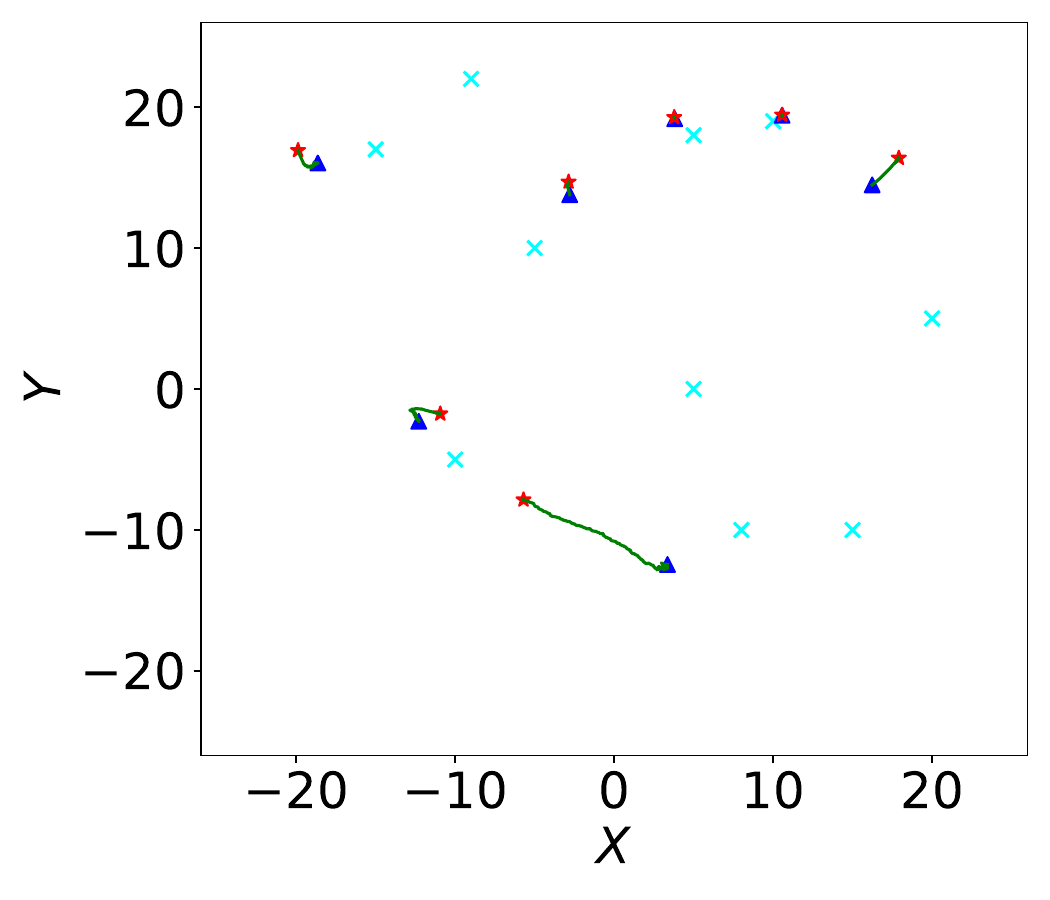}
         \caption{update of sensor locations, $J=2000$}
         \label{fig:random guess 3}
     \end{subfigure}
     \begin{subfigure}[b]{0.49\textwidth}
         \centering
         \includegraphics[width=0.8\textwidth]{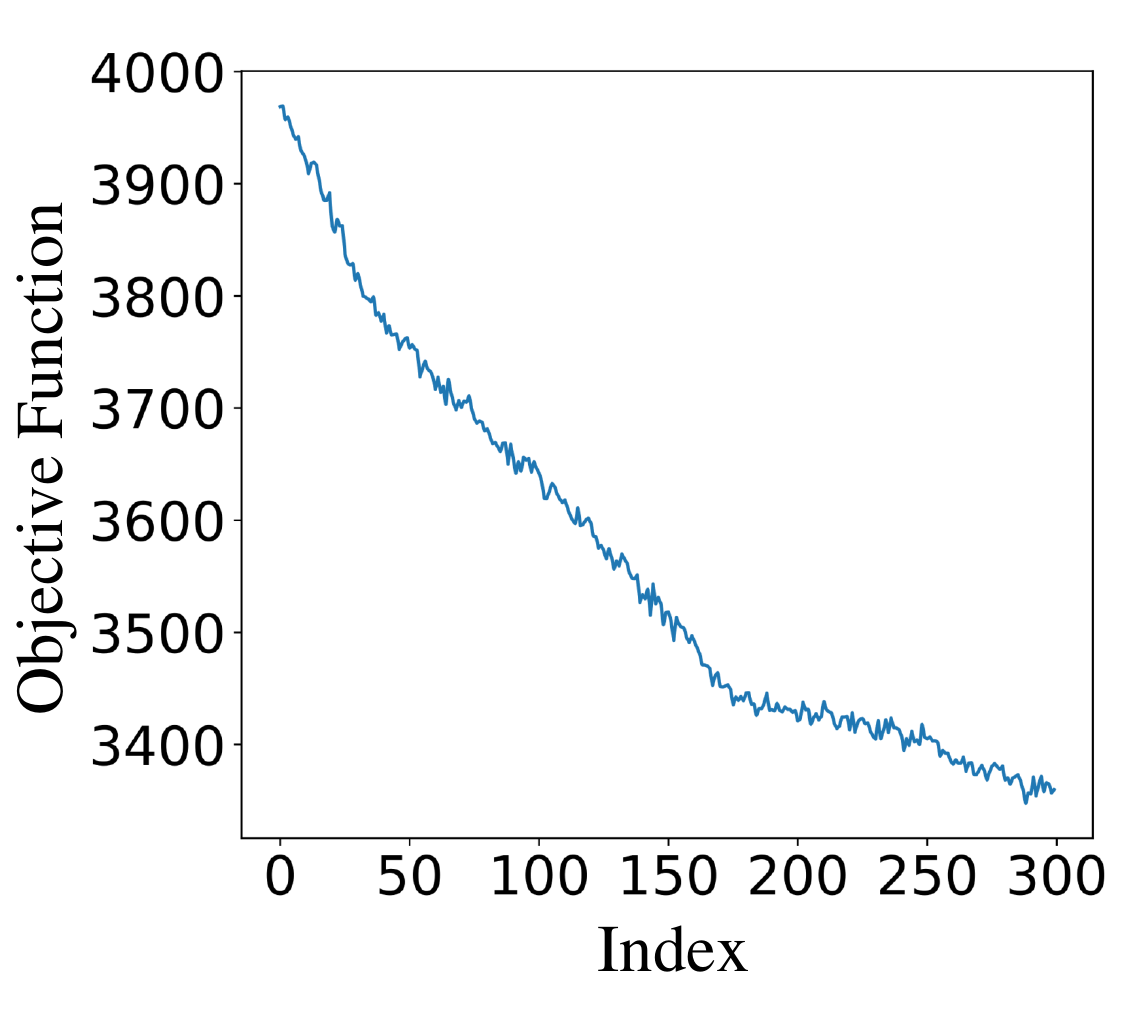}
         \caption{objective value along iterations, $J=2000$}
         \label{fig:random guess 4}
     \end{subfigure}
     \caption{Initial sensor locations by random guess (10 emission sources and 7 sensors)}
     \label{fig:random inital guess}
\end{figure}
\begin{figure}[!]
    \centering
    \includegraphics[width=1\linewidth]{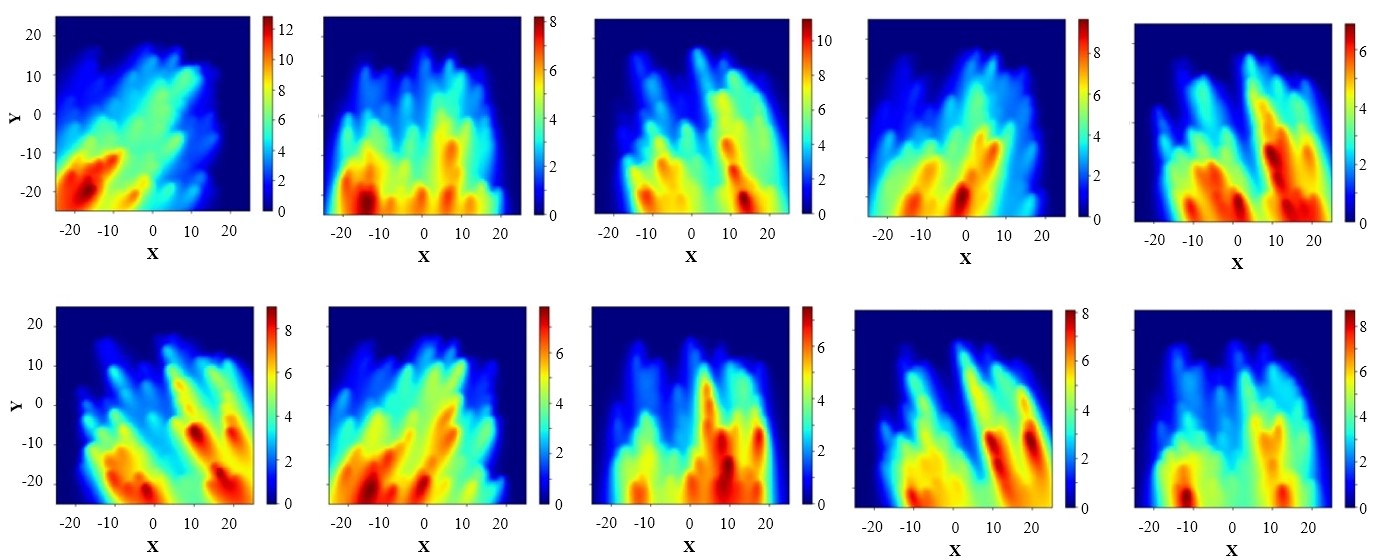}
    \vspace{-8pt}
    \caption{10 sampled scenarios using the Gaussian Plume model and parameters in the paper. It is seen that the concentration field from Gaussian Plume model is complex and depends on both emission parameters and wind conditions.}
    \label{fig:10 emission scenarios}
\end{figure}

\end{document}